\pdfoutput=1

\documentclass[11pt, reqno,preprint]{article}
\pdfoutput=1
\usepackage{jheppub}
\usepackage{epsfig}
\usepackage{amssymb}
\usepackage{amsmath}
\usepackage{mathrsfs}
\usepackage{hyperref}
\usepackage{multirow}
\newcommand{\eq}{\begin{equation}}
\newcommand{\eqe}{\end{equation}}
\newcommand{\eqa}{\begin{eqnarray}}
\newcommand{\eqae}{\end{eqnarray}}
\def\eqn#1{eq.~({\ref{#1}})}

\def\be{\begin{equation}}
\def\ee{\end{equation}}
\def\ba{\begin{eqnarray}}
\def\ea{\end{eqnarray}}

\def\d{\delta}

\def\s{\sigma}

\def\D{\Delta}

\def\L{\Lambda}
\def\nn{\nonumber}

\def\spa#1.#2{\left\langle#1\,#2\right\rangle}
\def\spb#1.#2{\left[#1\,#2\right]}
\def\Tab#1{Table~{\ref{#1}}}

\title{On Three-Algebra and Bi-Fundamental Matter Amplitudes and Integrability of Supergravity}
\author[a]{Yu-tin Huang,}
\author[b]{Henrik Johansson,}
\author[c,d,e]{Sangmin Lee} 

\affiliation[a]{Michigan Center for Theoretical Physics, %Randall Laboratory of Physics, 
University of Michigan, Ann Arbor, MI 48109, USA}
\affiliation[b]{Theory Division, Physics Department, CERN, CH-1211 Geneva 23, Switzerland}
\affiliation[c]{Department of Physics and Astronomy, Seoul National University, Seoul 151-747, Korea}
\affiliation[d]{Center for Theoretical Physics, Seoul National University, Seoul 151-747, Korea}
\affiliation[e]{College of Liberal Studies, Seoul National University, Seoul 151-742, Korea}

\emailAdd{yutinh@umich.edu, henrik.johansson@cern.ch, sangmin@snu.ac.kr}

\abstract{We explore tree-level amplitude relations for SU($N$)$\times$SU($M$) bi-fundamental matter theories. Embedding the group-theory structure in a Lie three-algebra, we derive Kleiss-Kuijf-like relations for bi-fundamental matter theories in general dimension. We investigate the three-algebra color-kinematics duality for these theories. Unlike the Yang-Mills two-algebra case, the three-algebra Bern-Carrasco-Johansson relations depend on the spacetime dimension and on the detailed symmetry properties of the structure constants.  We find the presence of such relations in three and two dimensions, and absence in $D>3$. Surprisingly, beyond six point, such relations are absent in the Aharony-Bergman-Jafferis-Maldacena theory for general gauge group, while the Bagger-Lambert-Gustavsson theory, and its supersymmetry truncations, obey the color-kinematics duality like clockwork. At four and six points the relevant partial amplitudes of the two theories are bijectively related, explaining previous results in the literature. In $D=2$ the color-kinematics duality gives results consistent with integrability of two-dimensional $\mathcal{N}=16$ supergravity: The four-point amplitude satisfies a Yang-Baxter equation; the six- and eight-point amplitudes vanish for certain kinematics away from factorization channels, as expected from integrability.}
\preprint{ MCTP-13-13~~~~~~~CERN-PH-TH/2013-098}

\begin{document}
\maketitle

\pagebreak
%%%%%%%%%%%%%%%%%%%%%%%%%%%%%%%%%%%%%%%%%%%%%%%%%%
\section{Introduction}
%%%%%%%%%%%%%%%%%%%%%%%%%%%%%%%%%%%%%%%%%%%%%%%%%%%%%%%%

In the quest to formulate an action of multiple M2 branes, Bagger, Lambert and Gustavsson (BLG)~\cite{BLG1,BLG2} realized that the gauge-group algebra of the maximally supersymmetric  $\mathcal{N}=8$ theory must have a novel structure given by a natural generalization of the Lie two-bracket, $[\bullet,\bullet]$, to a triple product $[\bullet,\bullet,\bullet]$. Such algebraic structures are called three-algebras (in this terminology two-algebras are ordinary Lie algebras), or triple systems in the mathematical literature.

The $\mathcal{N}=6$ theory of multiple M2 branes was constructed by Aharony, Bergman, Jafferis and Maldacena (ABJM)~\cite{ABJM} as a Chern-Simons-matter (CSm) theory with the physical degrees of freedom transforming in the bi-fundamental representation of a U($N$)$\times$U($N$) Lie-algebra gauge group. Subsequent work~\cite{Gustavsson, BaggerLambert,VanRaamsdonk:2008ft} revealed that such CSm theories, which can be generalized to SU($N$)$\times$SU($M$)~\cite{HLLLP, ABJ}, are equivalent to theories constructed using three-algebras whose structure constants enjoy lesser symmetry compared with that of the BLG theory. 

Recently, the utility of the three-algebra formulation of CSm theory has become apparent in the context of scattering amplitudes. In the work of Bargheer, He and McLoughlin~\cite{Till}, it was shown that for six-point amplitudes in BLG and ABJM theories there exists a three-algebra-based color-kinematics duality, in complete analogy with the two-algebra color-kinematics duality for Yang-Mills theory, discovered by Bern, Carrasco and one of the current authors (BCJ)~\cite{BCJ}. As a consequence of the duality, when the S-matrix of the BLG or ABJM theory is organized into diagrams constructed out of only quartic vertices, then one can find particular representations such that the kinematic numerators of these diagrams satisfy the the same symmetry properties and general algebraic properties as the color factors. In such a representation the numerators acts as if they were part of a kinematic three-algebra, which is dual to the gauge-group three-algebra. 

The BCJ color-kinematics duality for Yang-Mills theory~\cite{BCJ}, which is known to hold at tree-level~\cite{Tree,Tree2} and conjectured to be valid at loop-level~\cite{BCJLoop}, has several interesting consequences. At tree level, it generates non-trivial relations between color-ordered partial amplitudes, so-called BCJ amplitude relations. And, more importantly, once duality-satisfying numerators are found, gravity scattering amplitudes can be trivially constructed by simply replacing the gauge-theory color factors by kinematic numerators of the appropriate theory. This squaring or double-copy property of gravity was proven in ref.~\cite{Bern:2010yg}, for the case of squaring Yang-Mills theory. It has been argued that the color-kinematics duality and double-copy property are intimately tied to the improved ultraviolet behavior of maximal~\cite{BCJLoop, BCDJR}, as well as half-maximal~\cite{N=4SG} supergravity. Remarkably, the color-kinematics duality has interesting consequences and echoes in string theory~\cite{stringtheoryBCJ}.

For the three-algebra-based color-kinematics duality the evidence is still being collected at tree level. Thus far, only four- and six-point amplitudes have been analyzed in the literature.  In ref.~\cite{Till}, the authors obtained the first non-trivial amplitude relations among color-ordered six-point amplitudes of ABJM. Furthermore, by appropriately squaring the duality-satisfying numerators of the six-point amplitudes, they found gravity amplitudes that agree with those of $\mathcal{N}=16$ supergravity of Marcus and Schwarz~\cite{E8}.  In ref.~\cite{HenrikYt}, it was shown that the three-algebra BCJ-relations exist up to six-points for a large class of CSm theories with non-maximal supersymmetry, and each theory squares or double-copies to a corresponding supergravity theory. The fact that three-dimensional supergravity amplitudes can be obtained in this way is fascinating for a variety of reasons. As is already known, these three-dimensional supergravity theories can alternatively be constructed from double copies of three-dimensional super-Yang-Mills (sYM) theories, as follows from the two-algebra color-kinematics duality. Although bewildering, by uniqueness of gravity theories, one should expect that these two distinct constructions give the same answers, as was indeed shown in ref.~\cite{HenrikYt}. Furthermore, for the relevant CSm theories only even-multiplicity amplitudes are non-vanishing, while both even and odd amplitudes exist in three-dimensional sYM theory. Naively, this leads to a conflict between the two double-copy constructions; however,  it is resolved by realizing that odd-multiplicity amplitudes are killed by the enhancement of supergravity R-symmetry in the double copy~\cite{HenrikYt}. Lastly, since the work of Kawai, Lewellen and Tye (KLT)~\cite{KLT}, it has been known that supergravity amplitudes can be obtained from sYM via the relationship between closed and open string amplitudes, in the low energy limit. Interestingly, there is no string-theory understanding as to why such (weak-weak) relations should exist between supergravity and CSm theory. 

As mentioned, a theory with SU($N$)$\times$SU($M$) bi-fundamental matter can be naturally embedded in a three-algebra theory. Lessons learned from three-dimensional CSm theories show that three-algebra embeddings can be extremely useful for organizing the color structure of tree-level amplitudes, as well as exposing hidden structures therein. This calls for a systematic study of scattering amplitudes subject to such embeddings, in general classes of bi-fundamental theories. In this paper we proceed with this analysis. 

The four-indexed structure constants of the three-algebra famously satisfy a fundamental identity, which is the direct generalization of the two-algebra Jacobi identity. Once the color factors of bi-fundamental matter theories are embedded in a three-algebra, this identity allows us to find Kleiss-Kuijf-like partial amplitude relations. These relations are simply a reflection of the over-completeness of the color structures. Since the amplitude relations follow from the algebraic nature of the color factors, they are valid for arbitrary spacetime dimensions. For the special case of $D=3$ BLG theory, or any SU(2)$\times$SU(2) bi-fundamental theory with equal and opposite gauge couplings, there is an important enhanced antisymmetry of the structure constants. This color structure allows for a more refined notion of partial amplitudes, which are inherently non-planar, and satisfy their own type of amplitude relations. Note that, while it is known that SU(2)$\times$SU(2) is the unique finite-dimensional Lie algebra of BLG theory that is free of ghosts~\cite{uniqueBLG}, much of our analysis for the BLG theory will proceed without any assumption about the gauge group, other than the antisymmetry property and fundamental identity of the structure constants.

In this paper we search for evidence of color-kinematics duality in general bi-fundamental theories. Although simple counting at tree level reveals that one can always find kinematic numerators that are dual to the color factors in these theories, the miraculous and useful properties of color-kinematics duality, such as BCJ amplitude relations and double-copy construction of gravity, only emerge in special cases.
We find that three-algebra BCJ amplitude relations and corresponding double-copy formula for supergravity only exist for $D<4$; furthermore, the symmetry properties of the three-algebra structure constants plays a crucial role in $D=3$ and $D=2$ dimensions.  Contrary to previous expectations, we find that only BLG-like theories (totally antisymmetric structure constants) admit BCJ relations for general multiplicity, whereas general (three-dimensional) ABJM-like theories fail at this starting at eight points. The mismatch is surprising given the close relationship between the theories; as is well known, SO(4) BLG theory can be considered to be a special case of ABJM with SU(2)$\times$SU(2) Lie algebra~\cite{VanRaamsdonk:2008ft}. Proper analysis of the generalized-gauge-invariant~\cite{BCJ,BCJLoop} content reveals that the partial amplitudes of the two types of theories are drastically different starting at eight points, whereas the four- and six-point partial amplitudes are simply related. This explains the previous low-multiplicity results in the literature \cite{Till,HenrikYt}, which were simply observations that straightforwardly generalize for BLG-like theories, but not for ABJM-like theories in three dimensions. 

Nevertheless, since BLG amplitudes can always be obtained from the ABJM ones ({\it i.e.} by restricting the gauge group to SU(2)$\times$SU(2)), there is a direct path linking both theories with supergravity:  ABJM theory $\longrightarrow$ BLG theory $\longrightarrow$ $D=3$ supergravity. For BLG theory, and its supersymmetric truncations, we show that BCJ relations exists through at least ten points. And by squaring the duality-satisfying BLG numerators, we have verified that the resulting double-copy results give correct supergravity amplitudes up to at least eight points. 

For kinematics restricted to $D=2$ dimensions, the double copy of BLG theory gives scattering amplitudes of two-dimensional maximal $\mathcal{N}=16$ supergravity. While these generally suffer from severe infrared divergences, even at tree level, there are many finite tree amplitudes that we here consider. For two-dimensional supergravity theories, much like their three-dimensional parents, the bosonic degrees of freedom reside in the scalar sector, whose interactions are described by a non-linear sigma model. For the maximally supersymmetric theory, which is non-conformal, the target space is E$_{8(8)}$/SO(16) (same as its three-dimensional parent). It was realized long ago that the non-linear equations of motion of this theory are equivalent to integrability conditions for a system of linear equations~\cite{Nicolai:1987kz}, and the theory enjoys a hidden infinite-dimensional global E$_{9(9)}$ symmetry~\cite{Nicolai:1998gi}. 

At four and six points, we work directly with double copies of two-dimensional ABJM amplitudes, where the kinematics correspond to color-ordered alternating light-like momenta. Similarly, at eight points we use two-dimensional BLG amplitudes where we have correlated the lightcone direction and superfield chirality. This choice of kinematics allows us to obtain two-dimensional tree amplitudes without encountering explicit collinear and soft divergences. Observing that the two-dimensional four-point tree amplitude in ABJM theory satisfies the Yang-Baxter equation (even though two-dimensional ABJM is not integrable),  the supergravity amplitude inherits this property via the double copy. At six and eight point, even though the reduced ABJM and BLG amplitudes are non-vanishing, the gravity amplitudes obtained from the double-copy construction manifestly vanish. This is consistent with integrality, which implies that the S-matrix vanishes for all values of the momenta except for those corresponding to factorization channels of products of four-point amplitudes. Indeed, all our results are consistent with two-dimensional maximal supergravity theory being integrable. 

Finally, we note that there are a number of interesting amplitude relations that do not fit the usual pattern of such relations, Curiously, in $D=2$ novel BCJ relations emerge for ABJM theory, even beyond six points. Although, surprisingly, the ABJM double-copy prescription generally does not give $D=2$ supergravity amplitudes, since some of the resulting component amplitudes at eight points are nonvanishing, contrary to what the BLG double copy and SYM double copy give. This raises intriguing questions as to what is the role of those BCJ relations, and whether or not this suggest that two-dimensional $\mathcal{N}\ge12$ supergravity can be deformed, contrary to expectations.  Furthermore, we observe that the so-called bonus relations, which arise from improved asymptotic behavior of the amplitude under non-adjacent Britto-Cachazo-Feng-Witten (BCFW) deformations, may give relations beyond those of BCJ. For the six-point ABJM amplitudes, we identify one additional bonus relation that reduces the basis down to three independent amplitudes, the same count as in BLG theory.  Incidentally, via supersymmetry truncation of the six-point BLG amplitudes one recover the same ABJM amplitude identity in disguise as a BCJ relation valid for BLG. However, proper analysis reveals that the true basis is even smaller than what BCJ and bonus relations give. Moreover, the true basis of partial amplitudes is shown to be the of the same size in BLG and ABJM theories up to eight points, suggesting that the amplitudes can be bijectively mapped, contrary to irreversible relationship that is given by the gauge group structures.

The organization is as follows: we begin in section~\ref{SecReview} with a review of the color structure and partial amplitudes of Yang-Mills, bi-fundamental and three-algebra theories. In section~\ref{SecKK}, we discuss the Kleiss-Kuijf-like relations for general bi-fundamental matter theories. In section~\ref{SecBCJ}, we explore the BCJ relations for BLG- and ABJM-type theories, and in section~\ref{SecIntegrability}, we investigate the $D=2$ consequences, including integrability of supergravity.  In section~\ref{SecBonus}, we discuss additional amplitude relations that arises due to the improved large-$z$ BCFW behavior of ABJM.

%%%%%%%%%%%%%%%%%%%%%%%%%%%%%%%%%%%%%%
\section{Color structure and partial amplitudes of bi-fundamental theories}
%%%%%%%%%%%%%%%%%%%%%%%%%%%%%%%%%%%%%%
\label{SecReview}
Scattering amplitudes of gauge theories are given in terms of color-algebra factors tangled with functions of kinematic invariants. Although the color factor of an individual Feynman diagram is readily identified, its kinematic factor is not gauge invariant. As a remedy, it is useful to disentangle the color and kinematics, expressing the full amplitude as an expansion over a basis of color factors with coefficients that are gauge invariant kinematic factors -- referred to as partial amplitudes. The disentanglement is most often done using a basis that is larger than needed, leading to the existence of non-trivial relations among the partial amplitudes. In this section we discuss these issues in the context of bi-fundamental theories.    

%%%%%%%%%%%%%%%%%%%%%%%%%%%%%%%%%%%%%%
\subsection{Color structure of bi-fundamental theories\label{ColorIntro}}
%%%%%%%%%%%%%%%%%%%%%%%%%%%%%%%%%%%%%%

We begin with a brief review of the color structure of tree-level scattering amplitudes in Yang-Mills theory, with or without adjoint matter fields. All physical degrees of freedom are in the adjoint representation of a Lie algebra, implying that the group-theory factors entering an amplitude are built out of the three-indexed structure constants $f^{abc}={\rm Tr}([T^a,T^b]T^c)$. The structure constants are totally antisymmetric, and satisfy a three-term Jacobi identity,
\eq
f^{ca[b}f^{de]c}=0\,.
\eqe
An important consequence of this identity is that not all color factors are independent. It is known that for an $n$-point amplitude, there are only $(n-2)!$ independent color factors. This counting can be understood straightforwardly using a diagrammatic argument, as was done by Del Duca, Dixon and Maltoni~\cite{DDM}. They showed that, starting with the color factor of an arbitrary Feynman diagram, repeated use of the Jacobi identity allows one to rewrite it as a sum over color factors in the following multi-peripheral form:
$$\vcenter{\hbox{\includegraphics[scale=0.8]{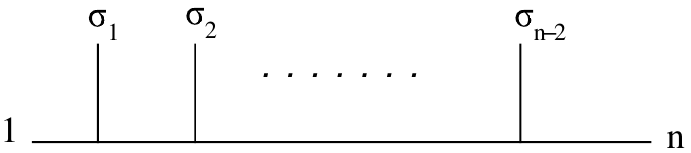}}}\quad \rightarrow \quad f^{a_1a_{\sigma_1}b_1}f^{b_1a_{\sigma_2}b_2}\cdots f^{b_{n-3}a_{\sigma_{n-2}}a_n}\,,$$
where the positions of legs 1 and $n$ are fixed and the $\sigma_i$ represent a permutation of the remaining $n-2$ legs.
For example, color diagrams that have a Y-fork extending from the baseline are reduced using the following diagrammatic Jacobi identity:     
$$\vcenter{\hbox{\includegraphics[scale=0.8]{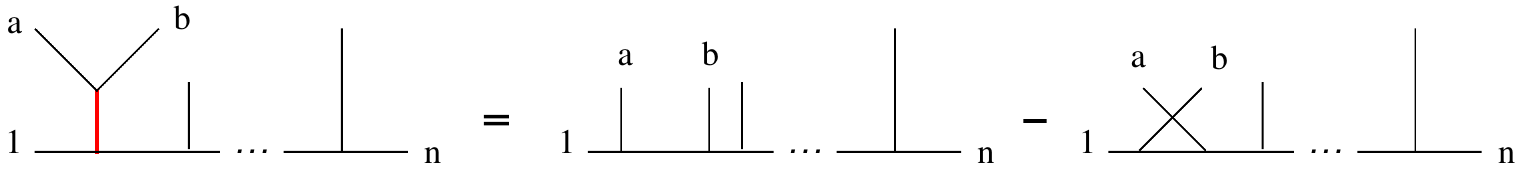}}}\,.$$
There are a total of $(n-2)!$ possible terms in the multi-peripheral representation thus implying the same number of independent color factors. Expanding all color factors in basis, the full color-dressed amplitude is given as~\cite{DDM}
\eq
\mathcal{A}_n=\sum_{\sigma \in S_{n-2}}f^{a_1a_{\sigma_1}b_1}f^{b_1a_{\sigma_2}b_2}\cdots f^{b_{n-3}a_{\sigma_{n-2}}a_n}\,A_{n}(1,\sigma_1,\sigma_2,\ldots,\sigma_{n-2},n)\,,
\eqe
where $A_n$ are partial color-ordered amplitudes, and the sum is over all permutations acting on $(2,\cdots, n-1)$. For convenience, we have suppressed the explicit coupling-constant dependence, as we will do frequently in this paper. The same partial amplitudes appear in an alternative, manifestly crossing symmetric, representation that uses trace factors of fundamental generators. In this trace-basis, the color-dressed amplitude is
\eq
\mathcal{A}_n=\sum_{\sigma \in S_{n-1}}{\rm Tr}(T^{a_{\sigma_1}}T^{a_{\sigma_2}}\cdots T^{a_{\sigma_{n-1}}}T^{a_n})A_{n}(\sigma_1,\sigma_2,\ldots,\sigma_{n-1},n)\,,
\label{traceDecomp}
\eqe
where one sums over all permutations acting on $(1,\cdots, n-1)$. Since this gives $(n-1)!$ terms, the trace-basis is over complete. It implies that the color-ordered partial amplitudes must satisfy special linear relations, known as the Kleiss-Kuijf relations~\cite{KK}. Under these, the color-ordered amplitudes reduce to $(n-2)!$ independent  ones; the same number as the number of independent color factors. For theories with fundamental matter, such as QCD, the color decomposition of the amplitude is more complicated and we will not cover it here (see e.g. ref.~\cite{Kemal} for a detailed discussion of  amplitudes with fundamental quarks).

More exotic matter representations are the focus of this paper.
In particular, we consider SU($N_1$)$\times$SU($N_2$) quiver gauge theories with two bi-fundamental matter fields, indicated by the following quiver diagram.
$$\includegraphics[scale=0.6]{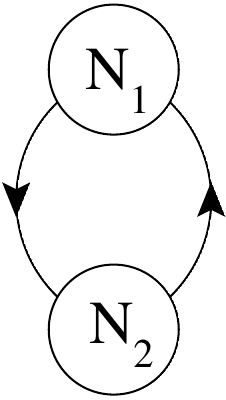}$$
For this discussion we do not restrict ourselves to any particular spacetime dimension. The dynamics of the vector field can be governed either by the usual Yang-Mills Lagrangian or by a Chern-Simons Lagrangian in three dimensions. In either case, our discussion will be restricted to amplitudes that have pure-matter external states. This setup implies that the matter carries conserved charges, and thus only even-multiplicity matter amplitudes exist. 

For bi-fundamental theories the color factors of Feynman diagrams consist of products of delta functions.  Using the notation that the fundamental and anti-fundamental indices of SU($N_1$) and SU($N_2$) are given by $(^\alpha,\; _\alpha)$ and $(^{\tilde\alpha},\; _{\tilde\alpha})$ respectively,  the color dressed amplitude of $k$ matter states $(\Phi_i)^{\tilde\alpha}\,_\alpha$ and $k$ anti-matter states $(\overline{\Phi}_i)^\alpha\,_{\tilde\alpha}$, with $n=2k$, is conveniently decomposed as~\cite{Bargheer:2010hn} 
\eq
\mathcal{A}_n((\overline{\Phi}_1)^{\alpha_{\bar{1}}}\,_{\tilde\alpha_{\bar{1}}}\cdots (\Phi_n)^{\tilde\alpha_n}\,_{\alpha_n})=\hskip-0.5cm\sum_{\sigma \in S_{k},~ \bar \sigma \in \bar S_{k-1}}\hskip-0.5cmA_n(\bar1,\sigma_1, \bar \sigma_1, \ldots, \bar \sigma_{k-1},\sigma_k) \, 
\delta^{\tilde{\alpha}_{\sigma_1}}_{\tilde\alpha_{\bar{1}}}
\cdots
\delta^{\tilde\alpha_{\sigma_k}}_{\tilde\alpha_{\bar \sigma_{k-1}}}\,
\delta^{\alpha_{\bar \sigma_1}}_{\alpha_{\sigma_1}}
\cdots 
\delta^{\alpha_{\bar{1}}}_{\alpha_{\sigma_k}}\,.
\label{BiFunOrder}
\eqe
Here, one sums over all distinct permutations $S_k$ and $\bar S_{k-1}$ acting on even $(2,4,\ldots,2k)$ and odd legs $(\bar3,\bar5,\ldots, \overline{2k-1})$, respectively. We have added a bar on the odd numbers, to emphasize that they are in the conjugate representation.
Partial amplitudes with only Bosonic external states satisfy two-site cyclic symmetry and flip symmetry as follows:
\eqa
A_n(\bar 1,2, \bar 3, 4,\ldots, \overline{n-1},n)&=&A_n(\bar 3, 4,\ldots, \overline{n-1},n,\bar 1,2)\,, \nn \\ 
A_n(\bar 1,2, \bar 3, 4,\ldots, \overline{n-1},n)&=&(-1)^{k-1}A_n(\bar 1, n,\overline{n-1},\ldots, 4, \bar 3,2)\,.
\label{cyclicAndFlip}
\eqae
The amplitude decomposition (\ref{BiFunOrder}) is quite similar to \eqn{traceDecomp} for adjoint amplitudes in Yang-Mills theory. Both the trace factors in \eqn{traceDecomp} and the Kronecker delta functions factors in \eqn{BiFunOrder} lead to a cyclic color-ordered structure of the partial amplitudes. The two-site-cyclic and reversal symmetry imply that there are $(k-1)!k!/2$ distinct color ordered amplitudes. If some of the matter fields satisfy fermonic statistics, the symmetries~(\ref{cyclicAndFlip}) are altered by signs, but the counts remain the same.

As we will demonstrate, the distinct $(k-1)!k!/2$ color-ordered amplitudes are not all independent. The origin of such redundancy is very similar to the redundancy present in Yang-Mills amplitudes: there is an additional structure in the color factors of the theory, which is not manifest in the Kronecker basis, or trace basis.  We will show that by embedding the color factors in a three-algebra construction, the amplitude relation that exposes the redundancy comes from the Jacobi identity (or fundamental identity) satisfied by the three-algebra structure constants.

As the three-algebra will play a central role in our analysis, we here give a lightening review of Lie three-algebras, following the notation of ref.~\cite{BaggerLambert}. Consider two complex vector spaces $V_1$ and $V_2$ with dimensions $N_1$ and $N_2$, respectively. We are interested in linear maps $(M^a)^{\tilde\alpha}\,_\alpha$, such that $M^a:V_1\rightarrow V_2$. Similarly the conjugate maps act as $\overline{M}^{\bar{a}}:V_2\rightarrow V_1$ (we may define $(\overline{M}^{\bar{a}})^{\alpha}\,_{\tilde\alpha}=((M^a)^{\tilde\alpha}\,_\alpha)^\dagger$). As the matrices $M^a$ and $\overline{M}^{\bar{a}}$ carry opposite bi-fundamental indices, the natural product that defines an algebra is the triple product: 
\eq
[M^a,M^b;M^{\bar{c}}] \equiv (M^aM^{\bar{c}}M^b-M^bM^{\bar{c}}M^a )^{\tilde\alpha}_{\phantom{\alpha'}\beta}\equiv f^{ab\bar{c}}_{\phantom{ab\bar{c}} d}\,(M^d)^{\tilde\alpha}_{\phantom{\alpha'}\beta}\,,
\eqe
where 
\eq
f^{ab\bar{c}\bar{d}}=f^{ab\bar{c}}_{\phantom{ab\bar{c}} e} \,h^{e\bar{d}}={\rm Tr}\left[(M^aM^{\bar{c}}M^b-M^bM^{\bar{c}}M^a)\overline{M}^{\bar{d}}\right]\,.
\label{TripleProd}
\eqe
In the above the last index of the four-indexed structure constants has been raised using the metric $ h^{a\bar{b}}={\rm Tr}(M^a\overline{M}^{\bar{b}})$. As shown in ref.~\cite{BaggerLambert} for Chern-Simons matter theory,  the closure of $\mathcal{N}=6$ supersymmetry algebra on the gauge field requires the following fundamental identity: 
\eq
f^{ab\bar{d}}_{\phantom{abd} g}\, f^{cg\bar{e}\bar{f}}
+f^{ba\bar{e}}_{\phantom{abd}g}\, f^{cg\bar{d}\bar{f}}
+f^{*\bar{d}\bar{e}b}_{\phantom{*abd}\bar{g}}\, f^{ca\bar{g}\bar{f}}
+f^{*\bar{e}\bar{d}a}_{{\phantom{*abd}}\bar{g}}\, f^{cb\bar{g}\bar{f}}=0\,,
\label{PreFund}
\eqe
where the $f^{ab\bar{c}\bar{d}}$ are subject to the constraints $f^{ab\bar{c}\bar{d}}=-f^{ba\bar{c}\bar{d}}$ as well as $f^{*\bar{c}\bar{d}ab}=f^{ab\bar{c}\bar{d}}$. Using these properties the fundamental identity can be rewritten as:
\eq
\framebox[11cm][c]{$f^{ab\bar{d}}_{\phantom{abd}g} \, f^{cg\bar{e}\bar{f}}
-f^{ab\bar{e}}_{\phantom{abd}g} \, f^{cg\bar{d}\bar{f}}
-f^{ca\bar{f}}_{\phantom{abd}g} \, f^{bg\bar{d}\bar{e}}
+f^{cb\bar{f}}_{\phantom{abd}g} \, f^{ag\bar{d}\bar{e}}=0\,.$}
\label{fundamental}
\eqe 
As we will shortly see, this will be the fundamental identity that is suitable for ABJM-type bi-fundamental theories.

To see how the usual color structure in bi-fundamental theories can be converted into the above three-algebra construction, let us begin with the first non-trivial amplitude: the four-point amplitude. As mentioned, the bi-fundamental matter fields give a natural ordering to partial amplitudes. Looking at the partial amplitude proportional to the color factor
\eq
c_{1234}=\delta^{\tilde{\alpha}_2}_{\tilde{\alpha}_{\bar{1}}}
\delta^{\alpha_{\bar{3}}}_{\alpha_2}
\delta^{\tilde{\alpha}_4}_{\tilde{\alpha}_{\bar{3}}}
\delta^{\alpha_{\bar{1}}}_{\alpha_4}\,.
\eqe
there are exactly two terms contributing, corresponding to the propagation of the channel with either the SU($N_1$) or SU($N_2$) gauge fields. Pictorially we have
\eq
\nonumber\vcenter{\hbox{\includegraphics[scale=0.9]{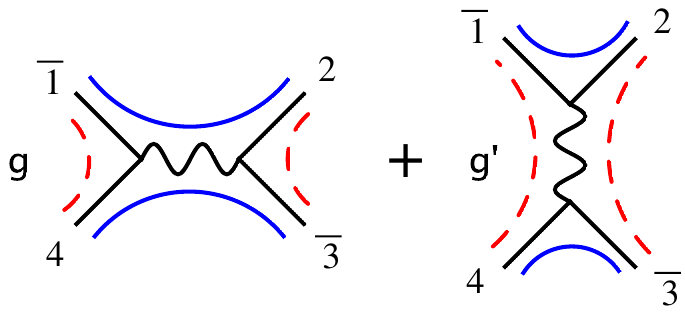}}} \,,
\eqe
where we have used colored dashed/un-dashed lines to indicate the contraction of two distinct color indices, and $(g,g')$ are the coupling constants of the two gauge group. We will assume that $(g,g')$ are the only coupling constants of theory, in which case any potential four-point contact terms can be naturally associated with the two diagrams according to their coupling constant assignment. The full amplitude is
\eq
\mathcal{A}_4(\bar{1},2,\bar{3},4)=c_{1234}\left(g\frac{n_s}{s}+g'\frac{n_t}{t}\right)+(2\leftrightarrow 4)\,.
\label{4ptKin}
\eqe
Now if we identify $g=-g'$, we obtain
\eq
\mathcal{A}_4(\bar{1},2,\bar{3},4)=g(c_{1234}-c_{1432})\left(\frac{n_s}{s}-\frac{n_t}{t}\right)\,.
\eqe
We may think of $(c_{1234}-c_{1432})\sim f^{a_{2}a_{4}\bar{a}_{1}\bar{a}_{3}}$ as the new elementary group theory factor. To make the identification exact, we promote the explicit pairs of fundamental indices into bi-fundamental (or three-algebra) indices by multiplying by conversion coefficients (Clebschs):
\eq
f^{a_{2}a_{4}\bar{a}_{1}\bar{a}_{3}}=
(M^{a_2})^{\tilde\alpha_2}\,_{\alpha_2}
(\overline{M}^{\bar{a}_1})^{\alpha_1}\,_{\tilde\alpha_1}
(M^{a_4})^{\tilde\alpha_4}\,_{\alpha_4}
(\overline{M}^{\bar{a}_3})^{\alpha_3}\,_{\tilde\alpha_3}
\left( 
\delta^{\tilde{\alpha}_2}_{\tilde{\alpha}_{\bar{1}}}
\delta^{\alpha_{\bar{3}}}_{\alpha_2}
\delta^{\tilde{\alpha}_4}_{\tilde{\alpha}_{\bar{3}}}
\delta^{\alpha_{\bar{1}}}_{\alpha_4}-
\delta^{\tilde{\alpha}_4}_{\tilde{\alpha}_{\bar{1}}}
\delta^{\alpha_{\bar{3}}}_{\alpha_4}
\delta^{\tilde{\alpha}_2}_{\tilde{\alpha}_{\bar{3}}}
\delta^{\alpha_{\bar{1}}}_{\alpha_2}\right)\,.
\eqe
This clearly coincides with the definition given in \eqn{TripleProd}. Hence we arrive at the following three-algebra representation for the four-point amplitude:
\eq
\mathcal{A}_4(\bar{1},2,\bar{3},4)=g f^{a_{2}a_{4}\bar{a}_{1}\bar{a}_{3}} \left(\frac{n_s}{s}-\frac{n_t}{t}\right)\,.
\eqe
From here on, we refer to bi-fundamental theories with $g=-g'$ as ABJM-type theories. 

For the another natural choice of couplings, $g=g'$, the color factor and kinematic factor each becomes $s$--$t$ symmetric,
\eq
\mathcal{A}_4(\bar{1},2,\bar{3},4)=g(c_{1234}+c_{1432})\left(\frac{n_s}{s}+\frac{n_t}{t}\right)\,.
\eqe
 This tells us that we should define four-index structure constants that are symmetric under exchange of the two barred indices, as well as the two un-barred ones:
\eq
h^{a_{2}a_{4}\bar{a}_{1}\bar{a}_{3}}=
(M^{a_2})^{\tilde\alpha_2}\,_{\alpha_2}
(\overline{M}^{\bar{a}_1})^{\alpha_1}\,_{\tilde\alpha_1}
(M^{a_4})^{\tilde\alpha_4}\,_{\alpha_4}
(\overline{M}^{\bar{a}_3})^{\alpha_3}\,_{\tilde\alpha_3}
\left( 
\delta^{\tilde{\alpha}_2}_{\tilde{\alpha}_{\bar{1}}}
\delta^{\alpha_{\bar{3}}}_{\alpha_2}
\delta^{\tilde{\alpha}_4}_{\tilde{\alpha}_{\bar{3}}}
\delta^{\alpha_{\bar{1}}}_{\alpha_4}+
\delta^{\tilde{\alpha}_4}_{\tilde{\alpha}_{\bar{1}}}
\delta^{\alpha_{\bar{3}}}_{\alpha_4}
\delta^{\tilde{\alpha}_2}_{\tilde{\alpha}_{\bar{3}}}
\delta^{\alpha_{\bar{1}}}_{\alpha_2}\right)\,.
\eqe
One can verify that the corresponding fundamental identity is given by:
\eq
h^{ab\bar{d}}_{\phantom{abd}g} \, h^{cg\bar{e}\bar{f}}
+h^{ab\bar{e}}_{\phantom{abd}g} \, h^{cg\bar{d}\bar{f}}
-h^{ca\bar{f}}_{\phantom{abd}g} \, h^{bg\bar{d}\bar{e}}
-h^{cb\bar{f}}_{\phantom{abd}g} \, h^{ag\bar{d}\bar{e}}=0\,.
\label{fundamental2}
\eqe 

Finally, for the gauge group SU(2)$\times$SU(2) and $g=-g'$, the structure constants enjoys extra symmetry due to the small rank. In particular, one can now map the bi-fundamental color factors into the four-dimensional Levi-Civita tensor:  
\eq
\epsilon^{\alpha_1\alpha_2}\epsilon^{\alpha_3\alpha_4}\epsilon^{\tilde{\alpha}_1\tilde{\alpha}_4}\epsilon^{\tilde{\alpha}_2\tilde{\alpha}_3}-\epsilon^{\alpha_1\alpha_4}\epsilon^{\alpha_2\alpha_3}\epsilon^{\tilde\alpha_1\tilde\alpha_2}\epsilon^{\tilde\alpha_3\tilde\alpha_4}=\epsilon^{\alpha_1\tilde{\alpha}_1; \alpha_2\tilde{\alpha}_2; \alpha_3\tilde{\alpha}_3;\alpha_4\tilde{\alpha}_4}\,.
\eqe
After soaking up the index pairs with $M$'s one can identify $f^{abcd}=\epsilon^{abcd}$, and the fundamental identity becomes
\eq
f^{fg[d}\,_ef^{abc]e}=0\,,
\label{BLGFun}
\eqe
where the indices are raised or lowered at will.
This is the $\rm{SO}(4)=\rm{SU}(2)\times \rm{SU}(2)$ three-algebra that was constructed by BLG~\cite{Gustavsson, BaggerLambert,VanRaamsdonk:2008ft}. More generally, we will refer to the theory with $g=-g'$ and totally antisymmetric $f^{abcd}$ as BLG-type theories. For SU(2)$\times$SU(2) theories with $g=g'$, there is no enhanced symmetry for the structure constant. 

In the above discussion, it was convenient to identify the coupling constants of the two-gauge field: $g=\pm g'$. Such an identification is most natural if the coupling constant is marginal. For example, in three dimensions we can simply consider supersymmetric Chern-Simons theories. In four dimensions we can consider $\mathcal{N}=2$ supersymmetric theory with $N_1=N_2$, which is superconformal. In any case, we observe the following three interesting scenarios for the bi-fundamental quiver theory:
\eq
\left\{\begin{array}{ll}   g=-g': & f^{ab\bar{c}\bar{d}}=-f^{ba\bar{c}\bar{d}}=-f^{ab\bar{d}\bar{c}}\,,\quad ({\rm ABJM\;type})\\   g=-g': & f^{abcd}=\frac{1}{4!} f^{[abcd]}\,, \hskip 2cm ({\rm BLG\;type})\\ 
g=g': & h^{ab\bar{c}\bar{d}}=h^{ba\bar{c}\bar{d}}=h^{ab\bar{d}\bar{c}}\,.  
\end{array}\right.
\eqe
As it turns out, in the absence of other constraints, the theories with symmetric structure constants $ h^{ab\bar{c}\bar{d}}$, will have parallel properties to the ABJM type theories.\footnote{We have explicitly verified up to eight points that both types of theories have the same number of independent color factors, and the same partial amplitude relations, up to overall signs of the amplitudes.} Therefore in this paper, we focus on the first two cases.

\begin{figure}
\begin{center}
\includegraphics[scale=0.85]{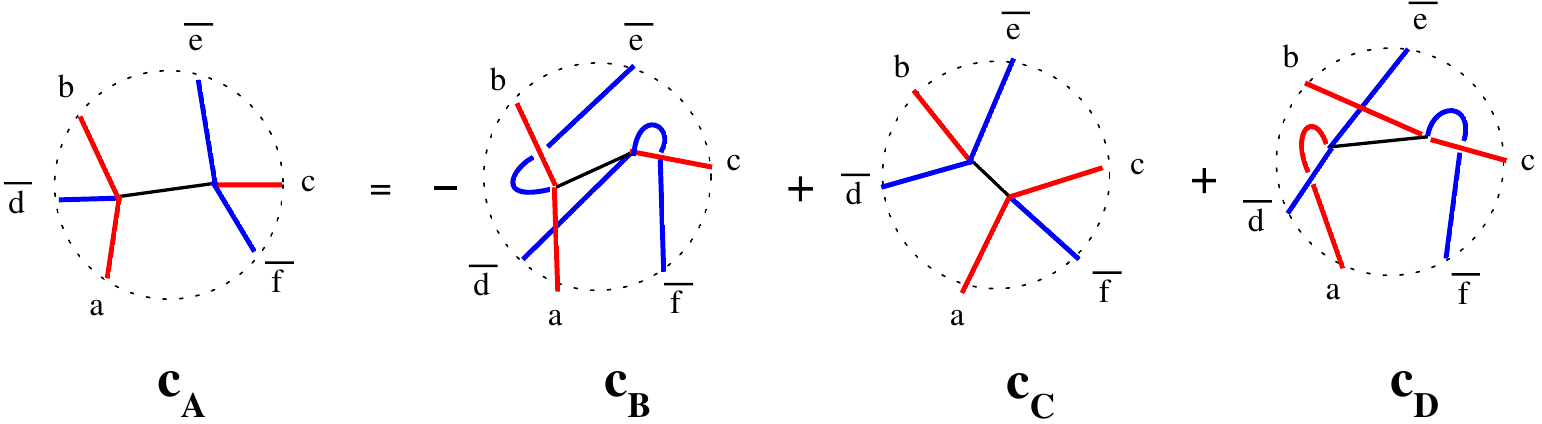}
\caption{A diagrammatic representation of the fundamental identity in \eqn{fundamental}.  The color factors of the diagrams are related through $c_{A}=-c_{B}+c_{C}+c_{D}$. }
\label{FunFig}
\end{center}
\end{figure}

The identification of the group theory structure in terms of three-algebra structure constants allows us to implement the fundamental identity in \eqn{fundamental} to identify the independent color structures. More precisely, since the color factor is now expressed in terms of four-indexed structure constants, from the color point of view it is more natural to use diagrams built out of quartic vertices. Now for each internal line in a given diagram, using \eqn{fundamental} we can relate the color factor of one diagram to three other distinct diagrams as shown in fig.~\ref{FunFig}, where the color factor for each diagram is given as:
\eq
c_{A}=f^{ab\bar{d}}_{\phantom{abd}g} \, f^{cg\bar{e}\bar{f}},\; c_{B}=f^{ab\bar{e}}_{\phantom{abd}g} \, f^{cg\bar{f}\bar{d}},\;c_{C}=
f^{ca\bar{f}}_{\phantom{abd}g} \, f^{bg\bar{d}\bar{e}},\;c_{D}=f^{bc\bar{f}}_{\phantom{abd}g} \, f^{ag\bar{d}\bar{e}}\,.
\eqe
Repeatedly applying such identities reduces the color factors to an independent basis. As we will see in section \ref{SecKK}, the number of independent color factors under the identity in fig.~\ref{FunFig} is smaller than the number of partial amplitudes, and this will lead to linear identities among them similar to the Kleiss-Kuijf identities for Yang-Mills theory.

%%%%%%%%%%%%%%%%%%%%%%%%%%%%%%%%%%%%%%
\subsection{Partial amplitudes for three-algebra theories \label{PartialAmpDef}}
%%%%%%%%%%%%%%%%%%%%%%%%%%%%%%%%%%%%%%
In the above we have introduced partial amplitudes for the bi-fundamental matter S-matrix with Kronecker delta functions as the color prefactor. As discussed, these theories can also be considered to be three-algebra theories, and in that formulation the definition of partial amplitudes becomes a more interesting problem. Given a three-algebra theory, we would like to work out the partial amplitudes using properties that do not rely on explicit matrix representations of the algebra, but only on the symmetry properties and fundamental identity of the four-indexed structure constant.

Here, we will consider a definition of partial amplitudes that utilizes the notion of ``generalized gauge invariance" introduced in refs.~\cite{BCJ,BCJLoop,Bern:2010yg}. Consider the following form of the color-dressed $n$-point tree amplitude:
\eq
\mathcal{A}_n= \sum_{i\in {\rm quartic}}\frac{c_in_i}{\prod_{\alpha_i}s_{\alpha_i}}\,,
\label{fullAmp}
\eqe
where the sums run over all distinct quartic tree diagrams, and the product in the denominator runs over the internal lines in a given diagram. For each internal line, there is a fundamental identity that relates the color structure of four distinct diagrams, as discussed in fig.\ref{FunFig}. This implies that the above representation is given in an overcomplete color basis, and hence there exists a redundancy in the $n_i$ factors, in particular they are gauge dependent. To see this one can deform the numerators using functions $\Delta_i$ satisfying 
\eq
n_i \rightarrow n_i + \Delta_i\,, ~~~{\rm such~that}~~~ \sum_{i\in {\rm quartic}}\frac{c_i \Delta_i}{\prod_{\alpha_i}s_{\alpha_i}}=0 \,.
\label{GaugeDeform}
\eqe
By construction, this ``generalized gauge transformation" will not alter the value of the amplitude in \eqn{fullAmp}. A partial amplitude, $A_n$, can then be defined as the combination of kinematic factors (numerators and propagators) such that $A_n$ is invariant under the transformation in \eqn{GaugeDeform}; that is, $A_n$ must be gauge invariant.

\subsubsection{ABJM-type partial amplitudes \label{PartialAmpDefsSubSubSection}}
  
 Let us first demonstrate that the bi-fundamental partial amplitudes defined in \eqn{BiFunOrder} satisfy the criterion of generalized gauge invariance. Consider the six-point amplitude of an ABJM-type theory; it contains nine quartic-diagram channels, 
 \eq %map old to new  {n1->n1,n3->n2, n4->n3, n5->n4, n9->n5, n7->n6, n8->n7, n6->n8,n2->n9}
 \mathcal{A}_6=
 \frac{c_1n_1}{s_{123}}+
 \frac{c_2n_2}{s_{126}}+
 \frac{c_3n_3}{s_{134}}+
 \frac{c_4n_4}{s_{125}}+
 \frac{c_5n_5}{s_{146}}+
 \frac{c_6n_6}{s_{136}}+ 
 \frac{c_7n_7}{s_{145}}+
 \frac{c_8n_8}{s_{124}}+
 \frac{c_9n_9}{s_{156}}
\,,
 \label{CDressedABJM}
 \eqe
 where $s_{ijl}=(k_i+k_j+k_l)^2$ and the color factors are
 \eqa
 \nonumber &&
 c_1=f^{13\bar{2}}_{\phantom{123} a} \, f^{a5\bar{4}\bar{6}}\,,\;\;
 c_2=f^{35\bar{4}}_{\phantom{235} a} \, f^{a1\bar{6}\bar{2}}\,,\;\;
 c_3=f^{13\bar{4}}_{\phantom{123} a} \, f^{a5\bar{6}\bar{2}}\,,\\
 \nonumber&& 
 c_4=f^{51\bar{2}}_{\phantom{123} a} \, f^{a3\bar{4}\bar{6}}\,,\;\;
 c_5=f^{35\bar{2}}_{\phantom{235} a} \, f^{a1\bar{4}\bar{6}}\,,\;\;
 c_6=f^{13\bar{6}}_{\phantom{123} a} \, f^{a5\bar{2}\bar{4}}\,,\\
 \nonumber&& 
 c_7=f^{51\bar{4}}_{\phantom{123} a} \, f^{a3\bar{6}\bar{2}}\,,\;\;
 c_8=f^{35\bar{6}}_{\phantom{235} a} \, f^{a1\bar{2}\bar{4}}\,,\;\;
 c_9=f^{51\bar{6}}_{\phantom{123} a} \, f^{a3\bar{2}\bar{4}}\,.
 \label{ff}
 \eqae
The numerators $n_i$ can, for example,  be built from Feynman diagrams: three-point vertices are combined to form non-local four-point vertices while the six-point contact terms are split up into two four-point vertices. Using \eqn{TripleProd} to convert  \eqn{CDressedABJM} to a trace basis one finds that the color-ordered partial amplitude is given by
 \eqa
 A^{\rm ABJM}(\bar 1,2,\bar 3,4,\bar 5,6)=\frac{n_1}{s_{123}}+\frac{n_2}{s_{126}}+\frac{n_9}{s_{156}}\,.
 \label{ABJM6pt}
 \eqae
As expected, the amplitude is simply the sum over the planar diagrams in the canonical color ordering. To show that this combination is gauge invariant, let us, for example, consider the the fundamental identity $c_1+ c_5 - c_7 - c_9  = 0$. Since we can freely add $(c_1+ c_5 - c_7 - c_9)\chi$ to \eqn{CDressedABJM}, it implies that any potential partial amplitude must be invariant under the following deformation:
\eq
n_i \rightarrow n_i + \Delta_i\,: ~~~\Delta_1=s_{123}\chi,\; \Delta_5=s_{146}\chi,\,\Delta_7=-s_{145}\chi,\;\Delta_9=-s_{156} \chi,\;\Delta_{2,3,4,6}=0\,. 
\eqe
It is straightforward to see that $A^{\rm ABJM}(\bar 1,2,\bar 3,4,\bar 5,6)$ is indeed invariant under the above transformation. Similarly, for all other such transformations the partial amplitude is invariant. From this it follows that the  color-ordered definition of partial amplitude is indeed invariant under generalized gauge transformations. 

For higher-point bi-fundamental amplitudes the details are exactly the same. The partial amplitudes that are invariant under generalized gauge transformations are precisely the color ordered ones, $A^{\rm ABJM}(\bar 1,2,\bar 3,\ldots, n)$, which can be expresses as a sum over distinct planar diagrams in the given color ordering.

\subsubsection{BLG-type partial amplitudes}
  
For BLG-type theories, the partial amplitudes can be defined in several ways.  Firstly, one can use color-ordered partial amplitudes that arise in the the bi-fundamental formulation of BLG. However, since the four-indexed structure constants enjoy more symmetry than is manifest in this formulation, such a representation will not be invariant under the generalized gauge transformation that arises from the BLG fundamental identity in \eqn{BLGFun}. Similarly, the bi-fundamental formalism does not take into account the relations of the finite-rank gauge group ${\rm SU}(2)\times{\rm SU}(2)={\rm SO}(4)$. 

Taking this into account, we can define two additional types of partial amplitudes for BLG-like theories:  Partial amplitudes that use the three-algebra formulation, taking into account the total antisymmetry and fundamental identity of the structure constants, or partial amplitudes that are directly defined for SO(4) theories. Up to six points, these two definitions will agree, but starting at eight points they lead to different partial amplitudes. 

Using the properties of the structure constants one can show that the simplest generalized-gauge-invariant partial amplitudes at six points have four channels. For example,
 \eqa
 A^{\rm BLG}(1,2,3,4,5,6)=\frac{n_1}{s_{123}}+\frac{n_2}{s_{126}}+\frac{n_9}{s_{156}}+\frac{n_{10}}{s_{135}}\,,
 \label{BLG6pt1}
 \eqae
 where the last term arose from a diagram $c_{10}n_{10}/s_{135}$, with $c_{10}=f^{135}_{\phantom{135} a} \, f^{a246}$,  that we added to the generic amplitude in~\eqn{fullAmp}.
Comparing this with \eqn{ABJM6pt}, we see that  $A^{\rm BLG}(1,2,3,4,5,6)$ contains one additional non-planar (with respect to the canonical ordering) channel. The absence of planar partial amplitudes is consistent with BLG being an inherently non-planar theory. 

Even though the partial amplitudes have distinct characteristics, the BLG and ABJM amplitude can be non-trivially related after proper identification of states and channels. Projecting the BLG states on chiral multiplets $(\bar1,2,\bar3,4,\bar5, 6)$ ({\it i.e.} supersymmetry truncation), one can set $n_{10}$ to zero. This is because the $s_{135}$ channel does not correspond to any physical propagating states (similarly $c_{10}$ is zero in a bi-fundamental formulation of BLG).  Since $n_{10}=0$ we can identify the amplitudes in \eqn{BLG6pt1} and  \eqn{ABJM6pt}: $A^{\rm BLG}(\bar1,2,\bar3,4,\bar5,6)= A^{\rm ABJM}(\bar1,2,\bar3,4,\bar5, 6)$. However, there are also other ways to assign chiralities to the external states. For example, the amplitude $A^{\rm BLG}(\bar1,2,4,\bar3,\bar5,6)$ does not have an alternating chiral pattern to its entries. In fact, this amplitude also contains four channels, but none of them correspond to $n_{10}/s_{135}$. So this BLG amplitude cannot be identified with a single ABJM amplitude after eliminating $n_{10}$. Instead it can be expressed as a sum over two ABJM amplitudes. Before writing the relation down, let us consider how many different partial BLG amplitudes there are at six points.

Using the symmetry properties of the $n_i$'s one can show that $A^{\rm BLG}(1,2,3,4,5,6)$ has a 48-fold permutation symmetry. Thus there are only $6!/48=15$ distinct partial amplitudes. We may rearrange the particle labels so that the symmetries are manifest. We define
 \eqa
 A^{\rm BLG}_{\rm SO(4)}(\{1,4\},\{2,5\},\{3,6\})\equiv A^{\rm BLG}(1,2,3,4,5,6),
 \label{BLG6pt2}
 \eqae
 where the amplitude is insensitive to the ordering inside the curly or round brackets, only the paring of the legs carry significance. This partial amplitude is exactly what one obtains in the SO(4) decomposition at six points, thus the subscript. Its color factor is precisely $\delta^{a_1}_{a_4}\delta^{a_2}_{a_5}\delta^{a_3}_{a_6}$, where the $a_i$ are SO(4) indices.

Having exposed the symmetries of the BLG partial amplitudes, it is clear that there are two distinct types of projections onto chiral states. For these, we have the two types of relations
 \eqa
 A^{\rm BLG}_{\rm SO(4)}(\{\bar 1,4\},\{2,\bar5\},\{\bar 3,6\})&=& A^{\rm ABJM}(\bar1,2,\bar3,4,\bar5, 6)\,, \nn \\
 A^{\rm BLG}_{\rm SO(4)}(\{\bar1,\bar3\},\{2,\bar5\},\{4,6\})&=&-A^{\rm ABJM}(\bar1,2,\bar3,4,\bar5, 6)-A^{\rm ABJM}(\bar1,2,\bar3,6,\bar5, 4)\,,
 \label{BLGfromABJM6pts}
 \eqae
and all other non-vanishing chiral projections are related to these by simple relabeling. Needless to say, these relations give a very convenient way of obtaining BLG partial amplitudes. 

For higher-point amplitudes, we can easily write down an SO(4) decomposition of BLG theory using the fact that the structure constants are given in terms of the Levi-Civita tensor $f^{abcd}=\epsilon^{abcd}$. The contraction of an even number of Levi-Civita tensors reduces to only Kronecker deltas, and the color factors are easy to enumerate in this case. 
This occurs for multiplicity $n=6,10,14,\ldots$, giving a decomposition into $(n-1)!!$ partial amplitudes,
\eq
\mathcal{A}_{4j+2}=\sum_{\sigma\in S_n/(Z_2^k S_k)}\delta^{a_{\sigma_1}}_{a_{\sigma_2}}\cdots \delta^{a_{\sigma_{n-1}}}_{a_{\sigma_{n}}}A^{\rm BLG}_{\rm SO(4)}(\{\sigma_1,\sigma_2\}, \cdots, \{\sigma_{n-1},\sigma_n\} )\,,
\eqe
where the sum is over all distinct pairings of legs. For multiplicity $n=2k=8,12,16,\ldots$ the  SO(4)  color factors are built out of an odd number of Levi-Civita tensors, which can be easily reduced to linear combinations of a single Levi-Civita tensor times a number of delta functions. However, the set of all such color factor satisfy further relations, making this overcomplete basis somewhat inconvenient for defining partial amplitudes. Nevertheless, for completeness of the discussion, we have counted the number of distinct partial amplitudes such a decomposition would generate, assuming a complete subset of these color factors would be used.  We find that the count is 91 at eight points. Furthermore, by analyzing the set of all $\epsilon^{abcd}(\delta^e_f)^{k-2}$ factors up to $k=5$, we find a pattern for the basis size that agrees with $C(k)(C(k)-1)/2$, where $C(k)=(2k)!/(k!(k+1)!)$ are the Catalan numbers. See Table \ref{BLGtable} for a summary of the counts.

\iffalse
For multiplicity $n=2k=4j$, with $j$ being integer, the color structure is somewhat less elegant because of the appearance of an explicit Levi-Civita tensor. Together with the Kronecker deltas one can identify $n(n-2)  (n-1)!!/4!$ distinct color factors; unfortunately, not all are independent {\color{red} (the independent count seems to be $C(2j)(C(2j)-1)/2=1,91,8646,\ldots$, where $C(k)$ are the Catalan numbers)}. For the decomposition in this basis the full amplitude is given as
 \eq
\mathcal{A}_{4j}=\hskip-6mm\sum_{\sigma\in S_n/(Z_2^{k-2} S_{k-2}S_4)}\hskip-6mm\epsilon^{a_{\sigma_1}a_{\sigma_2}a_{\sigma_3}a_{\sigma_4}} \delta^{a_{\sigma_5}}_{a_{\sigma_6}}\cdots \delta^{a_{\sigma_{n-1}}}_{a_{\sigma_{n}}}
A^{\rm BLG}_{\rm SO(4)}(\{\sigma_{1},\sigma_{2},\sigma_{3},\sigma_4\};\{\sigma_5,\sigma_6\}, \cdots, \{\sigma_{n-1},\sigma_{n}\})\,,
\eqe
where the first quadruplet entry of the partial amplitude is aligned with the Levi-Civita indices, and the remaining pairs with the Kronecker deltas. For both cases, the entires in the partition are orderless. 
\fi

Instead of relying on explicit SO(4) properties, we will in this paper use partial amplitudes derived from only the defining properties of the BLG three-algebra structure constants: total antisymmetry and the fundamental identity. As is  well known, the SO(4) group is a special case, and not the most generic group that obeys the BLG three-algebra. Albeit all other known examples are groups with Lorentzian signature. Nevertheless, for later applications to color-kinematics duality we will need this more general setup. 

Using generalized gauge invariance one can show that the simplest partial amplitude at eight points contains 30 channels. It is explicitly given as the 30-fold orbit of one quartic diagram
\eq
A^{\rm BLG}(1,2,3,4,5;6,7,8)=\sum_{Z_5 (1,2,3,4,5) \times S_3 (6,7,8)} \frac{n_{16;237;458}}{s_{237} s_{458}}\,,
\label{EightPointDef}
\eqe
where $n_{16;237;458}$ is the kinematic numerator that goes together with the $f^{16ab}f^{a237}f^{b458}$ color factor. The partial amplitude has manifest cyclic symmetry in the first five entries, and full permutation symmetry in the last three. Furthermore, it has a non-manifest flip antisymmetry $A^{\rm BLG}(1,2,3,4,5;6,7,8)=-A^{\rm BLG}(5,4,3,2,1;6,7,8)$ that follows from the symmetries of $n_{16;237;458}$. This implies that the  amplitude has a 60-fold symmetry, and that there are $8!/60=672$ distinct such partial amplitudes.

Like before, we can relate the chiral projections of these amplitudes with the ABJM ones. One can simply identify the 30 diagrams with channels that also appear in $A^{\rm ABJM}$.  Each ABJM partial amplitude contains 12 planar quartic channels, and appropriate linear combinations of these give the distinct projections of the BLG amplitudes. Using the symmetries of 
$A^{\rm BLG}$ one obtains six distinct chiral projections. Three of these are given by
\eqa
A^{\rm BLG}(\bar1\bar3\bar5\bar72;468) &=& 
 A^{\rm ABJM}(\bar14\bar36\bar58\bar72) + A^{\rm ABJM}(\bar14\bar38\bar56\bar72) + A^{\rm ABJM}(\bar16\bar34\bar58\bar72) \nn \\
 && \null+ A^{\rm ABJM}(\bar16\bar38\bar54\bar72) + A^{\rm ABJM}(\bar18\bar34\bar56\bar72) + A^{\rm ABJM}(\bar18\bar36\bar54\bar72)\,, \nn \\
A^{\rm BLG}(\bar32\bar5\bar78;\bar146) &=& 
 A^{\rm ABJM}(\bar12\bar34\bar56\bar78) + A^{\rm ABJM}(\bar12\bar36\bar54\bar78) + A^{\rm ABJM}(\bar12\bar54\bar36\bar78) \nn \\
 &&\null+ A^{\rm ABJM}(\bar12\bar56\bar34\bar78) - A^{\rm ABJM}(\bar18\bar34\bar76\bar52) - A^{\rm ABJM}(\bar18\bar36\bar74\bar52)\,,\nn \\
A^{\rm BLG}(2\bar3\bar5\bar78;\bar146) &=& -A^{\rm ABJM}(\bar12\bar34\bar56\bar78) - A^{\rm ABJM}(\bar12\bar36\bar54\bar78)\,,
\label{BLGfromABJM8pts}
\eqae
where we have suppressed the label delimiters for notational compactness, as we will do frequently in what follows. In addition to the above there are three more projections given by the chiral conjugates of \eqn{BLGfromABJM8pts}.

The existence of the relations (\ref{BLGfromABJM8pts}) show that the BLG and ABJM partial amplitudes can be mapped to each other in a surjective fashion. Simple diagrammatic analysis shows that the map cannot be straightforwardly inverted. ABJM amplitudes cannot be obtained by simple linear combinations of the BLG amplitudes with constant coefficients.\footnote{However, this does not preclude the existence of an inverse linear map that involves momentum dependent coefficients. In section~\ref{beyondBCJ} we argue that such relations exists.} In the following sections we explain that this is due to the fact that the bases under Kleiss-Kuijf-like relations are of different size for the two types of theories, starting at eight points. We will see that this property has important consequences for the color-kinematics duality. 

%%%%%%%%%%%%%%%%%%%%%%%%%%%%%%%%%%%%%%
\section{KK-like identities for SU($N$)$\times$SU($M$) bi-fundamental theories\label{SecKK}}
%%%%%%%%%%%%%%%%%%%%%%%%%%%%%%%%%%%%%%

In this section, we will discuss the Kleiss-Kuijf-type amplitude relations for bi-fundamental theories. The amplitude relations arise from the properties of the four-indexed structure constants. We have a number of situations to consider.

%%%%%%%%%%%%%%%%%%%%%%%%%%%%%%%%%%%%%%
 \subsection{ABJM type: $f^{ab\bar{c}\bar{d}}=-f^{ba\bar{c}\bar{d}}=-f^{ab\bar{d}\bar{c}}$}
%%%%%%%%%%%%%%%%%%%%%%%%%%%%%%%%%%%%%%
\label{ABJMcountingSection}
We begin by counting the number of distinct color factors that we encounter in the three-algebra formulation. This is equivalent to counting the number of quartic (four-valent) diagrams, for an $n$-point amplitude. Starting with a root, say leg $i$, the remaining parts of the diagram can be viewed as three lower-point branches of sizes $2m_1$, $2m_2$ and $(n-2m_1-2m_2+2)$. Pictorially, we have the following tree graph:
\eq
\includegraphics[scale=0.8]{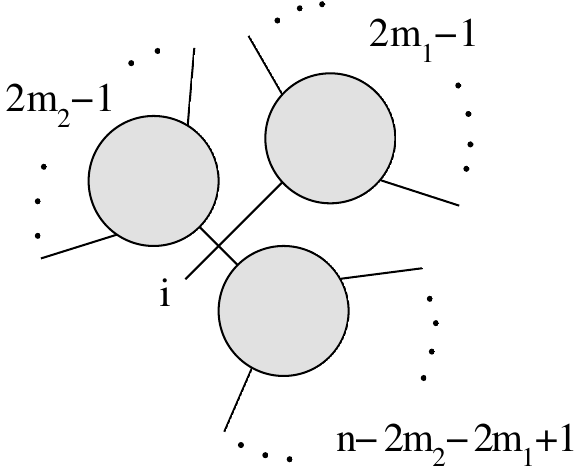}
\eqe
This organization allows us to iteratively express the number of diagrams in terms of the function $\nu(n)$, 
\eqa
\nonumber \nu(2k)&=&\frac{1}{2!}\sum_{m_1=1}^{k-1}\sum_{m_2}^{k-m_1}\left(\begin{array}{c}k-1 \\ m_1\end{array}\right)\left(\begin{array}{c}k \\ m_1-1\end{array}\right)\left(\begin{array}{c}k-m_1-1 \\ m_2-1\end{array}\right)\left(\begin{array}{c}k-m_1+1 \\ m_2\end{array}\right)\\
&&\hskip 2.2cm \null \times \nu(2m_1) \nu(2m_2) \nu(2k-2m_1-2m_2+2)\,.
\label{QuarticDia}
\eqae
with $\nu(2)=1$. The combinatorial factors in the first line correspond to distinct ways of distributing the bi-fundamental and anti-bi-fundamental fields on the first two branches.  A closed formula is given by
\eq
\nu(2k)=\frac{(3k-3)!k!}{(2k-1)! 2^{k-1}}\,.
\eqe 

Given that we know the total number of quartic diagrams, we can now simply count the number of such diagrams in each color-ordered partial amplitude. Trivially, this number is equal to the average count for all color-ordered amplitudes. In turn this average must be proportional to the total number of diagrams; thus we have the following relations:
\eq
\# [A_n(1,2,\cdots,n)]=\frac{2}{(k-1)!k!}\sum_{\rm \sigma} \# [A_n(\sigma)]=\frac{2}{(k-1)!k!} 2^{k-2} \# [\mathcal{A}_n]\,.
\eqe
where $\#[\cdots]$ counts the number of quartic diagrams in each amplitude, with  $\# [\mathcal{A}_n]=\nu(2k)$, and the sum runs over the $(k-1)!k!/2$ partial amplitudes. The factor of $2^{k-2}$  appears due to the overcount of identical diagrams; the overcount is two-fold for each vertex due to the antisymmetry property of the ABJM structure constants. This tells us that the number of quartic diagrams in a color-ordered $(2k)$-point amplitude is exactly $(3k-3)!/((2k-1)!(k-1)!)$.

The last count that we can simply deduce for ABJM theory, is the total number of fundamental identities.  
For this count, we observe that a fundamental identity acts on contractions of two structure constants, or equivalently two vertices connected by a propagator. Thus, we can associate the fundamental identities with the internal lines of the quartic diagrams. On one hand, this leads to an overcount by a factor of four since each identity relates four diagrams. On the other hand, we have not yet taken into account that there are several distinct fundamental identities that act on a given contraction of two structure constants. Careful counting gives that for the ABJM-type structure constants there are four distinct ways of obtaining a fundamental identity from a single $f^{ab\bar c \bar d}f^{de\bar g\bar h}$. Thus these two factors of four cancel out; and the total number of fundamental identities is equal to the number of propagators per diagram times the number of diagrams, that is, $(k-2)\nu(2k)$. 

Having counted the distinct four-term relations between different color factors, we would proceed by determining the number of independent identities. It is the number of independent fundamental identities that carries real significance. Using these we can reduce the color factors to a basis. The size of this basis tells us how many independent partial amplitudes there exists. Unfortunately, we have found no means for determining this count to all multiplicity, hence, we resort to case-by-case counting at low number of external legs.  After explicitly solving the $(k-2)\nu(2k)$ identities we obtain that the number of independent color factors for ABJM-type bi-fundamental matter theories are 1, 5, 57, 1144 for multiplicity 4, 6, 8, 10, respectively. 
We summarize the above discussion in Table~\ref{kk-table}.

\begin{table}[htbp]
\begin{tabular}{|l|c|c|c|c|c|c|}
\hline
% after \\ : \hline or \cline{col1-col2} \cline{col3-col4} ...
  external legs & 4 & 6& 8& 10&12& $n=2k$ \\ \hline
  quartic diagrams & 1 & 9& 216& 9900&737100&  $\nu(2k)=\frac{(3k-3)!k!}{(2k-1)!(2!)^{k-1}}$ \\
  partial amplitudes & 1 & 6& 72& 1440& 43200& $\frac{k! (k-1)!}{2}$ \\
  diagrams in partial amplitude & 1 & 3 & 12 & 55 & 273 & $\frac{(3k-3)!}{(2k-1)!(k-1)!}$ \\
  fundamental identities & 0 & 9 & 432 & 29700 & 2948400 & $(k-2) \nu(2k)$ \\
 independent color factors & 1 & 5 & 57 & 1144 & $*$ & $*$ \\
\hline
\end{tabular}
\caption{Counts of distinct diagrams, partial amplitudes and fundamental identities for ABJM theories. The count for the reduced color basis, or equivalently the basis under Kleiss-Kuijf-like relations, is given on the last line. An asterisk signify an undetermined quantity. }
\label{kk-table}
\end{table} 

An important message from Table~\ref{kk-table} is that, starting at six points, the number of independent color factors is less than the number of color-ordered partial amplitudes. As mentioned in section~\ref{ColorIntro}, this will lead to non-trivial amplitude identities for the color-ordered amplitudes which we now discuss.
 
%%%%%%%%%%%%%%%%%%%%%%%%%%%%%%%%%%%%%%
 \subsubsection{KK identities for ABJM-type bi-fundamental theories}
 %%%%%%%%%%%%%%%%%%%%%%%%%%%%%%%%%%%%%%
 \label{KKsection}
In Yang-Mills theory, the fact that the partial amplitudes are more prolific than the independent color factors leads to so-called Kleiss-Kuijf identities between the partial amplitudes. For the bi-fundamental matter theories we find similar types of relations. 

We now demonstrate that the partial amplitudes of ABJM-type theories satisfy the following KK-like amplitude relation:
 \begin{equation}
 \sum_{i\in S_k}A_{2k}(\bar{1},i_1,\bar{3},i_2,\bar{5},...,\overline{2k-1}, i_{k})=0
 \label{KK}
 \end{equation}
where the sum runs over all permutations of the even sites, and all the states are Bosonic.  For Fermionic states, one must properly weight the sum by the usual statistical signs. Note that, by conjugation and relabeling, a similar relation exists where the even legs are fixed and the odd legs are permuted. 

To see that such an identity arises from the purely group-theoretical structure, let us analyze the first nontrivial example: the six-point amplitude. We take the odd and even sites to coincide with the barred and un-barred representation respectively.  At six-point, as indicated in Table~\ref{kk-table}, there are a total of six independent partial amplitudes. Expressing them in terms of the kinematic factors defined in \eqn{CDressedABJM}, they are given as:
 \eqa
 \label{Rep}  \nonumber 
A_6(\bar 1,2,\bar 3,4,\bar 5,6)&=&\frac{n_1}{s_{123}}+\frac{n_2}{s_{126}}+\frac{n_9}{s_{156}}\,,\quad ~~ 
A_6(\bar 1,4,\bar 3,6,\bar 5,2)=\frac{n_3}{s_{134}}+\frac{n_4}{s_{125}}+\frac{n_8}{s_{124}}\,,\\
A_6(\bar 1,6,\bar 3,2,\bar 5,4)&=&\frac{n_5}{s_{146}}+\frac{n_6}{s_{136}}+\frac{n_7}{s_{145}}\,,\quad ~~ 
A_6(\bar 1,4,\bar 3,2,\bar 5,6)=-\frac{n_3}{s_{134}}-\frac{n_5}{s_{146}}-\frac{n_9}{s_{156}}\,, ~~~ \\ \nonumber  
A_6(\bar 1,6,\bar 3,4,\bar 5,2)&=&-\frac{n_2}{s_{126}}-\frac{n_4}{s_{125}}-\frac{n_6}{s_{136}}\,,\quad 
A_6(\bar 1,2,\bar 3,6,\bar 5,4)=-\frac{n_1}{s_{123}}-\frac{n_7}{s_{145}}-\frac{n_8}{s_{124}}\,,
 \eqae
 where the relative signs of the diagrams can be deduced from the definitions of the corresponding color factors in \eqn{CDressedABJM}.
From this representation one can immediately see that the identity~(\ref{KK}) is satisfied,
\eqa
\nonumber &&A_6(\bar 1,2,\bar 3,4,\bar 5,6)+A_6(\bar 1,4,\bar 3,6,\bar 5,2)+A_6(\bar 1,6,\bar 3,2,\bar 5,4)\\
&&\null+A_6(\bar 1,4,\bar 3,2,\bar 5,6)+A_6(\bar 1,6,\bar 3,4,\bar 5,2)+A_6(\bar 1,2,\bar 3,6,\bar 5,4)=0\,.
\label{Id}
\eqae
It may not be obvious to the reader that this example follows from pure group theory. However, note that the representation (\ref{Rep}) simply follows from the generic color-dressed amplitude in \eqn{CDressedABJM} after converting the three-algebra color factors into a trace basis, using \eqn{TripleProd}. Because of the generality of the derivation, the identity is valid for any bi-fundamental theory that admits ABJM-like three-algebra structure constants. 

We now prove the validity of \eqn{KK} for general multiplicity in the specific context of ABJM theory; however, we expect it to hold for generic ABJM-type bi-fundamental theories due the underlying group theoretic nature. For the proof we proceed in two different ways.  In the following, we will use a specific BCFW recursion developed for ABJM theory~\cite{Gang}. In the next section, we will give another proof based on the the twistor-string-like integral formula proposed in~\cite{SangminYt}.

The BCFW proof is established inductively, similar to what was done for Yang-Mills theory in~\cite{Feng:2010my}. The trivial inductive case is the four-point amplitude: it is simply the reflection symmetry of the partial amplitudes, 
 \eq
 A_4(\bar{1},2,\bar{3},4)+A_4(\bar{1},4,\bar{3},2)=0\,,
 \label{cyclic4pt}
 \eqe
which follows from the analogous symmetry relation of four-point color factors,  $f^{a_1 a_3 \bar a_2 \bar a_4}+f^{a_1 a_3 \bar a_4 \bar a_2}=0$. 

The general case in \eqn{KK} follows if we can relate the lower-multiplicity cases with the given case. This can always be done by expressing the individual partial amplitudes in their BCFW representations. For example, at six points, choosing legs 1 and 6 as the globally BCFW-shifted legs, we have:\footnote{Due to the quadratic dependence on the BCFW deformation parameter, the BCFW representation is schematically given as 
$A_n=\sum_{i} HA_LA_R/P^2_{ i}$,
where $H$ is a kinematic invariant that depends on the factorization channel $P^2_{\rm i}$~\cite{Gang}. Here, since we are collecting terms that have the same factorization channel, $H$ appears as a common factor and hence is suppressed throughout the discussion.  }
\eqa
\nonumber &&A_6(\bar{1}2\bar{3}4\bar{5}6)=\frac{A_4(\hat{\bar{1}}2\bar{3}\hat{P})A_4(\hat{\bar{P}}4\bar{5}\hat{6})}{s_{123}},\quad A_6(\bar{1}6\bar{3}2\bar{5}4)=\frac{A_4(\hat{\bar{P}}\hat{6}\bar{3}2)A_4(\bar{5}4\hat{\bar{1}}\hat{P})}{s_{145}}, \\
&&A_6(\bar{1}4\bar{3}2\bar{5}6)=\frac{A_4(\hat{\bar{1}}4\bar{3}\hat{P})A_4(\hat{\bar{P}}2\bar{5}\hat{6})}{s_{134}},\quad A_6(\bar{1}6\bar{3}4\bar{5}2)=\frac{A_4(\hat{\bar{P}}\hat{6}\bar{3}4)A_4(\bar{5}2\hat{\bar{1}}\hat{P})}{s_{125}}, \\
\nonumber &&A_6(\bar{1}4\bar{3}6\bar{5}2)=\frac{A_4(\hat{\bar{1}}4\bar{3}\hat{P})A_4(\hat{\bar{P}}\hat{6}\bar{5}2)}{s_{134}}+\frac{A_4(\hat{\bar{P}}2\hat{\bar{1}}4)A_4(\bar{3}\hat{6}\bar{5}\hat{P})}{s_{124}}+\frac{A_4(\bar{5}2\hat{\bar{1}}\hat{P})A_4(\hat{\bar{P}}4\bar{3}\hat{6})}{s_{125}}, \\
\nonumber &&A_6(\bar{1}2\bar{3}6\bar{5}4)=\frac{A_4(\hat{\bar{1}}2\bar{3}\hat{P})A_4(\hat{\bar{P}}\hat{6}\bar{5}4)}{s_{123}}+\frac{A_4(\hat{\bar{P}}4\hat{\bar{1}}2)A_4(\bar{3}\hat{6}\bar{5}\hat{P})}{s_{124}}+\frac{A_4(\bar{5}4\hat{\bar{1}}\hat{P})A_4(\hat{\bar{P}}2\bar{3}\hat{6})}{s_{145}}\,,
\eqae
where we use $P$ to denote the on-shell intermediate state in the factorization channel, and for notational brevity  we have suppressed the delimiters in the amplitude arguments. One can see that by combining the common propagators into pairs, each pair cancels precisely due to \eqn{cyclic4pt}. Thus the six-point KK identity, \eqn{KK}, is simply a consequence of \eqn{cyclic4pt}. 

We can now set up the inductive proof in more detail. We assume that \eqn{KK} holds for all $(n-2j)$-point amplitudes, with $0<j<n/2$. To prove the $n$-point identity, we shift legs $1$ and $n$ in \eqn{KK} and express all color ordered amplitudes in terms of the BCFW expansion. One can collect all terms that have the common BCFW channel, say $s_{1i_13i_2\cdots i_{j-1}(2j-1)}$, and a fixed ordering of the even labels $( i_1,i_2,\ldots, 2j-1)$ in each partial amplitude under consideration. Because of this fixed ordering, the contribution to the residue of this pole, in these amplitudes, is simply a common $A_{2j}$ factor multiplied by distinct $A_{n+2-2j}$ amplitudes of various orderings. The sum of these contributions then simply cancels due to the (\ref{KK}) identity that has been assumed for $A_{n+2-2j}$. Since $j$ and $(i_1,i_2,\cdot\cdot,i_{j-1})$ where kept generic in this argument, the vanishing holds for all terms in the BCFW representation, completing the proof of \eqn{KK}.

Might \eqn{KK} capture all the KK-like identities that one can deduce from the color structure of ABJM-type bi-fundamental theories? The answer is no. As explained, for an $(2k)$-point amplitude, there will be $k!(k-1)!/2$ independent amplitudes under reflection and cyclic permutation. Using up the $(k-1)(k-2)/2$ independent relations contained in \eqn{KK}, we are left with $(k-1)(k-2)(k!(k-3)!-1)/2$ superficially independent amplitudes. Comparing this with the true number of independent color factor, which was explicitly computed up to ten points using the fundamental identity (see Table \ref{kk-table}), we have a discrepancy starting at eight points:
\eqa
\left(\begin{array}{c|c|c|c}{\rm multiplicity\;}~ & 6 & 8 & 10 \\\hline {\rm independent\;}c_i & 5 & 57 & 1144\\\hline {\rm eq.}(\ref{KK}) & 5 & 69 & 1434\end{array}\right)\,.
\label{kk-counting-1}
\eqae
Even if we take into account the conjugate identities of \eqn{KK}, we only find three more independent ones at eight points, and this does not make up for the discrepancy of 12 identities. So it is clear that something new is required beyond six points. Indeed, we find the following new eight-point identity (for Bosonic external states):
\eqa
\nonumber &&\null-A_8(\bar{1}2\bar{3}4\bar{5}8\bar{7}6)-A_8(\bar{1}4\bar{3}2\bar{5}8\bar{7}6)+A_8(\bar{1}6\bar{3}8\bar{5}2\bar{7}4)+A_8(\bar{1}6\bar{3}8\bar{5}4\bar{7}2)\\
\nonumber &&\null+A_8(\bar{1}6\bar{7}8\bar{3}2\bar{5}4)+A_8(\bar{1}2\bar{7}6\bar{3}8\bar{5}4)+A_8(\bar{1}4\bar{7}6\bar{3}8\bar{5}2)+A_8(\bar{1}6\bar{7}8\bar{3}4\bar{5}2)\\
\nonumber &&\null+A_8(\bar{1}6\bar{7}2\bar{3}8\bar{5}4)+A_8(\bar{1}6\bar{7}4\bar{3}8\bar{5}2)+A_8(\bar{1}6\bar{3}2\bar{7}8\bar{5}4)+A_8(\bar{1}6\bar{3}4\bar{7}8\bar{5}2)\\
%\nonumber 
&&\null+A_8(\bar{1}6\bar{3}8\bar{7}2\bar{5}4)+A_8(\bar{1}6\bar{3}8\bar{7}4\bar{5}2)+A_8(\bar{1}4\bar{3}6\bar{7}8\bar{5}2)+A_8(\bar{1}2\bar{3}6\bar{7}8\bar{5}4)=0\,,%\\
\eqae
%
\iffalse
\eqa
\nonumber &&~~~\,A_8(\bar{1}2\bar{3}4\bar{5}8\bar{7}6)-A_8(\bar{1}4\bar{3}2\bar{5}8\bar{7}6)+A_8(\bar{1}6\bar{3}8\bar{5}2\bar{7}4)-A_8(\bar{1}6\bar{3}8\bar{5}4\bar{7}2)\\
\nonumber &&\null+A_8(\bar{1}6\bar{7}8\bar{3}2\bar{5}4)+A_8(\bar{1}2\bar{7}6\bar{3}8\bar{5}4)-A_8(\bar{1}4\bar{7}6\bar{3}8\bar{5}2)-A_8(\bar{1}6\bar{7}8\bar{3}4\bar{5}2)\\
\nonumber &&\null-A_8(\bar{1}6\bar{7}2\bar{3}8\bar{5}4)+A_8(\bar{1}6\bar{7}4\bar{3}8\bar{5}2)-A_8(\bar{1}6\bar{3}2\bar{7}8\bar{5}4)+A_8(\bar{1}6\bar{3}4\bar{7}8\bar{5}2)\\
%\nonumber 
&&\null+A_8(\bar{1}6\bar{3}8\bar{7}2\bar{5}4)-A_8(\bar{1}6\bar{3}8\bar{7}4\bar{5}2)-A_8(\bar{1}4\bar{3}6\bar{7}8\bar{5}2)+A_8(\bar{1}2\bar{3}6\bar{7}8\bar{5}4)=0\,,%\\
\eqae
\fi
%
along with 11 more similar ones. Starting at ten points the situation becomes more complicated, leaving us without general-multiplicity formulas for all KK-like relations.

%%%%%%%%%%%%%%%%%%%%%%%%%%%%%%%%%%%%%%%%%%%%%%%%%%%%%%%%%%%%%
\subsubsection{KK identities from amplitude-generating integral formula}
%%%%%%%%%%%%%%%%%%%%%%%%%%%%%%%%%%%%%%%%%%%%%%%%%%%%%%%%%%%%%
We will now take a step back and ask: if given a KK-like relation, are there other efficient ways for determining its validity? If so, these ways may give a path for determining the general formulas. For this purpose we will use the twistor-string-like formula for ABJM amplitudes. It has the advantage that the part of the amplitude that is not fully permutation invariant  is isolated to a very simple Park-Taylor-like factor, which allows us to extract any relation among distinct color orderings. 

Guided by the connected prescription for the twistor string theory \cite{RSVW} 
in four dimensions and the Grassmannian integral formula for the ABJM theory \cite{Lee:2010du}, 
two of the present authors recently proposed a twistor-string-like integral formula for the ABJM superamplitude \cite{SangminYt}:\footnote{This formula was recently shown to be equivalent to an alternative integral formula which satisfy all factorization properties, thus verifying it's validity ~\cite{Cachazo:2013iaa}.}
\begin{align}
A_{n}(\L) = \int\frac{d^{2\times n}\s}{{\rm vol}[{\rm GL}(2)]}
\frac{J \, \D \,\prod_{m=1}^k \d^{2|3}(C_{mi}[\s]\L_i)}{(1 2)(23) \cdots (n 1)}\,.
\label{twi3}
\end{align}
The integration variable $\s$ is a $(2 \times n)$ matrix, 
which is mapped to the $(k\times n)$ matrix $C[\s]$ by 
\be
\s = \begin{pmatrix}
a_1 \cdots a_n \\ b_1 \cdots b_n
\end{pmatrix}
\quad \rightarrow \quad
C_{mi}[\s] = a_i^{k-m}b_i^{m-1} \,.\label{vero-map}
\ee
The two-bracket in \eqref{twi3} is defined by $(i j)\equiv a_i b_j - a_j b_i$, 
and $\D$ is a delta-function constraint,
\be
\Delta = \prod_{j=1}^{2k-1} \d\left(\sum_i a_i^{2k-1-j} b_i^{j-1}\right)\,.
\label{null-vero-const}
\ee
Finally, the factor $J$ in eq.\eqref{twi3} is defined as a ratio $J={\rm (Num)}/{\rm (Den)}$ with
\begin{align}
{\rm Den} = \prod_{1\le i<j \le k} (2i-1 , 2j-1) \,,
\quad 
{\rm Num} = \det_{1\le i,j \le 2k-1} a_i^{2k-1-j} b_i^{j-1}
= \prod_{1\le i<j \le 2k-1} (i,j) \,.
\label{JJ}
\end{align}

We now want to show that the formula \eqref{twi3} 
satisfies the same KK identities for the ABJM amplitudes 
as found in section~\ref{KKsection} by studying the color factors. 
Since $\Delta$ and (Num) is completely invariant under arbitrary permutation, 
and (Den) is invariant  up to a sign under permutation of the odd sites, 
it is sufficient to focus on the Park-Taylor-like denominator,
\be
D_{2k}(\bar 1, 2,\ldots,2k) = \frac{1}{(12)(23)\ldots (2k, 1)}\,.
\ee
Let us first show that \eqn{KK} is indeed satisfied by \eqn{twi3}. First note that since the ABJM superamplitude has Fermionic external states on the odd sites, we can write the equivalent of the permutation sum in \eqn{KK}, acting on $D$'s, as 
\begin{align}
&S_{2k}
= \sum_{\rho \in S_k}D_{2k}(\bar{1},\rho_1,\bar{3},\rho_2,\bar{5},...,\overline{2k-1}, \rho_{k})(-1)^{\rho}= \det\Omega_{i,j}
= \det\bar{\Omega}_{i,j} \,,
\label{Did}
\end{align}
where $(-1)^{\rho}$ denotes the signature of the permutation $\rho$, and the $k \times k$ matrices $\Omega_{i,j}$ and $\bar{\Omega}_{i,j}$ are given as:
\eq
\Omega_{i,j}=\left( \frac{1}{(2i-1,2j)(2j,2i+1)}\right), \quad \bar{\Omega}_{i,j}=\left( \frac{1}{(2i,2j+1)(2j+1,2i+2)}\right) \,.
\label{OmegaDef}
\eqe
By definition, $S_{2k}$ is a homogenous function of $(-4k)$ powers of $\s$ variables. 
Collecting all the fractions using the obvious common denominator, 
we can write 
\be
S_{2k} = \frac{Q_{2k^2-4k}(\s)}{\prod_{i,j=1}^{k} (2i-1,2j)} \,,
\ee 
for some polynomial $Q$ of degree $(2k^2-4k)$. Now, from \eqn{Did} and \eqn{OmegaDef} it is easy to see if any two even legs are identified, for example $\s_{2k}=\s_{2k+2}$, then $S_{2k}$ must vanish due to the fact that two columns in $\Omega_{i,j}$ becomes identical. Similar conclusion can be reached for any two odd legs being identified. This implies that the polynomial $Q$ must contain the product of the following two factors:
\be
\prod_{i<j}^k (2i-1,2j-1) \quad {\rm and} \quad \prod_{i<j}^k (2i,2j) \,,
\ee 
each of which has degree $k^2-k$. The polynomial $Q$ has not enough degree 
to contain both factors, so the only consistent solution is that $S_{2k} $ is simply zero, thus completing the proof.  

From the previous discussion, we see that any non-trivial linear relations for permutated  ABJM amplitudes must be encoded as identities for the Park-Taylor-like factor $D_{2k}(1,2,\ldots,2k)$. This fact can be utilized to develop graphical tools to recursively generate all possible KK-like relations. To simplify computations, we use the homogeneity of $D_{2k}$ to 
pull out the `scale factor' 
from each two-bracket, 
\be
(ij) = a_i b_j - a_j b_i = b_i b_j \left( \frac{a_i}{b_i} - \frac{a_j}{b_j} \right) \equiv b_i b_j (x_i - x_j) \,,
\ee
and regard $(ij)$ as $(x_i - x_j)$ in what follows. 
Manipulations of $D_{2k}$ will involve two basic operations:
\begin{enumerate}
\item
Antisymmetry:
\be
\frac{1}{(\bar{a}b)}+\frac{1}{(b\bar{a})} = 0 \,. 
\ee
\item
Four-term identity: 
\be
\frac{1}{(\bar{a}b)(b\bar{c})(\bar{c}d)} 
+\mbox{(cyclic)} 
= 
\frac{(\bar{a}b)+(b\bar{c})+(\bar{c}d)+(d\bar{a})}{(\bar{a}b)(b\bar{c})(\bar{c}d)(d\bar{a})} = 0 \,.
\label{4Term}
\ee
\end{enumerate}
Again we have introduced bared indices to emphasize the connection to odd an even sites.

Next, we find it useful to introduce  a graphical representation for these operations 
as follows:
\be
\includegraphics[width=8cm]{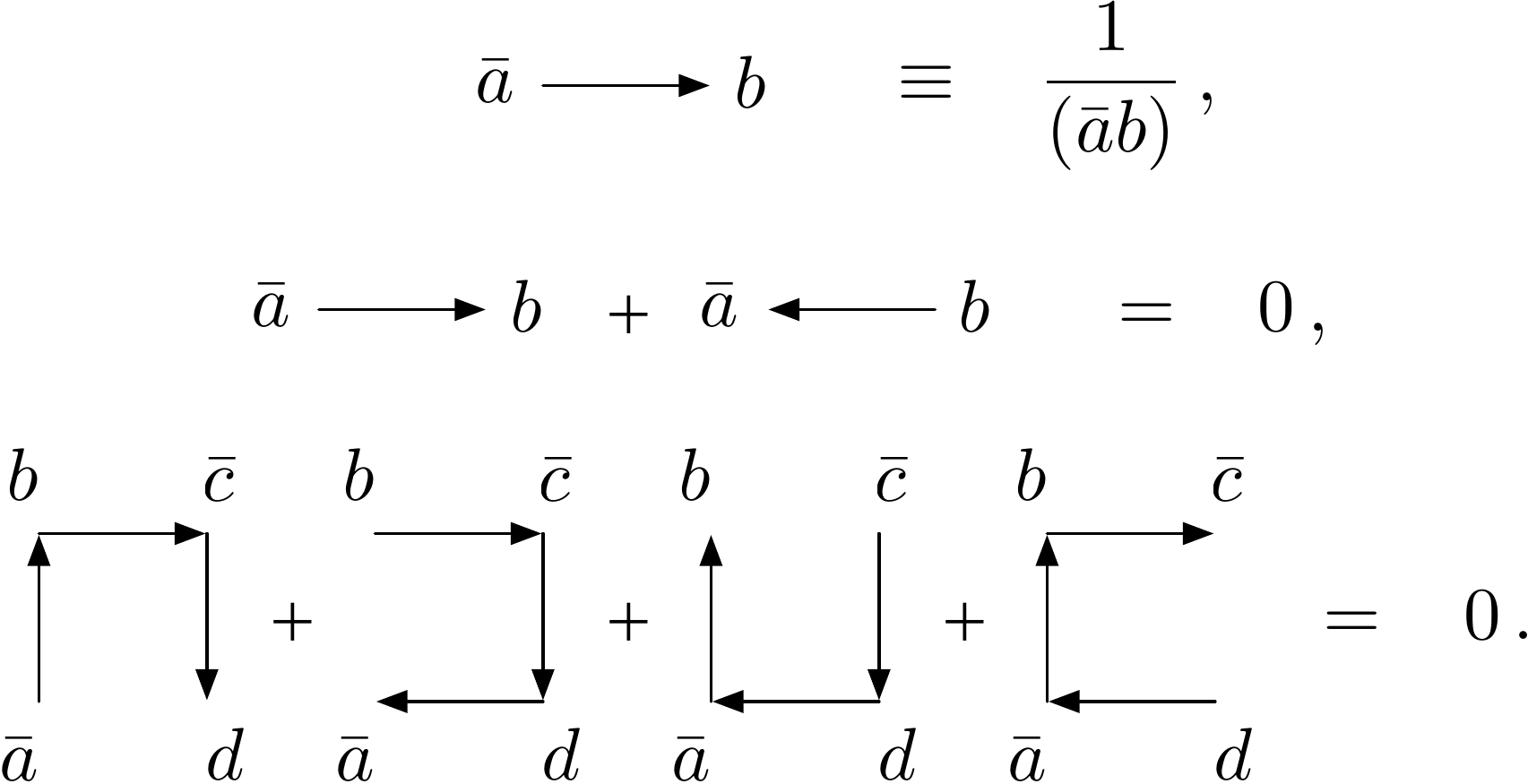}
\ee
Note that  $D_{2k}$ is simply a closed path in such representation. The graphical representation can be used to 
generate the generic KK identities recursively, deducing new identities for $D_{2k+2}$ from known identities for $D_{2k}$. 
We begin by attaching two ``open arrows" to $D_{2k}$, corresponding to adding two extra points. 
As depicted in the following diagram, adding the two arrows in three different ways allows us to ``close the path" using the four-term identity \eqn{4Term}
and produce a $D_{2k+2}$:
\be
\includegraphics[width=10cm]{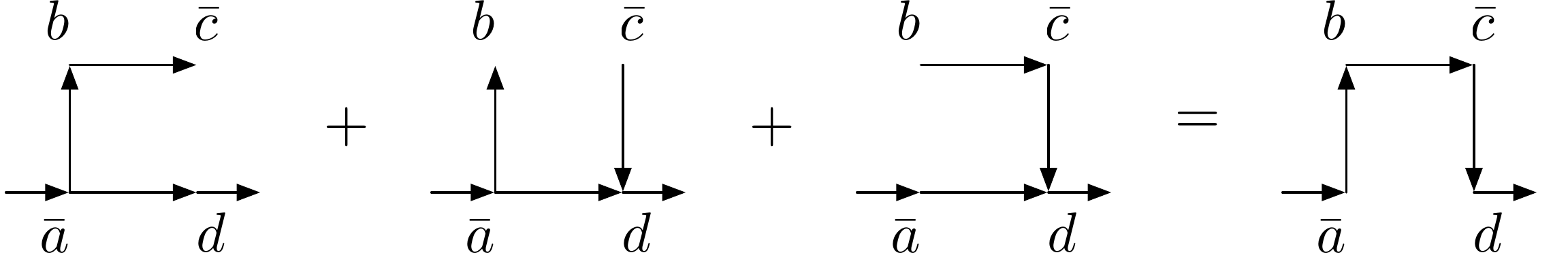}
\ee

To obtain a non-trivial recursive construction, 
we apply the basic operations repeatedly to shift around the open arrows  before closing the path, thereby generating a sum of many different terms. 
Eqs.~\eqref{ABJM-shift1}, \eqref{ABJM-shift2} present two simple shift operations. 
The relation \eqref{ABJM-shift1} is just a slight rewriting of the basic 
four-term identity. To derive the relation \eqref{ABJM-shift2}, 
we attach an extra arrow $(d\bar{e})$ to \eqref{ABJM-shift1} and 
apply the four-term identity to both terms on the right-hand side. 
Two out of the six terms thus generated cancel out. 

\begin{align}
&\includegraphics[width=9cm]{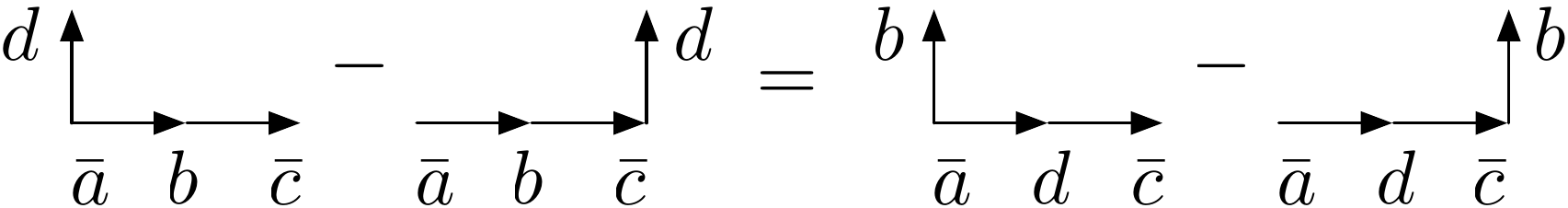}
\label{ABJM-shift1}
\\
&\includegraphics[width=13.5cm]{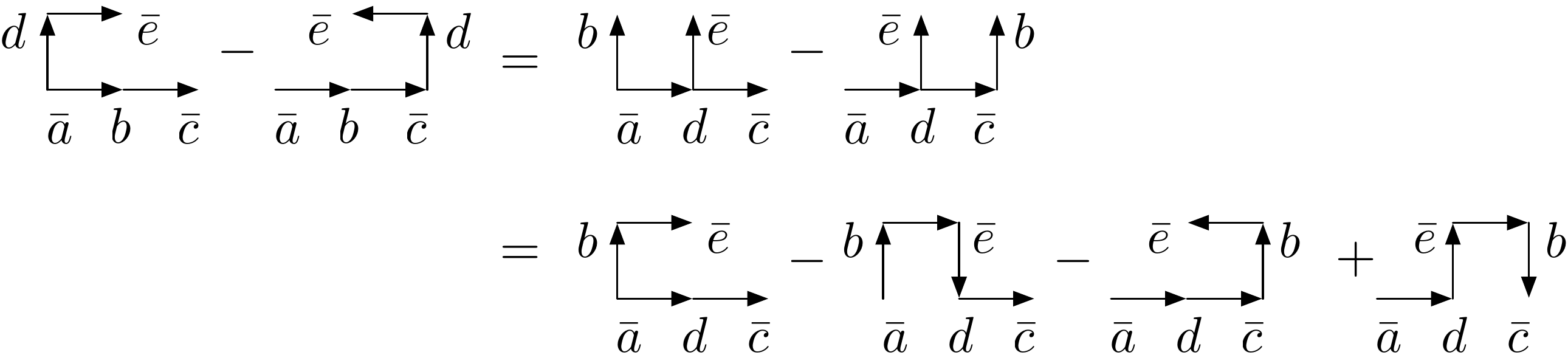}
\label{ABJM-shift2}
\end{align}

\noindent

\begin{figure}
\begin{center}
\includegraphics[scale=0.46]{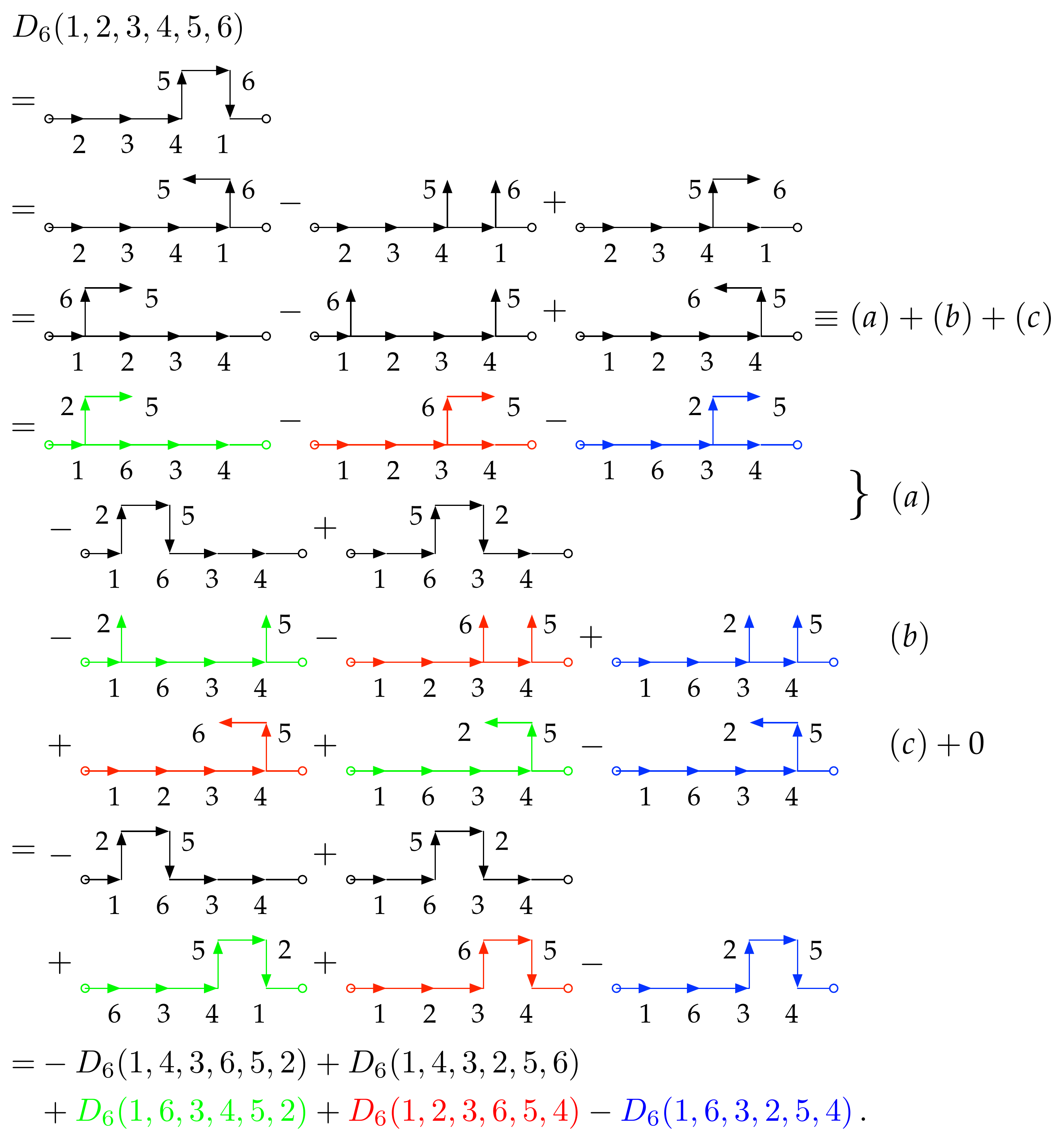}
\caption{A graphical derivation of the identity \eqref{ABJM-KK-6pt}. The two end points of each chain is identified to form a closed path. Applying the basic four-term identity to the second line gives the third line. The fourth line, $(a)+(b)+(c)$, is the same as the third, except 
that we turned the arrows to prepare for closing the path in the opposite direction. 
To obtain the fifth line, we apply \eqref{ABJM-shift2} to (a) and \eqref{ABJM-shift1} to (b). We leave (c) as it is, but add and subtract the same term next to it. 
Now, in addition to two closed paths, there are a total of nine diagrams with open arrows. Using the basic four-term identity, we group them into three closed paths. 
We colored the diagrams to show which terms are combined. The resulting five distinct closed paths on the right-hand-side and the original term on the left-hand-side together give the desired identity \eqref{ABJM-KK-6pt}.}
\label{D6-derivation}
\end{center}
\end{figure}

\noindent 
To illustrate the idea of this identity generating technique, we present some simple examples:

\begin{enumerate}

\item
Starting from the trivial identity $D_4(1,2,3,4)=D_4(1,2,3,4)$, 
attaching open arrows, shifting them around 
in two different ways, 
we reproduce the only KK identity for $D_6$,  
\begin{align}
& D_6(1,2,3,4,5,6) - D_6(1,6,3,4,5,2) + D_6(1,6,3,2,5,4) 
\nn \\
& \null - D_6(1,4,3,2,5,6)+ D_6(1,4,3,6,5,2) - D_6(1,2,3,6,5,4) = 0 \,,
\label{ABJM-KK-6pt}
\end{align}
where the signs can be traced back to the Fermionic nature of even sites of the superamplitude.
See Figure~\ref{D6-derivation} for a step-by-step derivation of this identity. 
\item 
Starting from the trivial identity $D_6(1,2,3,4,5,6)=D_6(1,2,3,4,5,6)$, 
attaching open arrows and shifting them around in different ways, 
we find a 24-term identity for $D_8$ that involves 
permutations of both even and odd labels, 
\begin{align}
& D_8(12345678)-D_8(18763452)+D_8(18763254)-D_8(14325678) \nn \\
\null +&D_8(18743652)-D_8(14783652)+D_8(18365274)-D_8(18365472) \nn \\ 
\null +&D_8(12345678)-D_8(18763452)+D_8(18763254)-D_8(14325678) \nn \\
\null +&D_8(18743652)-D_8(14783652)+D_8(18365274)-D_8(18365472) \nn \\ 
\null +&D_8(12783654)-D_8(18723654)+D_8(14387652)-D_8(18347652) \nn \\
\null +&D_8(18367452)-D_8(18367254)+D_8(18327654)-D_8(12387654)  = 0 \,.  
\label{ABJM-KK-8pt-24}
\end{align}
The derivation of this identity is a straightforward but lengthy generalization 
of figure~\ref{D6-derivation}. 
\item 
Starting from \eqref{ABJM-KK-6pt} and attaching open arrows 
on particle 1 and particle 6, we can produce a 16-term identity for $D_8$, 
%, 
%
\begin{align}
& D_8(12345678)-D_8(18763452)+D_8(18763254)-D_8(14325678) \nn \\
\null +& D_8(18743652)-D_8(14783652)+D_8(18365274)-D_8(18365472) \nn \\ 
\null +&D_8(12783654)-D_8(18723654)+D_8(14387652)-D_8(18347652) \nn \\
\null +&D_8(18367452)-D_8(18367254)+D_8(18327654)-D_8(12387654)  = 0 \,.  
\label{ABJM-KK-8pt-16}
\end{align}
\end{enumerate}
We have checked that, by taking linear combinations of these identities, 
we can exhaust all general KK identities at ten points, agreeing with the results obtained in section~\ref{KKsection} after taking into account that those formulas are for Bosonic states.

We conclude this subsection by noting that the factor $D_n$ is identical to that appearing in the twistor-string formula for $\mathcal{N}=4$ Yang-Mills theory~\cite{RSVW}. Since the remaining pieces in both theories are permutation invariant (up to statistical signs), this implies that all KK relations discussed here are also satisfied by Yang-Mills amplitudes, for adjoint particles. Therefore the KK-relations for ABJM type theories are simply a subset of that for Yang-Mills, such that even and odd sites do not mix, with proper identification of particle statistics.

%%%%%%%%%%%%%%%%%%%%%%%%%%%%%%%%%%%%%%
 \subsection{BLG type: $f^{abcd} \propto f^{[abcd]}$\label{BLGColor} }
%%%%%%%%%%%%%%%%%%%%%%%%%%%%%%%%%%%%%%

Parallel to the ABJM discussion, we start by counting the number of quartic graphs that appear in $n$-point BLG amplitudes, or distinct color factors built out of totally antisymmetric $f^{abcd}$'s. Using the same rooted diagrams as in section~\ref{ABJMcountingSection}, we can derive the corresponding iteration relation for the number of quartic BLG graphs, it is
\eq
 \nu(2k)=\frac{1}{3!}\sum_{m_1=1}^{k-1}\sum_{m_2}^{k-m_1}\left(\begin{array}{c}2k-1 \\ 2m_1-1\end{array}\right)\left(\begin{array}{c}2k-2m_1 \\ 2m_2-1\end{array}\right)\nu(2m_1) \nu(2m_2) \nu(2k-2m_1-2m_2+2)\,,
\label{QuarticDiaBLG}
\eqe
with $\nu(2)=1$. A closed formula is given by
\eq
\nu(2k)=\frac{(3k-3)!k!}{(k-1)! (3!)^{k-1}}\,.
\eqe

The color factors that correspond to the quartic diagrams satisfy four-term fundamental identities that we can write as $f^{abc[d}f^{egh]a}=0$. For each contraction $f^{abcd}f^{egha}$ we can choose $1+3$ out of $3+3$ free indices to antisymmetrize over, giving a total of six different possible fundamental identities. This implies that the total number of distinct fundamental identities is equal to the number of quartic graphs times the number of propagators, times six possible index antisymmetrizations, divided by an overcount of four, for counting each graph four times. The final count of BLG fundamental identities at $(2k)$ points is given by $3/2 (k-2) \nu(2k)$.

As for ABJM, beyond four points, the number of independent color factors is smaller than the number of partial amplitudes. This again implies linear amplitude identities. To see this let us again start with the six-point amplitude. The full color dressed BLG amplitude is given by:
\eq
 \mathcal{A}_6=\frac{c_1n_1}{s_{123}}+\frac{c_2n_2}{s_{126}}+\frac{c_3n_3}{s_{134}}+\frac{c_4n_4}{s_{125}}+\frac{c_5n_5}{s_{146}}+\frac{c_6n_6}{s_{136}}+\frac{c_7n_7}{s_{145}}+\frac{c_8n_8}{s_{124}}+\frac{c_9n_9}{s_{156}}+\frac{c_{10}n_{10}}{s_{135}}\,,
 \label{CDressedBLG}
 \eqe
where all but one of the color factors are defined in \eqn{ff}, dropping the bars on the indices; and the new one is $c_{10}=f^{135a}f^{a462}$. Now consider the following gauge invariant partial amplitudes:
\begin{align}
\nonumber A^{\rm BLG}_{\rm SO(4)}(\{1,4\},\{2,5\},\{3,6\})&=\frac{n_1}{s_{123}}+\frac{n_2}{s_{126}}+\frac{n_9}{s_{156}}+\frac{n_{10}}{s_{135}}\,,\\
\nonumber A^{\rm BLG}_{\rm SO(4)}(\{1,4\},\{2,3\},\{6,5\})&=-\frac{n_2}{s_{126}}-\frac{n_4}{s_{125}}-\frac{n_6}{s_{136}}-\frac{n_{10}}{s_{135}}\,,\\
A^{\rm BLG}_{\rm SO(4)}(\{1,4\},\{2, 6\},\{5,3\})&=-\frac{n_1}{s_{123}}-\frac{n_9}{s_{156}}+\frac{n_4}{s_{125}}+\frac{n_{6}}{s_{136}}\,.
\label{BLGPole}
\end{align}
One immediately sees that 
\eq
A^{\rm BLG}_{\rm SO(4)}(\{1,4\},\{2,5\},\{3,6\})+A^{\rm BLG}_{\rm SO(4)}(\{1,4\},\{2,3\},\{6,5\})+ A^{\rm BLG}_{\rm SO(4)}(\{1,4\},\{2, 6\},\{5,3\})=0\,.
\eqe
In general $A^{\rm BLG}_{\rm SO(4)}(\{i,j\},\{k,l\},\{m,n\})$ vanishes as one performs a cyclic sum over $i,j,k$. Repeated use of this identity, we arrive at five independent amplitudes,
\eqa
\nonumber &&A^{\rm BLG}_{\rm SO(4)}(\{1,4\},\{2,5\},\{3,6\}),\;A^{\rm BLG}_{\rm SO(4)}(\{1,6\},\{4,5\},\{3,2\}),\;A^{\rm BLG}_{\rm SO(4)}(\{1,2\},\{6,5\},\{3,4\})\\
 &&A^{\rm BLG}_{\rm SO(4)}(\{1,2\},\{4,5\},\{3,6\}),\;A^{\rm BLG}_{\rm SO(4)}(\{1,4\},\{6,5\},\{3,2\})\,.
\label{BLGInd}
\eqae
No more relations can be derived from the color structures alone. Using the fundamental identity one can show that there are exactly five independent color factors, matching the count above. Thus we conclude that \eqn{BLGInd} is a basis of partial amplitudes under all KK-like relations at six points in a BLG-like theory. 

%%%%%%%% TABLE %%%%%%%%%%%
\begin{table}
  \centering \begin{tabular}{|l|c|c|c|c|c|c|}
\hline
  external legs & 4 & 6& 8& 10& $n=2k$ \\ \hline
   quartic diagrams & 1 & 10& 280 & 15400 &   $\nu(2k)=\frac{(3k-3)!k!}{(k-1)!(3!)^{k-1}}$ \\
   partial ampls, general $f^{abcd}$  & 1 & 15 & 672 & 37800 &  $*$  \\
   partial ampls, SO(4) & 1 & 15 & 91 & 945 & $\bigg\{\begin{array}{c} (2k-1)!!\;{\rm for\;} k~{\rm odd}\; \\    C(k)(C(k)-1)/2 \;{\rm for\;} k~{\rm even}\;\end{array}$ \\
 fundamental identities & 0 & 15  & 840 & 69300 &   $\frac{3}{2}(k-2) \nu(2k)$ \\
   KK basis, general $f^{abcd}$& 1 & 5 & 56 &  1077 & $*$ \\
   KK basis, SO(4) & 1 & 5 & 56 &  552 & $*$  \\
   BCJ basis & 1 & 3 & 38 &  1029 & $*$ \\
\hline
\end{tabular}
  \caption{\label{BLGtable} Counts for BLG theory. First line gives the number of distinct color factors, or distinct quartic diagrams, in the full amplitude. The second line gives a count of distinct partial amplitudes of the simplest type (identified using generalized gauge invariance). The third line gives the same count in the case of SO(4) (color basis: products of multiple $\delta^{a}_b$ and at most one $\epsilon^{abcd}$). The  fourth counts the KK-independent amplitudes, or equivalently, the number of independent color factors. The fifth line counts the same quantity in the SO(4) case. The final line gives the basis of partial amplitudes independent under BCJ relations. $C(k)$ are the Catalan numbers. }
\end{table}
%%%%%%%%%%%%%%%%%%%%%%%

Proceeding to higher points, we can either find an exhaustive set of KK-like relations for partial amplitudes (such as $A^{\rm BLG}(1,2,3,4,5;6,7,8)$ in \eqn{EightPointDef}), or we can solve the overdetermined linear system of fundamental identities. Either task will result in a number that counts the basis size of KK-independent amplitudes, which has to be  equal to the number of independent color factors. Using the latter method, we obtain a count of exactly 56 independent color factors at eight points; and at ten points we find a basis size of 1044. Interestingly, both these numbers are lower than the corresponding ones in ABJM-like theories, despite the fact that BLG-like theories have a larger set of distinct color factors.  
For the partial amplitudes, this mismatch of KK-basis sizes can be connected to the observation in section~\ref{PartialAmpDefsSubSubSection} that the chirally projected BLG amplitudes can be written in terms of ABJM amplitudes, but the map is not invertible (assuming coefficients in the linear map are constants). 

In Table \ref{BLGtable}, we summarize all the determined counts of BLG quantities discussed in this section and in~\ref{PartialAmpDefsSubSubSection}. For completeness, this table also includes the KK-basis size for an amplitude decomposition that uses the explicit SO(4) Lie algebra. Up to eight points, it agrees with the count for general $f^{abcd}$ structure constants, but starting at ten points the SO(4) count is considerably smaller.  In the following section we will discuss the next layer of structure that can be imposed on general bi-fundamental amplitudes. For the purpose of BLG-like theories, we assume that the relevant KK-basis is the one obtained for the general $f^{abcd}$ structure constants. This is what is needed for color-kinematics duality.

%%%%%%%%%%%%%%%%%%%%%%%%%%%%%%%%%%%%%%
\section{BCJ color-kinematics duality~\label{SecBCJ}}
 %%%%%%%%%%%%%%%%%%%%%%%%%%%%%%%%%%%%%
The Kleiss-Kuijf identities in the previous sections are very general results that follow from the overcompleteness of the $f^{ab\bar c \bar d}$, $f^{abcd}$ and $f^{abc}$ expansions. Any quantum field theory whose interactions are dressed by such structure constants satisfy these identities. For further unfolding of the amplitude properties we must turn to the detailed kinematical structure of the theories.  

First we briefly review the color-kinematics duality proposed for Yang-Mills theories by Bern, Carrasco, and one of the current authors (BCJ)~\cite{BCJ}. The duality states that scattering amplitudes of Yang-Mills theory, and its supersymmetric extensions, can be given in a representation where the numerators $n_i$ reflect the general algebraic properties of the corresponding color factors $c_i$. More precisely, for an amplitude expressed using cubic diagrams, one can always find a representation such that the following parallel relations holds for the color and kinematic factors:
\eqa
\nonumber c_i\rightarrow -c_i~~&\Leftrightarrow&~~n_i\rightarrow -n_i\,\\
 c_i+c_j+c_k=0~~&\Leftrightarrow&~~n_i+n_j+n_k=0\,.
\label{dualityEqns}
\eqae
The first line signifies the antisymmetry property of the Lie algebra, and the second line signifies a Jacobi identity, schematically. The duality has several interesting consequences, both for gauge theory and gravity. On the gauge theory side, such representation leads to the realization that color-ordered amplitudes satisfy relations beyond the Kleiss-Kuijf ones. 

The construction of these BCJ relations are as follows: As already utilized in the previous sections, one may expand color-ordered amplitudes in terms of color-stripped diagrams that are planar with respect to appropriate ordering of external legs,
\eq
A_{(i)}= \sum_{{\rm planar\,w.r.t.}\,\sigma_i } \frac{n_j}{\prod_{\alpha_j}s_{\alpha_j}}\,,
\label{Aplanar}
\eqe
where $(i)$ is shorthand notation for a permutation $\sigma_i$; e.g. $(1)=\sigma_1=(1,2,3,\ldots,n)$, etc. The flip antisymmetry $n_i\rightarrow -n_i$ can then be used to identify cubic diagrams that are common in different partial amplitudes, and we may choose a KK-basis of $(n-2)!$ partial amplitudes.
Since the numerators $n_i$ satisfy the same Jacobi identity and symmetry properties as the color factors, there must be only $(n-2)!$ independent numerators. Choosing a particular set of independent numerators, \eqn{Aplanar} can be rewritten with the help of a $(n-2)!\times (n-2)!$  matrix $\Theta_{ij}$. It is defined by
\eq
A_{(i)}=\sum_{j=1}^{(n-2)!} \Theta_{ij}\hat n_{j}\,,
\eqe
where $\hat n_{j}$ are the independent numerators. The matrix $\Theta_{ij}$ is comprised solely of scalar $\phi^3$-theory propagators  (in~\cite{Vaman:2010ez} it was called propagator matrix).  The rank of the matrix $\Theta_{ij}$ is only $(n-3)!$, thus implying new amplitude relations beyond the Kleiss-Kuijf identities. The simplest type of such relations (sometimes called fundamental BCJ relations) can be nicely condensed to~\cite{BCJ}
\eq	
\sum_{i=3}^n\left(\sum_{j=3}^i s_{2j}\right)A_n(1,3,\cdots,i,2,i+1,\cdots,n) =0\,.
\eqe

Since the $\Theta_{ij}$ matrix is solely comprised of propagators, it can be straightforwardly continued to arbitrary spacetime dimension. Remarkably, the matrix has rank $(n-3)!$ in any dimension, but only for on-shell and conserved external momenta; off-shell the rank is $(n-2)!$. This can be interpreted as a non-trivial consistency check of the BCJ construction. Indeed, Yang-Mills theories exists in $D$ dimensions, and the S-matrix is well-defined only for physical on-shell and conserved momenta.

A more important consequence of the color-kinematics duality is the double-copy construction of gravity amplitudes~\cite{BCJ}. Once duality-satisfying numerators are found, a corresponding supergravity amplitude, whose spectrum is given by the tensoring of two Yang-Mills spectra, can be directly written as
\eq
\mathcal{M}_m=\Big(\frac{\kappa}{2}\Big)^m \sum_{i\in {\rm cubic}}\frac{n_i\tilde{n}_i}{\prod_{\alpha_i}s_{\alpha_i}}\,,
\label{Sugra}
\eqe
where at least one of the two sets of numerators must explicitly satisfy the duality (\ref{dualityEqns}).
This aspect of the conjecture as well as the existence of the duality-satisfying numerators have been proven at tree level. The double-copy aspect was proven in ref.~\cite{Bern:2010yg} for the cases of pure YM and ${\cal N}=4$ sYM, and the existence of numerators to all multiplicity that satisfy \eqn{dualityEqns} was exemplified in refs.~\cite{Tree} (see also refs.~\cite{Tree2}). The conjecture has been extended to loop level \cite{BCJLoop}, where duality satisfying numerators has been found for various amplitudes in different theories~\cite{BCJLoop,BCDJR,LoopBCJnumerators} and used in gravity constructions~\cite{N>=4SG,N=4SG}, though a formal proof is still an open problem.

%%%%%%%%%%%%%%%%%%%%%%%%%%%%%%%%%%%%%%
 \subsection{BCJ duality for three-algebra theories}
%%%%%%%%%%%%%%%%%%%%%%%%%%%%%%%%%%%%%%

Remarkably, color-kinematics duality exists also for other gauge theories that are not part of the family of Yang-Mills theories, but of Chern-Simons matter theories. In particular, the duality is believed to exist for certain gauge groups that are Lie three-algebras. For Lie three-algebra color-kinematics duality, one would as before require that the kinematical numerators respect the same symmetries and relations as the color factors,
\eqa
\label{dualityEq}
c_i\rightarrow -c_i~~&\Leftrightarrow&~~n_i\rightarrow -n_i\,\\
\nonumber c_i+c_j+c_k+c_l=0~~&\Leftrightarrow&~~n_i+n_j+n_k+n_l=0\,.
\label{3algebraCK}
\eqae
The first line signifies the antisymmetry properties of the three-algebra, and second line signifies the fundamental identity or generalized Jacobi identity. That these identities could be imposed on the kinematic numerators was first proposed by Bargheer, He and McLoughlin~\cite{Till} in the context of BLG and ABJM theories. Via the double-copy relation,
\eq
\mathcal{M}_m=\Big(\frac{\kappa}{2}\Big)^m \sum_{i\in {\rm quartic}}\frac{n_i\tilde{n}_i}{\prod_{\alpha_i}s_{\alpha_i}}\,,
\label{Sugra}
\eqe
they reproduced the four- and six-point amplitudes of $\mathcal{N}=16$ supergravity of Marcus and Schwarz. The same exercise was later shown to work for a large class of CSm and supergravity theories~\cite{HenrikYt}. Remarkably, the gravity amplitudes that are produced by the double copies of $D=3$ YM theories and that of CSm theories are identical, even though the two constructions are impressively distinct~\cite{HenrikYt}.

We should emphasize that all studies thus far~\cite{Till,HenrikYt} have been limited to four- and six-point amplitudes, which leaves open the possibility that the results do not generalize to multiplicities $n\ge8$.
Indeed, as we will explain, for ABJM-type theories with general gauge group, most of the expected color-kinematics properties are absent beyond six points. Before we get there, let us proceed by discussing BLG-type color-kinematics duality, which appears to work seamlessly.

 %%%%%%%%%%%%%%%%%%%%%%%%%%%%%%%%%%%%%%%
 \subsection{BCJ duality for BLG theory}
%%%%%%%%%%%%%%%%%%%%%%%%%%%%%%%%%%%%%%%
%
Let us now consider the BCJ relation for BLG-type theories. We will show the details of the six-point amplitude, and for eight and ten points we will only give the counts of relations and independent amplitudes. As discussed previously, BLG-type three-algebras allow one to reduce the color ordered amplitude to five independent ones. However, further reduction comes from color-kinematics duality.  The numerator must satisfy the same properties as the color factor. Using the six-point amplitude representation in \eqn{CDressedBLG} one would have, for example,
\eq
 c_3 - c_4 - c_5+c_6=0\quad\Leftrightarrow\quad  n_3 - n_4 - n_5+n_6=0\,.
\eqe
Imposing this numerator relation, together with 14 more similar relations (not all independent), leads to five independent numerators. By the duality, this number has to be the same as the number of KK-independent amplitudes (see \Tab{BLGtable}). Thus the KK-independent amplitudes can be expressed in terms of these five independent numerators
\eq
A_{(i)}=\sum_{j=1}^{5} \Theta_{ij} n_{j}\,, \hskip1cm i = 1,\cdots,5 \,,
\eqe
Naively, since $\Theta_{ij}$ is a square matrix, one would like to invert it and express the independent numerators in terms of color ordered amplitudes. However, upon deeper consideration this might not be a legal move.  Since, one should expect the numerators in a gauge theory to be gauge dependent, and thus not well defined in terms of S-matrix elements.
Indeed, just like the case of Yang-Mills theory, the $\Theta_{ij}$ matrix has lower rank than what is explicit. To show this in detail, we use $A_{(i)}$ with $i=1,\cdots,5$ defined in \eqn{BLGInd} as our independent basis,
\begin{align}
A_{(i)}= &\Big(A^{\rm BLG}_{\rm SO(4)}(\{1,4\},\{2,5\},\{3,6\}),\;A^{\rm BLG}_{\rm SO(4)}(\{1,6\},\{4,5\},\{3,2\}),\;A^{\rm BLG}_{\rm SO(4)}(\{1,2\},\{6,5\},\{3,4\}) \nn \\
 &~~ A^{\rm BLG}_{\rm SO(4)}(\{1,2\},\{4,5\},\{3,6\}),\;A^{\rm BLG}_{\rm SO(4)}(\{1,4\},\{6,5\},\{3,2\})\Big)\,,
 \label{BLGbasisChoice}
\end{align}
and the numerator basis $n_i$ for $ i=1,\ldots,5$. The reduction of the numerators is given by the dual fundamental identities; the independent content of these are
\eqa
&&n_6 = -n_3 + n_4 + n_5\,,~
n_7 = n_1 - n_2 - n_3 + n_4 + n_5\,,~
n_8 = n_1 - n_2 + n_4\,,\nn \\&&
n_9 = n_2 + n_3 - n_4\,,~~
n_{10} = n_1 + n_4 + n_5\,.
\eqae
Using the above bases, the matrix $\Theta_{ij}$ is then given as 
\eqa
\nonumber&&\Theta^{\rm  BLG}_{ij}=\\
\nonumber&&\left(\begin{array}{ccccc}
\frac{1}{s_{123}}+\frac{1}{s_{135}}~&~ \frac{1}{s_{126}}+\frac{1}{s_{156}}& \frac{1}{s_{156}}& \frac{1}{s_{135}}-\frac{1}{s_{156}}& \frac{1}{s_{135}}\\      
\frac{1}{s_{124}}+\frac{1}{s_{135}}~&~ -\frac{1}{s_{124}}& \frac{1}{s_{134}}& \frac{1}{s_{124}}+\frac{1}{s_{125}}+\frac{1}{s_{135}}& \frac{1}{s_{135}}\\      
\frac{1}{s_{135}}+\frac{1}{s_{145}}~&~ -\frac{1}{s_{145}}& -\frac{1}{s_{136}}-\frac{1}{s_{145}}& \frac{1}{s_{135}}+\frac{1}{s_{136}}+\frac{1}{s_{145}}& \frac{1}{s_{135}}+\frac{1}{s_{136}}+\frac{1}{s_{145}}+\frac{1}{s_{146}}\\      
-\frac{1}{s_{135}}& -\frac{1}{s_{156}}~&~ -\frac{1}{s_{134}}-\frac{1}{s_{156}}& -\frac{1}{s_{135}}+\frac{1}{s_{156}}& -\frac{1}{s_{135}}-\frac{1}{s_{146}}\\      
-\frac{1}{s_{135}}& -\frac{1}{s_{126}}~&~ \frac{1}{s_{136}}& -\frac{1}{s_{125}}-\frac{1}{s_{135}}-\frac{1}{s_{136}}& -\frac{1}{s_{135}}-\frac{1}{s_{136}}
 \end{array}\right)\,.\\
 \eqae
Imposing momentum conservation and on-shell constraints one sees that, while the determinant of this matrix does not vanish in generic spacetime dimension, it does vanish for three-dimensional kinematics. Thus in three-dimensional BLG-type theories, the color-kinematics duality leads to further amplitude relations beyond the KK-relations. This critical dimension-dependence of $\Theta$ was first observed in ref.~\cite{HenrikYt} for the ABJM six-point amplitude. Here we see the same phenomenon for BLG theory. More explicitly, $\Theta^{\rm BLG}_{ij}$ has rank three in $D=3$, and thus we have two additional amplitude relations, which reduces the number of independent amplitudes to exactly three. The apparent mismatch between independent amplitudes and independent numerators (three versus five) implies that the numerators are gauge dependent. In fact, to make up for the mismatch, the gauge dependence can be pushed into two redundant numerators; one can think of them as ``pure gauges". Choosing $n_3$ and $n_4$ as the redundant numerators, one can explicitly solve $n_i$ in terms of $A_{(2)},A_{(4)},A_{(5)}$ as well as $n_3$ and $n_4$; that is, $n_j^{*}=n_j^{*}(A_{(i)},n_3,n_4)$. Substituting the solution $n_j\rightarrow n_j^{*}$ into 
\eqa
A_{(1)}&=&\frac{n_1}{s_{123}}+\frac{n_2}{s_{126}} + \frac{n_1 + n_4 + n_5}{s_{135}} + \frac{n_2 + n_3 + n_4}{s_{156}}\,, \nn \\
A_{(3)}&=&\frac{n_5}{s_{146}}+\frac{n_1 + n_4 + n_5}{s_{135}} + \frac{n_4 + n_5-n_3}{s_{136}} + \frac{n_1-n_2-n_3 + n_4 + n_5}{s_{145}}\,,
\label{preRelation2} 
\eqae
we find that the ``pure gauges" $n_3$ and $n_4$ drop out, and \eqn{preRelation2} becomes two relations between color ordered amplitudes. After multiplying by common denominators, the two relations become
\eq
0=\sum_{i=1}^5S_iA_{(i)}=\sum_{i=1}^5\tilde{S}_iA_{(i)}\,,
\eqe
where $S_i$ and $\tilde{S}_i$ are degree-four polynomials of momentum invariants. Explicitly they are given by
\begin{align}
S_1&=0\,, \nonumber  \\ 
S_2&=s_{124}(s_{156}(s_{145}s_{146}-s_{135}s_{136})+s_{126}(s_{146}(s_{135}+s_{156})-s_{136}(s_{145}+s_{156})))\,, \\ \nonumber 
S_3&=s_{145}(s_{156}(s_{136}(2 s_{35} - s_{146})+s_{146}(s_{136}-s_{126}))-s_{126}s_{146}(2 s_{24} - s_{156}))\,, \\ \nonumber 
S_4&=s_{156}(s_{136}s_{145}(2 s_{35} - s_{146})+s_{146}(s_{136}(s_{126}+s_{135})+s_{145}(s_{135}+s_{136})+s_{124}(s_{126}+s_{145})))\,, \\ \nonumber 
S_5&=s_{126}(s_{145}s_{146}(s_{136}-2 s_{24})-s_{136}(s_{135}(s_{145}+s_{146})+s_{124}(s_{145}+s_{156})+s_{146}(s_{145}+s_{156})))\,,
\end{align}
and
\begin{align}
\tilde{S}_1&=s_{123}(s_{156}(s_{136}( 2 s_{35} - s_{146})+s_{146}(s_{136}-s_{126}))-s_{126}s_{146}(2 s_{24} - s_{156}))\,, \nonumber  \\ 
\tilde{S}_2&=s_{124}(s_{126}(s_{123}s_{136}+s_{146}(s_{135}+s_{136}))-s_{156}(s_{135}s_{136}+s_{146}(s_{123}+s_{136})))\,, \\ \nonumber 
\tilde{S}_3&=0\,, \\ \nonumber 
\tilde{S}_4&=s_{146}(s_{123}s_{126}(s_{156}-2 s_{24})-s_{156}(s_{123}(s_{126}+s_{135})+s_{124}(s_{123}+s_{136})+s_{126}(s_{135}+s_{136})))\,, \\ \nonumber   
\tilde{S}_5&=s_{136}(s_{126}(s_{124}s_{146}+s_{156}(s_{135}+s_{146}))+s_{123}(s_{126}(s_{124}+s_{135})+2s_{156}s_{35}))\,.
\end{align}
%
%
%
%
\iffalse % Not sure if this is so interesting when there are other problems with ABJM to worry about?
Note that in ref.~\cite{Till} it was reported that the number of independent amplitudes is four in both ABJM and BLG theories at six point. As shown, we find one further relation for BLG. The discrepancy of the results can be explained by the fact that the analysis done in ref.~\cite{Till} used only color-ordered partial amplitudes. Such partial amplitudes are present for any SU($N_1$)$\times$SU($N_2$) bi-fundamental matter theory, and thus the analysis done in ref.~\cite{Till} is formally valid for generic bi-fundamental matter theory.  However, for SU(2)$\times$SU(2) BLG theory a more physical partial amplitude is defined in terms of contracted SO(4) indices, as used in the present analysis. This gives a definition of partial amplitudes distinct from the color ordered amplitudes, which is unique to the SO(4) BLG theory, and that satisfies further relations than those derived from color-ordered amplitudes. 
\fi

Next we present the double-copy result of the six-point gravity amplitude using the BLG three-algebra color-kinematics duality. We find a relatively compact expression if we solve the numerators $n_i$ in terms of $A_{(1)},A_{(4)},A_{(6)}$ and $\tilde n_i$ in terms of $\tilde{A}_{(2)},\tilde{A}_{(3)},\tilde{A}_{(5)}$. We have
\eqa
\nonumber \mathcal{M}_6 &=& \frac{1}{B} \Big\{s_{145}s_{146}\tilde{A}_{(3)} \big[A_{(1)} s_{126} ( s_{134}-s_{124} ) + 
      A_{(4)} s_{134} (2 s_{35} - s_{146}) - 
     A_{(6)} s_{124} (2 s_{26} - s_{145})\big] \\
  \nonumber   && \null ~\, +s_{124} s_{134}  \tilde{A}_{(2)}\big[A_{(1)} s_{126} (s_{145} - s_{146}) - 
      A_{(4)} s_{146}(2 s_{26} - s_{134})  + 
      A_{(6)} s_{145} (2 s_{35} - s_{124})\big] \\
 \nonumber     &&\null ~\, + 
   s_{126} \tilde{A}_{(5)} \big[ A_{(1)} (s_{134}s_{145}(s_{135}+s_{146})+ s_{124}(s_{134} s_{145}- (2 s_{26} - s_{126}) s_{146})) \\
    && \null ~\, + A_{(6)} s_{124} s_{145} (s_{134} - s_{146}) + A_{(4)} s_{134} s_{146} (s_{145}-s_{124})     \big]\Big\}\,,
\label{BLGGravAmp}         
 \eqae
 where 
 \eq
B=2( s_{124} s_{146} s_{26} - s_{134} s_{145} s_{35})\,.
 \eqe
The tilde notation emphasizes that Grassmann-odd parameters should be tensored, not squared. At convenience one may replace the partial amplitudes in \eqn{BLGGravAmp} by their ${\cal N}=6$ supersymmety truncated counterparts. Using \eqn{BLGfromABJM6pts} we can map these to ABJM partial amplitudes that are conveniently accessible in the literature~\cite{Bargheer:2010hn,Gang}, and thus obtain explicit gravity amplitudes.  We have checked that these agrees with $D=4$ supergravity amplitudes dimensionally reduced to $D=3$, verifying the entire construction.

Going beyond six points, we find a multitude of BCJ relations. We have worked out the $\Theta_{ij}$ matrix and amplitude relations at eight and ten points explicitly. As before, the results are rather elaborate so we avoid explicit formulas, and instead present the counts of independent amplitudes. At eight points we find that the 56-dimensional basis of KK-independent partial amplitudes gets further reduced to 38 amplitudes that are independent under the BCJ relations. That is, the eight-point 56-by-56 $\Theta^{\rm BLG}$ matrix has rank 38 in $D=3$ dimensions (in $D>3$ it has the expected full rank 56, and in $D<3$ it diverges). For the ten-point case we find that the 1077-dimensional KK basis is reduced to a 1029 dimensional BCJ basis. So the 10-point 1077-by-1077 $\Theta^{\rm BLG}$ matrix has rank 1029 in $D=3$ dimensions (in $D>3$ it has the expected full rank 1077, and in $D<3$ it diverges). These counts are summarized in Table~\ref{BLGtable}.

Turning to supergravity at eight points: we have explicitly solved the $56$ independent numerators in terms of $38$ BLG partial amplitudes and a remaining set of $18$ pure gauge degrees of freedom. Altogether, we have 280 numerators (see Table~\ref{BLGtable}) that are linearly dependent on the $38$ chosen partial amplitudes as well as the $18$ pure-gauge numerators. For the BLG partial amplitudes we use \eqn{BLGfromABJM8pts} to map these to ABJM partial amplitudes, which we in turn compute using three-dimensional BCFW recursion. 
After taking double copies of the 280 numerators, the pure-gauge numerators drop out, and we obtain an expression for the eight-point supergravity amplitude. We have numerically checked that the resulting amplitude indeed matches the supergravity amplitude obtained from three-dimensional BCFW recursion as well as direct dimensional reduction of the four-dimensional amplitude. This concludes the verification of color-kinematics duality at eight points.

 %%%%%%%%%%%%%%%%%%%%%%%%%%%%%%%%%%%%%%
 \subsection{BCJ duality for ABJM theories}
%%%%%%%%%%%%%%%%%%%%%%%%%%%%%%%%%%%%%%
\label{ABJMandBCJsection}
 
We now consider BCJ duality for ABJM-type theories at six points. Recall that at six-point, a set of five independent amplitudes under the Kleiss-Kuijf relations was given in \eqn{CDressedABJM},
\eq
A_{(i)}=\Big(A_6(\bar{1}2\bar{3}4\bar{5}6),\;A_6(\bar{1}4\bar{3}6\bar{5}2),\;A_6(\bar{1}6\bar{3}2\bar{5}4),\;A_6(\bar{1}4\bar{3}2\bar{5}6),\;A_6(\bar{1}6\bar{3}4\bar{5}2),\;A_6(\bar{1}2\bar{3}6\bar{5}4)\Big)\,.
\label{ABJMPartial}
\eqe
Following the previous analysis, assuming BCJ duality one can reduce the number of kinematic numerators down to five. Again choosing $ n_i, i=1,\ldots,5$ as the independent numerators, the reduction relations are
\eq
n_6 = -n_3 + n_4 + n_5\,,~
n_7 = n_1 - n_2 - n_3 + n_4 + n_5\,,~
n_8 = n_1 - n_2 + n_4\,,~
n_9 = n_2 + n_3 - n_4\,.
\eqe
one finds that the matrix $\Theta_{ij}$ is given by 
\eq
\Theta^{\rm ABJM}_{ij}=
\left(\begin{array}{ccccc}
\frac{1}{s_{123}}~&~\frac{1}{s_{126}}+\frac{1}{s_{156}}~&~\frac{1}{s_{156}}~&~-\frac{1}{s_{156}}~&~0\\
\frac{1}{s_{124}}~&~-\frac{1}{s_{124}}~&~\frac{1}{s_{134}}~&~\frac{1}{s_{124}}+\frac{1}{s_{125}}~&~0\\
\frac{1}{s_{145}}~&~-\frac{1}{s_{145}}~&~-\frac{1}{s_{136}}-\frac{1}{s_{145}}~&~\frac{1}{s_{136}}+\frac{1}{s_{145}}~&~\frac{1}{s_{136}}+\frac{1}{s_{145}}+\frac{1}{s_{146}}\\
0~&~-\frac{1}{s_{156}}~&~-\frac{1}{s_{134}}-\frac{1}{s_{156}}~&~\frac{1}{s_{156}}~&~-\frac{1}{s_{146}}\\
0~&~-\frac{1}{s_{126}}~&~\frac{1}{s_{136}}~&~-\frac{1}{s_{125}}-\frac{1}{s_{136}}~&~-\frac{1}{s_{136}}
\end{array}\right)
\eqe
As before, after imposing momentum conservation and on-shell constraints the determinant vanishes for three-dimensional kinematics, but not for $D>3$~\cite{HenrikYt}. More explicitly, $\Theta^{\rm ABJM}_{ij}$ has rank four in $D=3$, and thus we have one additional amplitude relation, which reduces the number of independent amplitudes to exactly four. Choosing  $n_5$ as the redundant ``pure gauge'' numerator, one can explicitly solve $n_1,n_2,n_3,n_4$ in terms of $A_1,A_2,A_3,A_4$ and $n_5$; that is, $n_j^{*}=n_j^{*}(A_{(i)},n_5)$. Substituting the solution $n_j\rightarrow n_j^{*}$ into
\eq
A_{(5)}=A_6(\bar{1}, 6, \bar{3}, 4, \bar{5}, 2)=-\frac{n_4}{s_{125}} - \frac{n_2}{s_{126}}+\frac{n_3 - n_4 - n_5}{s_{136}} \,,
\label{preRelation}
\eqe
we find that the ``pure gauge" $n_5$ droops out, and \eqn{preRelation} is now a relation between color ordered amplitudes. We get
\eq
0=\sum_{i=1}^5S_i A_{(i)}\,,
\label{BCJ3D}
\eqe 
where $S_i$ are given by
\eqa
S_1&=&s_{123}(s_{126}s_{134}(s_{136}+s_{145})+s_{125}(s_{126}s_{134}+s_{134}s_{136}-s_{145}s_{156})+s_{124}(s_{125}(s_{126}+s_{136})\nn \\&&\null
+s_{126}(s_{134}+s_{136})+s_{136}(s_{134}+s_{156}))) \,, \nn \\
S_2&=&s_{124}(s_{125}(s_{136}+s_{145})s_{156}+s_{126}((s_{125}+s_{136})s_{156}-s_{134}s_{145})\nn \\&&\null
+s_{123}(s_{136}(2 s_{34} - s_{125})+s_{125}(2 s_{16} - s_{146}))) \,, \nn \\
S_3&=&-s_{145}(s_{123}(s_{136}(2 s_{34} - s_{125})+s_{125}(2 s_{16} - s_{146}))+(s_{125}(s_{126}+s_{136})+s_{126}s_{136})\nn \\&&\null \times (s_{134}+s_{156})+s_{124}(s_{125}(s_{126}+s_{136})+s_{126}(s_{134}+s_{136})+s_{136}(s_{134}+s_{156}))) \,, \nn \\
S_4&=&(s_{134}(s_{126}(s_{136}+s_{145})+s_{125}(2 s_{45}-s_{123} ))+s_{124}(s_{126}(s_{134}+s_{136})+s_{134}(s_{136}+s_{145})\nn \\&&\null
+s_{125}(2 s_{45}-s_{123} )))s_{156}+s_{123}(s_{124}(s_{134}s_{136}+s_{126}(s_{134}+s_{136})+s_{134}s_{156}\nn \\&&\null
+s_{136}s_{156}+s_{125}(2 s_{16} - s_{146}))+s_{134}((s_{136}+s_{145})(s_{126}+s_{156})+s_{125}(2 s_{16} - s_{146})))\,, \nn \\
S_5&=&-s_{126}(s_{125}s_{145}(s_{134}+s_{156})+s_{124}(s_{125}s_{145}+s_{134}s_{145}-s_{136}s_{156}))\nn \\&&\null
+s_{123}(2s_{124}s_{136}s_{34}+s_{125}(s_{134}s_{136}-s_{145}(s_{126}+s_{156})))\,.
\eqae

As stated, for dimensions $D>3$ the matrix $\Theta_{ij}$ is of full rank and BCJ amplitude relations are absent. However, for $D=2$ the matrix is in fact of rank three, giving further relations for the two-dimensional S-matrix. We will discuss the two-dimensional case in further detail in the next subsection.

Going beyond six points, we find that $\Theta^{\rm ABJM}_{ij}$ has full rank in $D=3$ as well as $D>3$, as explicitly verified up to ten points.\footnote{This result has been independently verified at eight points in ref.~\cite{allic}.} This implies that, in the absence of further constraints imposed on the amplitude numerators, there are no BCJ relations for three-dimensional ABJM amplitudes beyond six points. Even so, because the $\Theta^{\rm ABJM}_{ij}$ matrix is full rank we can invert it and obtain numerators that by construction satisfy the same properties as the color factors. Yet these numerators do not seem to have the desirable properties that one expects of a color-dual representation. We have explicitly verified that the double copy of these eight-point numerators do not give the correct supergravity amplitude, as obtained from recursion or dimensional reduction. Given that it has been shown that three-dimensional $\mathcal{N}=12$ supergravity is unique~\cite{deWit:1992up}, the double copy cannot compute an amplitude in any other meaningful theory. Hence, this is an interesting example of a situation when the double-copy procedure does not work even though duality-satisfying numerators can be obtained. The result is surprising considering the close relationship between ABJM and BLG tree-amplitudes. 

As discussed in section~\ref{PartialAmpDef}, at six points one can obtain the ABJM partial amplitudes from the BLG partial amplitudes via supersymmetry truncation, as given by \eqn{BLGfromABJM6pts}. Thus any BCJ relation or double-copy formula that is valid for BLG partial amplitudes have a corresponding relation for ABJM partial amplitudes. However, beyond six points one can no longer obtain ABJM partial amplitudes via supersymmetry truncation of BLG theory; {\it e.g.} the map in \eqn{BLGfromABJM8pts} is not invertible.  Thus the previous success in obtaining the correct supergravity amplitudes, at six points~\cite{Till, HenrikYt}, from either three-algebra color-kinematic duality, can be viewed as a consequence of the ability to identify the partial amplitudes of ABJM-type theories with that of BLG.

As a possible resolution of this puzzle, one might wonder if there are additional constraints beyond those of \eqn{3algebraCK} that needs to be imposed on the ABJM numerators starting at eight points. For example, the fundamental identities that the ABJM numerators satisfy are always a subset of the BLG fundamental identities.   One can wonder whether imposing the full set of BLG fundamental identities on ABJM numerators cures the observed problem. At six points, this works well: the number of quartic diagrams for BLG theory is exactly ten, while the number for ABJM theory is nine. Even though on the outset, it appears that the ABJM theory lacks one channel compared to BLG one can use generalized gauge freedom to set the numerator of the offending channel to zero, $n_{10}=0$ in \eqn{BLG6pt1}. With these constraints, the ABJM and BLG numerators satisfy exactly the same algebraic properties. At eight points, the same procedure does not work: there are 280 quartic diagrams in BLG theory, compared to 216 for ABJM. The generalized gauge freedom for BLG gives that there are $56-38=18$ free numerators. But the discrepancy to $280-216=64$ is too large, hence one cannot choose a gauge such that the BLG numerator constraints can be directly transferred to ABJM. Nevertheless, there might be other constraints that can be imposed on the ABJM numerators at eight points. In section~\ref{beyondBCJ} we explain that there exists many amplitude relations for ABJM theory (as well as BLG theory) whose origin are not understood. These relations could support the existence of new constraints that can consistently be imposted on ABJM numerators.

%%%%%%%%%%%%%%%%%%%%%%%%%%%%%%%%%%%%%%
 \section{Supergravity integrability and $D=2$ BCJ duality}
%%%%%%%%%%%%%%%%%%%%%%%%%%%%%%%%%%%%%%
\label{SecIntegrability}
We now consider ABJM, BLG and supergravity amplitudes living in two-dimensional spacetime.  To be specific, we take the supergravity theory to be either the maximally supersymmetric ${\cal N}=16$ theory, or the reduced version with ${\cal N}=12$ supersymmetries. However, at tree level, all pure supergravity theories are simple truncations of the maximal theory.  

We obtain the $D=2$ gauge-theory amplitudes by analytically reducing the three-dimensional ones, and the supergravity amplitudes from the double-copy procedure. Because of the highly constrained on-shell kinematics, special attention is needed to avoid collinear and soft divergences, even at tree level. While it should be possible to compute sensible physical quantities ({\it e.g.} cross sections) for any momenta, here we restrict ourself to kinematical configurations where the massless tree-level S-matrix is finite. As we will see, this is sufficient for a number of interesting observations. 

For the purpose of maintaining a finite tree-level S-matrix, we initially choose the momenta of the external color-ordered particles to be pointing in alternating light-like directions: $k_{2i+1}^\mu \parallel \bar{e}^\mu \equiv (1,1)$ and $k_{2i}^\mu \parallel e^\mu \equiv (1,-1)$, where $e,\bar{e}$ are two light-like basis vectors. In particular, we have
\eq
k_i=\bar{\kappa}_i \bar{e}~~{\rm for}~i~{\rm odd},\quad\quad k_j=\kappa_j e~~{\rm for}~j~{\rm even}\,,
\label{KenDef1}
\eqe
where $\kappa_i,\bar{\kappa}_j$ are scaling factors for the momenta. Momentum conservation implies that $\sum_i \kappa_i=\sum_j \bar{\kappa}_j=0$.
Thus, for this setup, the momentum direction is correlated with the chirality of the particles. In fact, this correlation is responsible for the absence of collinear and soft divergences, as such would require that on-shell chiral particles could evolve or split into one or several on-shell antichiral particles, $1\rightarrow \bar 2+ \bar 3+\ldots+ \bar n$.  This is forbidden by supersymmetry (note that, for supergravity, chirality corresponds to helicity from the $D=4$ theory perspective).

We begin with the six-point ABJM amplitude. At six points, for kinematics (\ref{KenDef1}), the matrix $\Theta_{ij}^{\rm ABJM}$ has rank three. This implies that there are at most three independent six-point amplitudes. The two BCJ relations can be obtained by recycling the details and notation used in section~\ref{ABJMandBCJsection}:  we solve the numerators $n_i$ in terms of $A_{(2)},A_{(4)},A_{(5)}$ and $n_3,n_4$, then substitute the solution back into $A_{(1)}$ or $A_{(3)}$, similar to above.  The two independent amplitude relations, valid for alternating two-dimensional momenta, are given by
\begin{align}
0&=(A_{(1)} s_{25} - A_{(2)} s_{16}) (s_{14} s_{12} - s_{56} s_{36}) 
+ (A_{(4)} s_{12} - A_{(5)} s_{56}) (s_{14} s_{36} + s_{25} s_{14} + s_{25}s_{36}), \nn \\
0&=(A_{(2)} s_{16} - A_{(3)} s_{34}) (s_{14} s_{12} - s_{56} s_{36})
+ (A_{(4)} s_{36} - A_{(5)} s_{14}) (s_{34} s_{12} + s_{56} s_{34} + s_{56} s_{12})\,. 
\label{BCJ2D}
\end{align}
Interestingly, the coefficients of the amplitudes are only degree-three polynomials of momentum invariants, and moreover the relations are significantly simpler than the corresponding three-dimensional ABJM relation (\ref{BCJ3D}).  Unlike the three-dimensional case, $\Theta_{ij}^{\rm ABJM}$ continues to have less-than-full rank even beyond six points. We determined the rank up to ten points; for each multiplicity we find novel BCJ relations.  The counts of independent ABJM amplitudes, subject to the two-dimensional BCJ relations and kinematics (\ref{KenDef1}),  are
\eq
  \centering \begin{tabular}{|c|c|c|c|c|c|c|}
\hline
  external legs & 4 & 6& 8& 10 \\ \hline
  $D=2$ BCJ basis & 1 & 3 & 38 &  987  \\
\hline
\end{tabular}\,\,.
 \eqe

We can now demonstrate some interesting applications of color-kinematics duality to $D=2$ supergravity amplitudes. Using the double-copy formula (\ref{Sugra}), we can now derive a gauge-invariant expression that gives six-point two-dimensional  supergravity amplitudes (with manifest $\mathcal{N}=12$ supersymmetry) in terms of the two-dimensional ABJM amplitudes. We have
\eqa
\nonumber\mathcal{M}_6(\bar{1},2,\bar{3},4,\bar{5},6)&=&\frac{s_{12}s_{34}s_{56}}{(s_{23}-s_{14})(s_{36}-s_{12})(s_{34}-s_{16})} \Big((s_{34}-s_{16}) A_{(1)} {\tilde A}_{(1)}\\
&&\hskip 2cm \null+(s_{36}-s_{12})A_{(2)}{\tilde A}_{(2)}+(s_{23}-s_{14})A_{(3)}{\tilde A}_{(3)} \Big)\,,\quad
\label{D=2DC}
\eqae
where the formula is only valid for the kinematics in \eqn{KenDef1}, and the gravity states correspond to chiral and antichiral $\mathcal{N}=12$ supermultiplets. The $A_{(i)}$ and ${\tilde A}_{(i)}$ are ABJM amplitudes with color ordering defined in section~\ref{ABJMandBCJsection}. The explicit two-dimensional form of the ABJM amplitudes can be obtained by direct dimensional reduction of the three-dimensional amplitudes. For the kinematics in \eqn{KenDef1}, the superamplitude takes a very simple form
\eqa
\nonumber A^{\rm ABJM}_{D=2}(\bar 1,2, \bar 3, 4, \bar 5,6)&=&i \frac{\delta^{(3)}(\sum_{\rm even} \lambda_{i} \eta_i)
\delta^{(3)}(\sum_{\rm odd}  \bar \lambda_i \eta_i)}{\bar \lambda_1 \lambda_2 \bar \lambda_3 \lambda_4 \bar \lambda_5 \lambda_6} \\
&&\times \sum_{s=\pm}\delta^{(3)} \left(s\frac{\bar \lambda_3 \eta_1 - \bar \lambda_1 \eta_3}{ \bar\lambda_5}+ i \frac{ \lambda_6 \eta_4 - \lambda_4 \eta_6}{ \lambda_2}\right)\,,
\eqae
where $\lambda$, $\bar \lambda$ are scalar-valued spinor-helicity variables, which are related to the lightcone momenta: $\kappa_i=(\lambda_i)^2$ and $\bar \kappa_i=(\bar \lambda_i)^2$. Upon close inspection we note that $A^{\rm ABJM}_{D=2}(\bar 1,2, \bar 3, 4, \bar 5,6)$, is, in fact, totally symmetric in the (1,3,5) labels and totally antisymmetric in the (2,4,6) labels. Compensating for the Fermionic statistics on the even sites, the six distinct orderings of the amplitudes give identical result $A_{(i)}=A_{(j)}$ (note we have tacitly assumed Bosonic amplitudes in \eqn{BCJ2D} and \eqn{D=2DC}). As a consequence, after pulling out the common factor $A_{(1)}{\tilde A}_{(1)}$, and accounting for momentum conservation $\sum_{j}s_{ij}=0$, \eqn{D=2DC} becomes manifestly zero! Is it a coincidence that the six-point supergravity amplitude vanishes for this alternating kinematics? No, as we will see, it vanishes for all kinematics that is not plagued by collinear or soft divergences.

For non-alternating two-dimensional kinematics the ABJM superamplitudes as well as the entries of the $\Theta_{ij}$ matrix become divergent. However, one can typically remedy the situation by imposing external state choices that eliminates the divergent channels. The kinematics in \eqn{KenDef1} was fortunate to have this property automatically satisfied. For other kinematics one can proceed more carefully. We use the double-copy representation in \eqn{BLGGravAmp}, and regulate the infrared divergences using $D=3$ momenta $k_i^M=(k_i^\mu,m_i)$, where $k_i^\mu$ are two-dimensional momenta and $m_i^2$ are very small parameters. The three-dimensional momenta are taken to be massless, so the two-dimensional ones are massive $k_i^2-m_i^2=0$. Let us be explicit and consider the limit $m_i  \rightarrow 0$, such that
 \eq
 k_i \rightarrow \bar{\kappa}_i \bar{e}~~{\rm for}~~i=1,2,5, \quad\quad  k_i \rightarrow \kappa_i e~~{\rm for}~~i=3,4,6\,.
 \label{KenDef2}
 \eqe
Note that for a supergravity amplitude $\mathcal{M}_6(\bar{1},2,\bar{3},4,\bar{5},6)$, with manifest ${\cal N}=12$ supersymmetry, any kinematics with three particles in each light-like direction is related to either \eqn{KenDef1} or \eqn{KenDef2} via trivial relabeling. For the kinematics (\ref{KenDef2}), the channel $n_4\tilde{n}_4/s_{125}$ naively diverges, since $s_{125}\rightarrow 0$. However, after choosing a component amplitude that does not have such pole, we are allowed to set $n_4$ or $\tilde{n}_4$ to zero. For example, the ABJM component amplitude proportional to $(\eta_1^1\eta_2^1\eta_5^1)(\eta_2^2\eta_4^2\eta_6^2)(\eta_1^3\eta_3^3\eta_5^3)$, where the superscripts are SU(3) indices, has a vanishing $s_{125}$ pole, so we set  $n_4=0$. The state choices of the first parenthesis ensures that all partial amplitudes are finite for configuration (\ref{KenDef2}). The particular form of the ABJM partial amplitudes can be conveniently obtained of from the ``cyclic gauge" of ref.~\cite{Gang}. At leading order, they are given as
\eqa
\nonumber 
&&A_{(1)}=\mathcal{O}(m^2)\,,\quad 
A_{(3)}=\mathcal{O}(m^2)\,,\quad 
A_{(4)}=\mathcal{O}(m^2)\,,\quad 
A_{(6)}=\mathcal{O}(m^2)\,,\\
&&A_{(2)}=-i\frac{\kappa_3\bar{\kappa}_2}{\lambda_6\lambda_4\bar{\lambda}_1\bar{\lambda}_5}+\mathcal{O}(m)\,,\quad A_{(5)}=-A_{(2)}+\mathcal{O}(m)\,.
\label{CyclicGauge}
\eqae
Note that we cannot drop the $\mathcal{O}(m)$ and $\mathcal{O}(m^2)$ terms immediately, because if we use the double-copy formula in \eqn{BLGGravAmp} the denominator $B$ also vanishes in this limit, $B =\mathcal{O}(m^2)$. 
Remarkably, after plugging amplitudes (\ref{CyclicGauge}) into \eqn{BLGGravAmp}, and taking the limit  (\ref{KenDef2}),  the supergravity amplitude again vanishes! Thus we see evidence of the general pattern: as long as the ABJM amplitudes are not contaminated by divergent channels, the corresponding supergravity amplitude always vanishes.

The spectacular vanishing is perhaps not unexpected; it can be understood as a statement supporting integrability of the theory. Integrability of maximal $D=2$ supergravity, which is a theory that has a dimensionless coupling yet should be non-conformal, has been argued in~\cite{Nicolai:1987kz, Nicolai:1998gi}. Integrability of the two-dimensional S-matrix should imply that all higher-point amplitudes vanish for generic kinematics, except for momenta that allows for factorization into products of the four-point S-matrix~\cite{ZZ}. However, as the massless S-matrix is contaminated by infrared divergences, indicating that its asymptotic states are not properly identified, we can expect to see some deviation from this statement. Nevertheless, for gravity amplitudes that are manifestly free of infrared problems, we confirm that they behave as expected from integrability.

We will now discuss the properties found to be consistent with integrability, starting with the four-point amplitude. For kinematical reasons, the four-point amplitude in $D=2$ is nonvanishing only for elastic scattering, {\it e.g.} on $\delta^{(2)}(k_1+k_3)\delta^{(2)}(k_2+k_4)$ support, in any massless theory.  For the alternating momentum configuration (\ref{KenDef1}), the four-point ABJM tree superamplitude is given by
\eq
A^{\rm ABJM}_{D=2}(\bar 1,2, \bar 3, 4)=i \frac{\delta^{(3)}(\sum_{\rm even} \lambda_{i} \eta_i)
\delta^{(3)}(\sum_{\rm odd}  \bar \lambda_i \eta_i)}{\bar \lambda_1 \lambda_2 \bar \lambda_3 \lambda_4}\,,
\eqe
and a corresponding supergravity amplitude, with ${\cal N}=12$ supersymmetry manifest, is given by $M_{D=2}(\bar 1,2, \bar 3, 4)=A^{\rm ABJM}_{D=2}(1,\bar 2, 3, \bar 4) \tilde{A}^{\rm ABJM}_{D=2}(1,\bar 2, 3, \bar 4)$. For other kinematics, some of the component amplitudes are divergent, and thus needs to be regulated, or otherwise carefully treated. We will not discuss the the divergent component amplitudes in any detail. 

In order to have a consistent factorization of higher-multiplicity amplitudes, the integrable S-matrix should satisfy the Yang-Baxter equation. In three-dimensions, it is known~\cite{ArkaniHamed:2012nw} that the sewing of three four-point tree amplitudes in ABJM theory indeed satisfies a ``Yang-Baxter-like" identity. Pictorially it is
\eq
\includegraphics[scale=0.75]{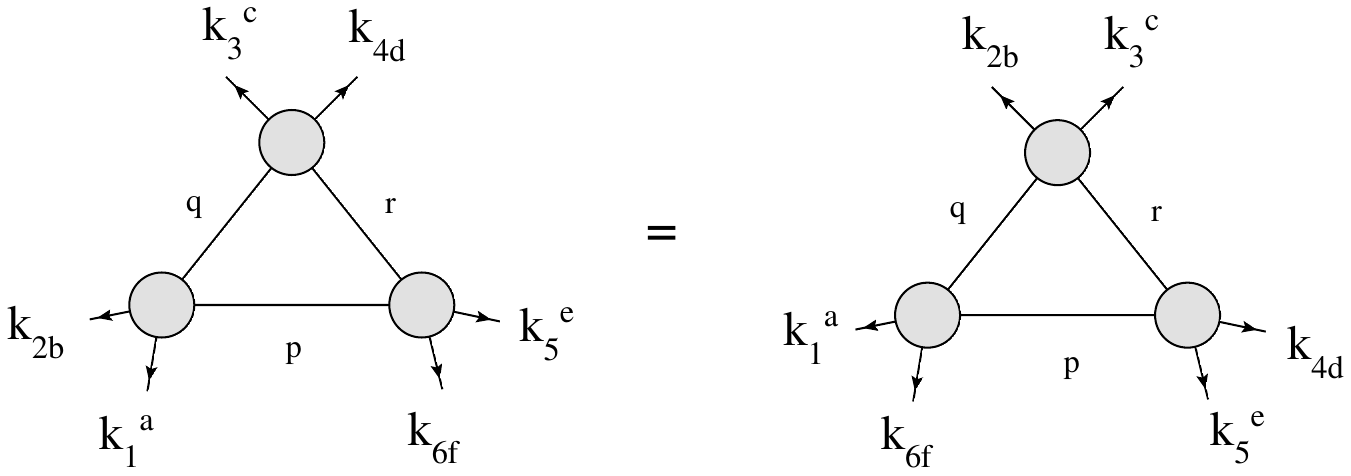}\,,
\eqe
where each blob represents a four-point amplitude and each internal line indicates a sum over the spectrum as well as an on-shell phase space integral. In fact, these diagrams are usually called unitarity triple cuts. The indices $a,b,\cdots,f$ on each external lines indicate the particle species of each leg, with lower and upper indices being chiral and antichiral states, respectively. This identity stems from the fact that the two triple cuts can be mapped to two different BCFW representations of the six-point ABJM amplitude, which by consistency of the BCFW recursion has to be identical~\cite{ABJMBCFW}.  However, the identity is not precisely the Yang-Baxter equation; the external and internal momenta do not correspond to elastic scattering for generic $D=3$ kinematics.  This is cured by taking the two-dimensional limit, where the four-point amplitudes forces the momenta into this configuration. More precisely, the kinematics degenerates into only three momentum lines, $k_1, k_2, k_3$, which are diagonally identified across each four-point amplitude.  Straightening out the lines in the diagrams, the identity looks like
\eq 
\includegraphics[scale=0.7]{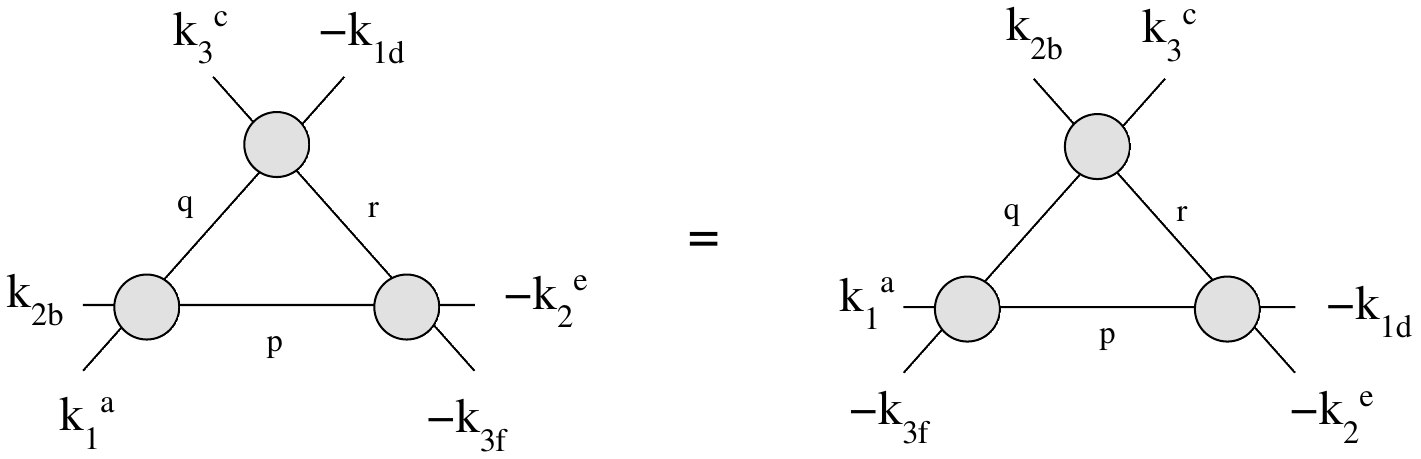}\,.
\eqe
This is precisely the Yang-Baxter equation:
%
\iffalse
\eqa
\nonumber &&\mathcal{S}^{ab}(k_1,k_2)\mathcal{S}^{cd}(k_3,-k_1)\mathcal{S}^{ef}(-k_2,-k_3)\\
\nonumber&\equiv&\sum_{ {\rm states}}A_4(\ell_1,k_1^a,k_2^b,-\ell_2)A_4(\ell_2,k^c_3,-k^d_1,-\ell_3)A_4(\ell_3,-k^e_2,-k^f_3,-\ell_1)\\
&=&\mathcal{S}^{fa}(-k_3,k_1)\mathcal{S}^{bc}(k_2,k_3)\mathcal{S}^{de}(-k_1,-k_2)\,.
\eqae
\fi
%
\iffalse
\eq
\mathcal{S}^{ap}_{\phantom{ap}bq}(k_1,-k_2)\mathcal{S}^{cq}_{\phantom{cq}dr}(k_3,-k_1)\mathcal{S}^{er}_{\phantom{er}fp}(k_2,-k_3)
=\mathcal{S}^{ap}_{\phantom{ap}fq}(k_1,-k_3)\mathcal{S}^{eq}_{\phantom{eq}dr}(k_2,-k_1)\mathcal{S}^{cr}_{\phantom{cr}bp}(k_3,-k_2)\,.
\eqe
%
\fi
%
%
\eq
\mathcal{S}^{ap}_{bq}(k_1,-k_2)\mathcal{S}^{cq}_{dr}(k_3,-k_1)\mathcal{S}^{er}_{fp}(k_2,-k_3)
=\mathcal{S}^{ap}_{fq}(k_1,-k_3)\mathcal{S}^{eq}_{dr}(k_2,-k_1)\mathcal{S}^{cr}_{bp}(k_3,-k_2)\,.
\label{YangBaxterEq}
\eqe
To see this more clearly we subtract out the trivial part and only consider the transfer matrix $i\mathcal{T}=\mathcal{S}-1$, and let $T=\mathcal{T}^{(0)}$ be the tree-level contribution, we can then write this object directly as a unitarity cut. It is given by
\eqa
&&T^{ap}_{bq}(k_1,-k_2)T^{cq}_{dr}(k_3,-k_1)T^{er}_{fp}(k_2,-k_3)\equiv 
\\ \nonumber  && \hskip1.1cm \null J\hskip -2mm \sum_{p,q,r \in {\rm states}} \hskip -4mm A_4(\ell_1^p,k_1^a,(-k_2)_b,(-\ell_2)_q)\, A_4(\ell_2^q,k^c_3,(-k_1)_d,(-\ell_3)_r)\,A_4(\ell_3^r,k^e_2,(-k_3)_f,(-\ell_1)_p)\,,
\eqae
where the internal momenta is constrained by on-shell conditions $\ell_i^2=0$ and momentum conservation $\ell_2=\ell_1+k_1-k_2$, $\ell_3=\ell_1+k_3-k_2$, and $J=\int d^DÊ\ell_1 \delta(\ell_1^2)\delta(\ell_2^2)\delta(\ell_3^2)$ is the Jacobian factor from the phase-space integration, and for convenience we have dropped factors of $2,\pi, i$ and $g$.
Thus, we observe that the four-point tree amplitude in ABJM, evaluated in two dimensions, is a solution to the Yang-Baxter equation at lowest nontrivial order in the perturbative expansion! Note that at least one of the four-point amplitudes in the equivalence (\ref{YangBaxterEq}) will diverge for $D=2$ kinematics, since at least two of the light-light momenta $k_1,k_2,k_3$ necessarily become collinear in the $D=2$ limit, resulting in a soft-exchange singularity. However both sides will diverge in the same fashion. The simplest way to see this, is to regulate the divergence using three-dimensional kinematics and amplitudes, treating the extra-dimensional momenta as a mass regulator for the $D=2$ divergence. Since the three-dimensional version already satisfies the equality, so will the regulated $D=2$ result.
 
Finally, as the supergravity four-point tree amplitude is simply a double-copy of the corresponding ABJM amplitude, we may simply square the $T$-matrix part of \eqn{YangBaxterEq}, it also satisfies the Yang-Baxter equation,
\eq
[T^{ap}_{bq}(k_1,-k_2)T^{cq}_{dr}(k_3,-k_1)T^{er}_{fp}(k_2,-k_3)]^2
=[T^{ap}_{fq}(k_1,-k_3)T^{eq}_{dr}(k_2,-k_1)T^{cr}_{bp}(k_3,-k_2)]^2\,.
\eqe
If we are to be careful, we must divide both sides of the equation by the Jacobian factor $J$ and take care of factors of $2\pi$ and $i$, in order to obtain properly normalized supergravity $T$-matrices, but it does not matter for the validity of the equivalence.
So while the $D=2$ ABJM theory Yang-Baxter equation is an interesting curiosity, it is the $D=2$ supergravity double-copy version of this identity that has real bearing on the integrability of the theory, since only the latter theory satisfies the integrability factorization property at six points.

As stated, in an integrable theory, all higher-point amplitudes should vanish unless the kinematics correspond to a factorization channel given by products of four-point amplitudes. At six points, we confirmed that the alternating light-like momenta (\ref{KenDef1}) give vanishing supergravity tree amplitudes. This momenta do not correspond to elastic scattering for any of the values of $\kappa_i$ and $\bar{\kappa}_i$, since such would require an even number of particles on each of the two lightcones directions. And such elastic scattering is needed for a non-vanishing four-point amplitude in the factorization channels. This implies that the six-point amplitude should vanish identically on kinematics (\ref{KenDef1}), as we indeed find.  The six-point amplitudes in ABJM and BLG theories are non-vanishing for two-dimensional momenta (\ref{KenDef1}); this shows that neither $D=2$ ABJM theory nor $D=2$ BLG theory are integrable, at least not in the normal sense of two-dimensional integrability. 

The same analysis applies for the eight-point tree amplitudes. For supergravity, the amplitude $M_{D=2}(\bar 1,2, \bar 3, 4,\bar 5, 6,\bar 7, 8)$ again vanishes on kinematics (\ref{KenDef1}), providing non-trivial support for integrality. The corresponding two-dimensional ABJM and BLG amplitudes are non-vanishing. The construction of the supergravity eight-point amplitude was done as follows:  We used BCFW recursion to obtain the eight-point ABJM partial amplitudes, then we applied the linear map in \eqn{BLGfromABJM8pts} to obtain the corresponding eight-point BLG partial amplitudes. Using the BLG partial amplitudes we solved for the numerators in a color-kinematics dual representation. By squaring these numerators, we obtain three-dimensional supergravity amplitudes. After confirming that these give the correct answer in $D=3$  (by comparing to three-dimensional BCFW recursion of supergravity) we used the three-dimensional momenta as a regulator when approaching the $D=2$ kinematics (\ref{KenDef1}) in a limiting procedure. The limit is well behaved; the supergravity amplitude vanishes, as expected. 

As explained in the beginning of this section, the $D=2$ ABJM theory enjoys novel BCJ relations beyond six points. One might expect that the presence of BCJ relations is indicative of the existence of a ABJM double-copy formula for supergravity (without going through BLG amplitudes as an intermediate step).  Indeed, at six points this works without problems. However, at eight points the situation is somewhat unclear. We have verified that the double-copy result is overall non-vanishing for kinematics (\ref{KenDef1}), suggesting that it does not compute the correct $D=2$ supergravity amplitudes. However, many of the component amplitudes of the ABJM double copy do exhibit non-trivial vanishings (even when the corresponding ABJM component amplitudes are non-zero), suggesting that the the double copy may compute some meaningful quantity. One might ask if the non-vanishing double-copy amplitudes could correspond to some deformation of the supergravity theory Lagrangian. But since the amplitudes have manifest $\mathcal{N}=12$ supersymmetry, the room for deformations are small, even possibly non-existent, similar to the corresponding higher-dimensional supergravitites. Thus this puzzle is an first example of a case where, although BCJ-amplitude relations exists, the corresponding double-copy formula does not give the expected gravity amplitudes.  Further studies of $D=2$ ABJM theory BCJ relations and double-copy amplitude are likely needed to bring full clarity into this.

%%%%%%%%%%%%%%%%%%%%%%%%%%%%%%%%%%%%%%
 \section{Bonus relations and more~\label{SecBonus}}
 %%%%%%%%%%%%%%%%%%%%%%%%%%%%%%%%%%%%%%
 
We end the discussion on tree-level amplitude relations by exposing additional hidden relations in ABJM  and BLG theory. As mentioned in section~\ref{PartialAmpDef}, one can obtain six-point ABJM partial amplitudes from the BLG theory simply via supersymmetry truncation. From this it follows that ABJM should obey two BCJ-like relations at six points, one more than observed, since that is the count for BLG theory.  As we show below, this extra relation can be manifested as a bonus relation due to the improved asymptotic behavior of the BCFW shift. However, even after talking into account BCJ and bonus relations, there are still more unexplored structure that relates different partial amplitudes.

 %%%%%%%%%%%%%%%%%%%%%%%%%%%%%%%%%%%%%%
 \subsection{Bonus relations from large-$z$ behavior}
 %%%%%%%%%%%%%%%%%%%%%%%%%%%%%%%%%%%%%%
It is well known Yang-Mills-theory amplitudes enjoys improved large-$z$ falloff as the BCFW shifted legs become non-adjacent. For any non-adjacent shift the Yang-Mills amplitude vanishes as $1/z^2$. This behavior can be shown straightforwardly~\cite{CachazoZ} using the RSVW twistor string formula of $\mathcal{N}=4$ SYM~\cite{RSVW}. Feng, Huang and Jia shown that the BCJ amplitude relations can be cast as bonus relations that emerge from this improved large-$z$ behavior~\cite{Feng:2010my}. 

It was discussed in~\cite{Gang} (see also appendix~\ref{LargeZ}) that ABJM also enjoys improved large-$z$ behavior. Unlike $\mathcal{N}=4$ SYM, the large-$z$ falloff continues to improve as the shifted legs are taken further apart in the color ordering. For example, for $n=4,6,8$ one has
\eqa
\nonumber n=4&:& (1,2)\rightarrow \frac{1}{z},\;(1,3)\rightarrow \frac{1}{z^2}\,,\\
\nonumber n=6&:& (1,2)\rightarrow \frac{1}{z},\;(1,3)\rightarrow \frac{1}{z^2},\;(1,4)\rightarrow \frac{1}{z^3}\,,\\
n=8&:& (1,2)\rightarrow \frac{1}{z},\;(1,3)\rightarrow \frac{1}{z^2},\;(1,4)\rightarrow \frac{1}{z^3}\,\;(1,5)\rightarrow \frac{1}{z^2}\,,
\eqae
where we use $(i,j)$ to indicate the shifted legs. 
 
The fact that the amplitude enjoys improved large-$z$ falloff beyond that necessary for BCFW recursion, $1/z$, can be utilized to extract non-trivial linear relations between amplitudes. In ref.~\cite{Feng:2010my} the authors started with the KK relations, multiply an inverse propagator and then apply the standard BCFW shift, which can become non-adjacent shift depending on the ordering of the amplitude since the KK relations includes amplitudes of different ordering.  Although the inverse propagator introduces an extra power of $z$ at large $z$, due to the improve large-$z$ behavior of non-adjacent shift some of the amplitudes appearing in a given KK relation will not contribute at $z\rightarrow\infty$. As the KK relation holds irrespective of the value of $z$, this implies non-trivial linear relations among the amplitudes that do contribute at $z\rightarrow\infty$. In the following we will use the six-point ABJM amplitude to illustrate this analysis.

We begin again by BCFW shifting legs 1 and 6 for the six-point KK identity. For simplicity, we will consider the pure scalar amplitude $A_6(\bar{\phi}_i\phi_j\bar{\phi}_k\phi_l\bar{\phi}_m\phi_n)$: 
\eqa
\nonumber&&A_6(\hat{\bar{1}}2\bar{3}4\bar{5}\hat{6})+A_6(\hat{\bar{1}}4\bar{3}\hat{6}\bar{5}2)+A_6(\hat{\bar{1}}\hat{6}\bar{3}2\bar{5}4)+A_6(\hat{\bar{1}}4\bar{3}2\bar{5}\hat{6})+A_6(\hat{1}\hat{\bar{6}}\bar{3}4\bar{5}2)+A_6(\hat{\bar{1}}2\bar{3}\hat{6}\bar{5}4)=0\,.\\
\label{6ptkk}
\eqae
Picking the un-barred scalars to carry the same SU(4) R-index and the barred scalars carrying the conjugate one, the six-point color ordered amplitude is given by:
\eqa
&&A(\bar{\phi}_{1I}\,\phi_2^{\phantom{2}I}\, \bar{\phi}_{3I}\, \phi_4^{\phantom{4}I}\, \bar{\phi}_{5I}\, \phi_6^{\phantom{6}I})=Y^{z}(\bar{\hat{1}}2\bar{3}4\bar{5}\hat{6})+Y^{z*}(\hat{\bar{1}}2\bar{3}4\bar{5}\hat{6})\\
\nonumber &&~~~\equiv\frac{i}{(\langle 2| p_{246}|5\rangle+i\langle 46\rangle\langle31\rangle) (\langle 4| p_{246}|1\rangle+i\langle 62\rangle\langle53\rangle) (\langle 6| p_{246}|3\rangle+i\langle 24\rangle\langle15\rangle) }\,+\, {\rm c.c.}\,,
\eqae
where we have used the function $Y^{z,z*}$ to denote their origin as the two terms in the BCFW recursion, with legs $1$ and $6$ shifted. Due to the propagator's quadratic dependence on the BCFW deformation parameter~\cite{Gang}, there are two solutions on each factorization channel, denoted by $z$ and $z^*$ respectively.

Multiplying \eqn{6ptkk}) by $s_{\hat{1}23}(z)$, we consider the following $z$-integral: 
\eqa
\nonumber\oint_{z=\infty}\frac{s_{\hat{1}23}(z)}{1-z} &&\bigg(A_6(\hat{\bar{1}}2\bar{3}4\bar{5}\hat{6})+A_6(\hat{\bar{1}}4\bar{3}\hat{6}\bar{5}2)+A_6(\hat{\bar{1}}\hat{6}\bar{3}2\bar{5}4)+A_6(\hat{\bar{1}}4\bar{3}2\bar{5}\hat{6})\\
&&~\null +A_6(\hat{\bar{1}}\hat{6}\bar{3}4\bar{5}2)+A_6(\hat{\bar{1}}2\bar{3}\hat{6}\bar{5}4)\bigg)=0\,.
\label{LargeZInt}
\eqae
Let us look at which of these terms contribute to the integral. At large $z$, the terms in the integrand shifts as:
\eqa
\nonumber s_{\hat{1}23}(z)\rightarrow z^2,\;A_6(\hat{\bar{1}}2\bar{3}4\bar{5}\hat{6})\rightarrow\frac{1}{z},\;A_6(\hat{\bar{1}}4\bar{3}\hat{6}\bar{5}2)\rightarrow\frac{1}{z^3},\;A_6(\hat{\bar{1}}\hat{6}\bar{3}2\bar{5}4)\rightarrow\frac{1}{z}\,,\\
A_6(\hat{\bar{1}}4\bar{3}2\bar{5}\hat{6})\rightarrow\frac{1}{z},\;A_6(\hat{\bar{1}}\hat{6}\bar{3}4\bar{5}2)\rightarrow\frac{1}{z},\;A_6(\hat{\bar{1}}2\bar{3}\hat{6}\bar{5}4)\rightarrow\frac{1}{z^3}\,.
\eqae
The terms $s_{\hat{1}23}(z)A_6(\hat{\bar{1}}4\bar{3}\hat{6}\bar{5}2)$ and $s_{\hat{1}23}(z)A_6(\hat{\bar{1}}2\bar{3}\hat{6}\bar{5}4)$ scales as $1/z$ thus vanishes as $z\rightarrow\infty$. For these terms, the residue at $z=1$ cancels with that at finite $z$ and thus do not contribute to the integral. For the others the pole at infinity is given by the sum of the residues of the finite poles as well as that of $z=1$. Explicitly the integral gives
\eqa
\nonumber &&s_{123}A_6(\bar{1}2\bar{3}4\bar{5}6)+\left[s_{123}A_6(\bar{1}6\bar{3}2\bar{5}4)-\sum_{z,z^*}s_{\hat{1}23}(z_{145})Y^{z,z^*}_6(\hat{\bar{1}}\hat{6}\bar{3}2\bar{5}4)\right]\\
\nonumber&&+\left[s_{123}A_6(\bar{1}4\bar{3}2\bar{5}6)-\sum_{z,z^*}s_{\hat{1}23}(z_{143})Y^{z,z*}_6(\hat{\bar{1}}4\bar{3}2\bar{5}\hat{6})\right]\\
&&+\left[s_{123}A_6(\hat{\bar{1}}\hat{6}\bar{3}4\bar{5}2)-\sum_{z,z^*}s_{\hat{1}23}(z_{125})Y^{z,z*}_6(\hat{\bar{1}}\hat{6}\bar{3}4\bar{5}2)\right]=0\,.
\label{BCJ0}
\eqae
We have used square brackets to indicate the contribution coming from each of the non-vanishing terms in \eqn{LargeZInt}). For the first term in \eqn{LargeZInt}), there is only a pole at $z=1$ since the finite pole, $1/s_{123}(z)$, was canceled by the pre factor. The third, fourth and fifth term in \eqn{LargeZInt}) is given by a sum of the residue at $z=1$ and the residue at the factorization pole, i.e.$s_{\hat{1}45}$, $s_{\hat{1}43}$ and $s_{\hat{1}25}$ respectively. We use the notation $s_{\hat{1}23}(z_{145})$ to indicate that it is the shifted invariant $s_{\hat{1}23}(z)$, with $z$ evaluated at the solution of $s_{\hat{1}45}(z)=0$. Note that since $s_{123}-s_{\hat{1}23}(z_{145})=-s_{145}\,$, the first square bracket in \eqn{BCJ0}) can be rewritten as:
$$s_{123}A_6(\bar{1}6\bar{3}2\bar{5}4)-\sum_{z,z^*}s_{\hat{1}23}(z_{145})Y^{z,z^*}_6(\hat{\bar{1}}\hat{6}\bar{3}2\bar{5}4)=-s_{145}A_6(1\bar{6}3\bar{2}5\bar{4})\,.$$
For the remaining two square brackets, the the $z$ dependence in the combination $-s_{123}+s_{123}(z_{134})$ and $-s_{123}+s_{123}(z_{125})$ does not drop out. This leads to the result that the two $Y$ functions are weighted differently. As we will soon see, it is convenient to write the sum as: 
\eqa
\nonumber\sum_{z,z^*}s_{123}(z_{143})Y^{z,z*}_6(\hat{\bar{1}}4\bar{3}2\bar{5}\hat{6})&=&\frac{s_{123}(z_{143})+s_{123}(z^*_{143})}{2}\left(Y^z_6(\hat{\bar{1}}4\bar{3}2\bar{5}\hat{6})+Y^{z^*}_6(\hat{\bar{1}}4\bar{3}2\bar{5}\hat{6})\right)\\
\nonumber&&+\frac{s_{123}(z_{143})-s_{123}(z^*_{143})}{2}\left(Y^z_6(\hat{\bar{1}}4\bar{3}2\bar{5}\hat{6})-Y^{z^*}_6(\hat{\bar{1}}4\bar{3}2\bar{5}\hat{6})\right)\\
\nonumber&\equiv&S[s_{123}(z_{143})]\left(Y^z_6(\hat{\bar{1}}4\bar{3}2\bar{5}\hat{6})+Y^{z^*}_6(\hat{\bar{1}}4\bar{3}2\bar{5}\hat{6})\right)\\
\nonumber&+&AS[s_{123}(z_{143})]\left(Y^z_6(\hat{\bar{1}}4\bar{3}2\bar{5}\hat{6})-Y^{z^*}_6(\hat{\bar{1}}4\bar{3}2\bar{5}\hat{6})\right)\,.
\eqae
 An important property of  the $Y$ functions is that while the sum gives the purely scalar tree amplitude, the difference gives the purely Fermionic tree amplitude (see eq.~(5.31) of~\cite{Gang} ). i.e.:
\eqa
\nonumber &&Y^z_6(\hat{\bar{1}}2\bar{3}4\bar{5}\hat{6})+Y^{z^*}_6(\hat{\bar{1}}2\bar{3}4\bar{5}\hat{6})=A(\bar{\phi}_{1I}\phi_2\,^I\bar{\phi}_{3I}\phi_4\,^I\bar{\phi}_{5I}\phi_6\,^I)\equiv A_{6\phi}(\bar{1}2\bar{3}4\bar{5}6)\,,\\
\nonumber &&Y^z_6(\hat{\bar{1}}2\bar{3}4\bar{5}\hat{6})-Y^{z^*}_6(\hat{\bar{1}}2\bar{3}4\bar{5}\hat{6})=-iA(\bar{\psi}_{1}\,^I\psi_{2I}\bar{\psi}_{3}\,^I\psi_{4I}\bar{\psi}_{5}\,^I\psi_{6I})\equiv -iA_{6\psi}(\bar{1}2\bar{3}4\bar{5}6)\,.
 \eqae
Thus we have finally arrived at the following linear relations for amplitudes:
\eqa
\nonumber &&s_{123}A_{6\phi}(\bar{1}2\bar{3}4\bar{5}6)-s_{145}A_{6\phi}(\bar{1}6\bar{3}2\bar{5}4)+s_{123}A_{6\phi}(\bar{1}4\bar{3}2\bar{5}6)+s_{123}A_{6\phi}(\bar{1}6\bar{3}4\bar{5}2)\\
\nonumber&&-S\bigg[s_{123}(z_{143})\bigg]A_{6\phi}(\bar{1}4\bar{3}2\bar{5}6)+iAS\bigg[s_{123}(z_{143})\bigg]A_{6\psi}(\bar{1}4\bar{3}2\bar{5}6)\\
&&-S\bigg[s_{123}(z_{125})\bigg]A_{6\phi}(\bar{1}6\bar{3}4\bar{5}2)+iAS\bigg[s_{123}(z_{125})\bigg]A_{6\psi}(\bar{1}6\bar{3}4\bar{5}2)=0\,.
\label{BCJ1}
\eqae

The identity in \eqn{BCJ1}) relates purely Bosonic amplitudes with purely Fermionic ones. If we were to start with the KK identity of the super amplitude, then the corresponding identity would related the amplitude whose Fermionic multiplets on the even sites, to those with Fermionic multiplet on the odd sites. One can multiply \eqn{6ptkk}) with $s_{134}(z)$ and repeat the above steps to obtain another linear relation:
\eqa
\nonumber &&s_{134}A_{6\phi}(\bar{1}2\bar{3}4\bar{5}6)+s_{134}A_{6\phi}(\bar{1}6\bar{3}2\bar{5}4)+s_{134}A_{6\phi}(\bar{1}4\bar{3}2\bar{5}6)-s_{125}A_{6\phi}(\bar{1}6\bar{3}4\bar{5}2)\\
\nonumber&&-S\bigg[s_{134}(z_{123})\bigg]A_{6\phi}(\bar{1}2\bar{3}4\bar{5}6)+iAS\bigg[s_{134}(z_{123})\bigg]A_{6\psi}(\bar{1}2\bar{3}4\bar{5}6)\\
&&-S\bigg[s_{134}(z_{145})\bigg]A_{6\phi}(\bar{1}6\bar{3}2\bar{5}4)+iAS\bigg[s_{134}(z_{145})\bigg]]A_{6\psi}(\bar{1}6\bar{3}2\bar{5}4)=0\,.
\label{BCJ2}
\eqae
One now has two equations that relate $A_{6\phi}(\bar{1}2\bar{3}4\bar{5}6)$, $A_{6\phi}(\bar{1}6\bar{3}2\bar{5}4)$, $A_{6\phi}(\bar{1}4\bar{3}2\bar{5}6)$ and $A_{6\phi}(\bar{1}6\bar{3}4\bar{5}2)$, which we will call our basis amplitudes denoted by $(\tilde{A}_1,\tilde{A}_2,\tilde{A}_3,\tilde{A}_4)$ respectively, to their Fermionic counter part. One can obtain another two sets of equations by repeating the same steps as before, but starting with the KK relations of the purely Fermionic amplitudes. In principle these four sets of linear relations should allow us to express the four Fermionic amplitudes, $A_{6\psi}(\bar{1}2\bar{3}4\bar{5}6)$, $A_{6\psi}(\bar{1}6\bar{3}2\bar{5}4)$, $A_{6\psi}(\bar{1}4\bar{3}2\bar{5}6)$ and $A_{6\psi}(\bar{1}6\bar{3}4\bar{5}2)$, to the four basis amplitude. However, the four linear relation has only rank $3$ instead of $4$!  This immediately leads to the fact that one has a new amplitude relation! 
\eq
\sum_{i=1}^4a_i\tilde{A}_i=0\,,
\eqe
where $a_i$ are coefficient functions that only depend on the kinematic invariants.  As discussed in the previous subsection \ref{ABJMandBCJsection}, the BCJ amplitude relations can be used to reduce the number of independent amplitudes down to the basis amplitudes discussed above. Thus any new relations among these basis amplitudes are beyond that implied by the color kinematic duality of ABJM. Indeed one finds that this is precisely the extra relation obtained from the BCJ relations of BLG theory. 

\subsection{Structure beyond BCJ and bonus relations}
\label{beyondBCJ}
As a last remark, we should address the existence of amplitude relations that goes beyond those of BCJ and bonus relations. While these seem to have no clear purpose in the color-kinematics duality of BLG theory, one could speculate that they might have some role to play in the various puzzles that occur when applying color-kinematics duality to ABJM theory. At six and eight points, we find that there exists further relations in both BLG and and ABJM theory, such that the true basis of partial amplitudes is the same for both theories, implying the existence of a bijective non-trivial map between the two types of amplitudes. This is a highly unexpected result that goes against the intuition that BLG is a special case and ABJM is the general case. For example, setting the gauge group ranks $N_1=N_2=2$ seems like an irreversible operation that converts ABJM to BLG. Likewise BLG has only non-planar tree partial amplitudes, making it difficult to imagine how these can be converted into planar ABJM amplitudes. Nevertheless, new relations exists through eight points and most likely to all multiplicity. Whether they give rise to identical basis sizes for ABJM and BLG at ten points and beyond, we leave as an open problem.

We will not give the extra relations here, as we do not have analytical formulas for them, only numerical proof of their existence. For this numerical proof we must introduce a new concept, or object, that measures what we define as the ``true basis'' of partial amplitudes. Just like the BCJ relations are determined from the $\Theta$ matrix, the true basis of partial amplitudes are determined from the (state)$\times$(partial amplitude) matrix. That is, we rewrite the color-dressed superamplitude\footnote{For non-supersymmetric theories one can similarly consider  $\mathbb{A}_h^{\phantom{h}\sigma}$, only the collective state index $h$ is not contacted with $ \chi^h$.)} as
\eq
{\cal A}_n=\sum_{\sigma,h} \mathbb{A}_h^{\phantom{h}\sigma} c_\sigma \chi^h\,,
\eqe
where the index $\sigma$ runs over all distinct partial amplitudes and $h$ runs over all possible external state configurations, $c_\sigma$ are all the color factors of the corresponding partial amplitudes, and $\chi^h$ collects all non-vanishing products of the Grassmann-odd parameters of the external states. The matrix $\mathbb{A}_h^{\phantom{h}\sigma}$ is simply a table of the individual component amplitudes organized according to their (state)$\times$(partial amplitude) structure. Any relations that the partial amplitudes satisfy must translate into linear relations on the column vectors of $\mathbb{A}_h^{\phantom{h}\sigma}$, for a fixed kinematical point. Thus the rank of this matrix is the true basis of partial amplitudes. The rank is straightforward to compute numerically; for both ABJM and BLG, we find a rank of 2 and 14 for the three-dimensional six- and eight-point amplitudes, respectively. (At four-points the rank is trivially one.) Note that these bases are smaller than those derived from either the BCJ or bonus relations.  Since the rank is the same for both theories, and since we have the surjective maps (\ref{BLGfromABJM6pts}) and (\ref{BLGfromABJM8pts}), the information content of the ABJM and BLG partial amplitude must be the same up to eight points. The corresponding amplitude relations, as well as the map from BLG to ABJM partial amplitudes, should be simple linear identities of the amplitudes with coefficients that depend only on the kinematic invariants.

%%%%%%%%%%%%%%%%%%%%%%%%%%%%%%%%%%%%%%%
 \section{Conclusion}
%%%%%%%%%%%%%%%%%%%%%%%%%%%%%%%%%%%%%%

In this paper we have studied the structure of pure bi-fundamental matter amplitudes in gauge theories, with or without propagating gauge fields on internal lines. We followed the known procedure~\cite{Gustavsson, BaggerLambert,VanRaamsdonk:2008ft}  of embedding the bi-fundamental color structure in a three-algebra formulation. The fundamental identitiy, or three-algebra Jacobi identity, allowed us to construct non-trivial identities for the pure matter partial amplitudes, in close analogy to the Kleiss-Kuijf relations~\cite{KK} in Yang-Mills theory. Since these relations are solely due to the algebraic properties of the structure constants, they hold for bi-fundamental matter amplitudes in general gauge theories. The relations depend only on the symmetries and fundamental identitiy of the corresponding three-algebra structure constants, but not on the detailed Lagrangian nor on the dimension of spacetime. For a particular simple class of these identities two independent proofs are given in the context of the $\mathcal{N}=6$ ABJM theory. The first proof uses a three-dimensional variant~\cite{Gang} of BCFW recursion, and the second uses the twistor-string-like amplitude representation of ref.~\cite{SangminYt}. In addition, we construct graphical representations and operations that can in principle be used to prove any given Kleiss-Kuijf-like identity in ABJM theory.

Using the three-algebra construction, we explored the possible existence of color-kinematics duality in a general setting. We find that for bi-fundamental matter amplitudes, BCJ amplitude relations~\cite{BCJ} exist in three and two dimensions; and not in $D>3$. Furthermore, the number of independent amplitudes under such relations critically depends on the symmetry properties of the three-algebra structure constants. Contrary to previous expectations, we find that only three-algebra theories with totally antisymmetric structure constants, such as the $\mathcal{N}=8$ BLG theory, admit BCJ relations for general multiplicity, whereas general three-dimensional bi-fundamental theories, such as ABJM theory, fail at this starting at eight points. This result was unexpected since SO(4) BLG theory can be considered to be a special case of ABJM with SU(2)$\times$SU(2) Lie algebra~\cite{VanRaamsdonk:2008ft}. We use generalized gauge transformations~\cite{BCJ,BCJLoop} to show that the invariant partial amplitudes of the two types of theories are simply related at four and six points, but starting at eight points this is no longer true. The previous low-multiplicity results in the literature \cite{Till,HenrikYt} were observations that generalize for BLG-like theories, but not for general bi-fundamental theories in three dimensions. 

We have explicitly verified that the double-copy results obtained from BLG theory through eight points indeed matches with the supergravity amplitudes obtained from either BCFW recursion or dimensional reduction. Note that while BCJ amplitude relations are absent for ABJM amplitudes beyond six points, numerators exists that satisfy proper symmetries and fundamental identity. However, by squaring these duality-satisfying numerators, one does not obtain correct gravity amplitudes at eight points. Interestingly, for Yang-Mills theory, the formal proof showing that gravity amplitudes are obtained by the double-copy of duality-satisfying numerators~\cite{Bern:2010yg}, did not rely on the existence of BCJ amplitude relations. This suggest that a detailed study of an analog proof for three-algebra theories would be rewarding, as identifying the subtle difference between the two cases might lead to a remedy for the double copy of ABJM theory.

Novel BCJ relations for bi-fundamental theories of ABJM-type theories re-emerges upon dimensional reduction down to $D=2$. We obtain S-matrix elements of this theory by dimensionally reducing the ABJM amplitudes to $D=2$, mainly working with kinematics corresponding to alternating light-like momenta. This choice of kinematics allows us to obtain two-dimensional tree amplitudes without encountering explicit collinear and soft divergences.  At six points, even though the reduced ABJM amplitudes are non-vanishing, the gravity amplitudes obtained from the BCJ double-copy manifestly vanish.  At eight points, using the fact that BLG partial amplitudes can be obtained as a linear combination of ABJM ones, we have explicitly verified that the double-copy results derived from the BLG color-kinematics duality vanish. The results are cross-checked using higher-dimensional supergravity amplitudes evaluated near $D=2$ kinematics. These vanishings are expected in an integrable theory, where $(n>4)$-point amplitudes should vanish unless they are evaluated on kinematics corresponding to a factorization channel. Moreover, we find that the four-point $D=2$ ABJM amplitude satisfies the Yang-Baxter equation. A corresponding $D=2$ supergravity Yang-Baxter equation is obtained from the ABJM one via the double copy. Thus our results support the existence of integrability in two-dimensional maximal  ${\cal N}=16$ supergravity~\cite{Nicolai:1987kz, Nicolai:1998gi}. Since the observed vanishings are tree-level results, we find the same vanishings in any supersymmetric truncation of supergravity. However, the check of the Yang-Baxter equation is a loop-level result, and thus we only considered the ${\cal N}=12$ and ${\cal N}=16$ theories.

Note that while $D=2$ BCJ amplitude relations can be found for ABJM theory beyond six points, the corresponding double-copy result at eight points does not vanish, contrary to the correct behavior of pure supergravity tree amplitudes. It is interesting to ask whether such discrepancy is indicative of the need to have further structure imposed on the kinematic numerators in order to obtain the correct supergravity amplitude. Certainly, further hidden amplitude relations exists, supporting this idea. Another interesting question to ask is whether or not the double-copy of ABJM theory is computing an amplitude in a deformed version of $\mathcal{N}=12$ supergravity. Note that while uniqueness of $\mathcal{N}>8$ supergravity theories in $D=3$ has been proven~\cite{deWit:1992up}, to our knowledge a similar statement has not been proven for $D=2$.

At six points, we initially find that BLG theory has one more independent BCJ relation than what is observed in ABJM theory. This halfway result is rather surprising, since the six-point ABJM partial amplitudes can be obtained from the corresponding BLG ones via supersymmetry truncation. Indeed, we find that there exists a hidden six-point relation in ABJM theory that can be seen as a bonus relation arising from the improved asymptotic behavior of the BCFW deformation. Interestingly, such improved behavior, and hence the presence of bonus relations, are present beyond eight points, even though corresponding BCJ relations in ABJM are absent. The precise structure of higher-multiplicity bonus relations, and whether or not they can be related to the BCJ relations for BLG theory, is an interesting open problem. Going beyond the BCJ and bonus relations there are further unexplored structures in ABJM and BLG theories. By introducing a (states)$\times$(partial amplitudes) matrix we find that the true number of independent partial amplitudes is smaller that what the known amplitude relations give. Surprisingly, up to eight points the counts of truly independent partial amplitudes are the same for BLG and ABJM, suggesting a bijective relationship between the partial tree amplitudes of the two theories, which do not follow from the respective gauge group structures. 

In this paper we have demonstrated the usefulness of the three-algebra formulation of bi-fundamental matter in the context of scattering amplitudes. This suggests that a study of more general amplitudes and theories that admit three-algebra structures may be fruitful. An example close to the current considerations is to decouple one of the gauge fields in the bi-fundamental theory. This gives a conventional gauge theory with a simple gauge group and fundamental matter. Hence, one can also use three-algebra structure constants as a bookkeeping device for quark amplitudes. The fact that the three-algebra BCJ relations are only valid in $D\le3$ dimensions is a peculiar feature. It should be better understood in from the perspective that Chern-Simons-matter theory is a theory of membranes. In particular, the (weak-weak) double-copy formula that relates BLG theory to supergravity could lead to new insights in string theory.

%%%%%%%%%%%%%%%%%%%%%%%%%%%%%%%%%%%%%%%%%%%%%%%%
\section{Acknowledgement}
%%%%%%%%%%%%%%%%%%%%%%%%%%%%%%%%%%%%%%%%%%%%%%%%
We thank Zvi Bern, Heng-Yu Chen, Gregory Korchemsky, Neil Lambert and Radu Roiban for helpful discussions on these topics. We thank Joonho Kim for independent numerical checks of KK identities and critical reading of the manuscript. Y-t.H. is thankful for the hospitality of the Perimeter Institute, where part of this work was done. Research at the Perimeter Institute is supported in part by the Government of Canada through NSERC and by
the Province of Ontario through MRI. The work of Y-t.H. is supported by the US DoE grant DE-SC0007859. The work of S.L. is supported by the National Research Foundation of Korea (NRF) Grants 
2012R1A1B3001085 and 2012R1A2A2A02046739.

 \appendix
 %%%%%%%%%%%%%%%%%%%%%%%%%%%%%%%%%%%%%%
 \section{Large-$z$ behavior of Grassmannian integrals \label{LargeZ}}
 %%%%%%%%%%%%%%%%%%%%%%%%%%%%%%%%%%%%%%
 Let us begin by analyzing the improved large-$z$ behavior of the Grassmannian integral. We begin by reviewing the result in the setup of orthogonal Grassmannian.
  
 Under a BCFW deformation, the two selected spinors $\Lambda_1$ and $\Lambda_i$ are rotated by an SO(2) matrix $R(z)$, with $R^{T}R=1$. Under this deformation, the Grassmannian integral becomes 
 \eq
 \int \frac{dC^{2k^2}}{\prod_{i=1}^kM_i(C)}\delta(CC^T)\delta(CR\Lambda)=\int \frac{d\tilde{C}^{2k^2}}{\prod_{i=1}^kM_i(\hat{C}(z))}\delta(\tilde{C}\tilde{C}^T)\delta(\tilde{C}\Lambda)\,,
 \eqe
where we have redefined $\tilde{C}=C*R$ and $\hat{C}=\tilde{C}R^T$. Thus the only $z$-dependence is now in the minors of the Grassmannian integral, and they enter in two columns of the Grassmannian. Defining $C^\pm\equiv C_1\pm iC_i$:
\eq
\hat{C}_1(z)=\frac{z}{2}C^++\frac{1}{2z}C^-,\;\hat{C}_i(z)=-\frac{zi}{2}C^++\frac{i}{2z}C^-\,.
\eqe
One can easily see that any minor that contains either $\hat{C}_1(z)$ or $\hat{C}_i(z)$ scales linearly in $z$ as $z\rightarrow\infty$, with the exception for the case that both are present, such as $M_1$, for which it scales as a constant. Thus generically the $z$-dependence of the Grassmannian integral scales as $z^{-(i-1)}$ at large $z$. 

Since for $n=4,6$ the Bosonic delta functions of the Grassmannian completely fixes the integral, there are no explicit integration to be done and the large-$z$ analysis can be done straight forwardly:
\eqa
\nonumber n=4&:& (1,2)\rightarrow \frac{1}{z},\;(1,3)\rightarrow \frac{1}{z^2}\,,\\
n=6&:& (1,2)\rightarrow \frac{1}{z},\;(1,3)\rightarrow \frac{1}{z^2},\;(1,4)\rightarrow \frac{1}{z^3}\,.
\eqae

For $n=8$ and beyond, the Grassmannian integral becomes a contour integral. Since the integral is localized on the zeroes of the minor, the solution might alter the large-$z$ dependence of the remaining minors. As an example, consider the shifting $(1,5)$, the large-$z$ behavior of each minor is given as 
\eqa
\nonumber M_1(z)&=&\frac{z}{2}{\rm det}[C_+C_2\cdots C_4]+\frac{1}{2z}{\rm det}[C_-C_2\cdots C_4]\,,\\
\nonumber  M_2(z)&=&-i\frac{z}{2}{\rm det}[C_2\cdots C_4C_+]+\frac{i}{2z}{\rm det}[C_2\cdots C_4C_-]\,.\\
\eqae
One sees that when evaluated on the zero of $M_1(z)$ at $z\rightarrow\infty$, the leading $\mathcal{O}(z)$ pieces of the minor $M_2(z)$ vanishes identically, {\it i.e.} at large $z$ the zeroes of $M_1(z)$ and $M_2(z)$ degenerates, and the overall large-$z$ behavior becomes $z^{-2}$ if the residue in question correspond to the zero of $M_1(z)$ or $M_2(z)$. Thus we have 
\eqa
n=8&:& (1,2)\rightarrow \frac{1}{z},\;(1,3)\rightarrow \frac{1}{z^2},\;(1,4)\rightarrow \frac{1}{z^3}\,\;(1,5)\rightarrow \frac{1}{z^2}\,.
\eqae 
On the other hand, if the number of integration is more than one, then the above analysis does not hold either. From the formula $(k-2)(k-3)/2$ we see that at 10 points, more care needs to be taken.


\begin{thebibliography}{99}

 \bibitem{BLG1} 
  A.~Gustavsson,
  %``Algebraic structures on parallel M2-branes,''
  Nucl.\ Phys.\ B {\bf 811}, 66 (2009)
  [arXiv:0709.1260 [hep-th]].
  %%CITATION = ARXIV:0709.1260;%%
  %586 citations counted in INSPIRE as of 12 Apr 2013
  
\bibitem{BLG2} 
  J.~Bagger and N.~Lambert,
  %``Gauge symmetry and supersymmetry of multiple M2-branes,''
  Phys.\ Rev.\ D {\bf 77}, 065008 (2008)
  [arXiv:0711.0955 [hep-th]].
  %%CITATION = ARXIV:0711.0955;%%
  %637 citations counted in INSPIRE as of 12 Apr 2013
   \bibitem{ABJM}
  O.~Aharony, O.~Bergman, D.~L.~Jafferis and J.~Maldacena,
  %``N=6 superconformal Chern-Simons-matter theories, M2-branes and their
  %gravity duals,''
  JHEP {\bf 0810}, 091 (2008) 
  [arXiv:0806.1218 [hep-th]].
  %%CITATION = JHEPA,0810,091;%%
  \bibitem{Gustavsson}
  A.~Gustavsson,
  %``Selfdual strings and loop space Nahm equations,''
  JHEP {\bf 0804}, 083 (2008)
  [arXiv:0802.3456 [hep-th]].
  %%CITATION = ARXIV:0802.3456;%%  
 
 \bibitem{VanRaamsdonk:2008ft} 
  M.~Van Raamsdonk,
  %``Comments on the Bagger-Lambert theory and multiple M2-branes,''
  JHEP {\bf 0805}, 105 (2008)
  [arXiv:0803.3803 [hep-th]].
  %%CITATION = ARXIV:0803.3803;%%
  %196 citations counted in INSPIRE as of 04 Apr 2013 
  
\bibitem{BaggerLambert} 
  J.~Bagger and N.~Lambert,
 % ``Three-Algebras and N=6 Chern-Simons Gauge Theories,''
  Phys.\ Rev.\ D {\bf 79}, 025002 (2009)
  [arXiv:0807.0163 [hep-th]].
  %%CITATION = ARXIV:0807.0163;%%


  
 \bibitem{HLLLP} 
  K.~Hosomichi, K.~-M.~Lee, S.~Lee, S.~Lee and J.~Park,
  %``N=5,6 Superconformal Chern-Simons Theories and M2-branes on Orbifolds,''
  JHEP {\bf 0809}, 002 (2008)
  [arXiv:0806.4977 [hep-th]]. 
 
\bibitem{ABJ} 
  O.~Aharony, O.~Bergman and D.~L.~Jafferis,
  %``Fractional M2-branes,''
  JHEP {\bf 0811}, 043 (2008)
  [arXiv:0807.4924 [hep-th]].
  %%CITATION = ARXIV:0807.4924;%%
  %239 citations counted in INSPIRE as of 22 Apr 2013

  
\bibitem{Till} 
  T.~Bargheer, S.~He and T.~McLoughlin,
  %``New Relations for Three-Dimensional Supersymmetric Scattering Amplitudes,''
  Phys.\ Rev.\ Lett.\  {\bf 108}, 231601 (2012)
  [arXiv:1203.0562 [hep-th]].
  %%CITATION = ARXIV:1203.0562;%%
  %14 citations counted in INSPIRE as of 20 Apr 2013  
  
\bibitem{BCJ} 
  Z.~Bern, J.~J.~M.~Carrasco and H.~Johansson,
 % ``New Relations for Gauge-Theory Amplitudes,''
  Phys.\ Rev.\ D {\bf 78}, 085011 (2008)
  [arXiv:0805.3993 [hep-ph]].
  
\bibitem{Tree}
 M. Kiermaier, Amplitudes 2010, http://www.strings.ph.qmul.ac.uk/$\sim$theory/Amplitudes2010/;\\
%
 N.~E.~J.~Bjerrum-Bohr, P.~H.~Damgaard, T.~Sondergaard and P.~Vanhove,
  %``The Momentum Kernel of Gauge and Gravity Theories,''
  JHEP {\bf 1101}, 001 (2011)
  [arXiv:1010.3933 [hep-th]].
  %%CITATION = ARXIV:1010.3933;%%
  %41 citations counted in INSPIRE as of 20 Apr 2013
  %
  
 \bibitem{Tree2}
  C.~R.~Mafra, O.~Schlotterer and S.~Stieberger,
  %``Explicit BCJ Numerators from Pure Spinors,''
  JHEP {\bf 1107}, 092 (2011)
  [arXiv:1104.5224 [hep-th]];\\
  %%CITATION = ARXIV:1104.5224;%%
  %31 citations counted in INSPIRE as of 20 Apr 2013
  %
 C.~-H.~Fu, Y.~-J.~Du and B.~Feng,
  %``An algebraic approach to BCJ numerators,''
  JHEP {\bf 1303}, 050 (2013)
  [arXiv:1212.6168 [hep-th]];\\
  %%CITATION = ARXIV:1212.6168;%%
  %4 citations counted in INSPIRE as of 20 Apr 2013
   %  
  M.~Tolotti and S.~Weinzierl,
  %``Construction of an effective Yang-Mills Lagrangian with manifest BCJ duality,''
  arXiv:1306.2975 [hep-th].
  %%CITATION = ARXIV:1306.2975;%%
  
\bibitem{BCJLoop}
Z.~Bern, J.~J.~M.~Carrasco and H.~Johansson,
%``Perturbative Quantum Gravity as a Double Copy of Gauge Theory,''
Phys.\ Rev.\ Lett.\  {\bf 105}, 061602 (2010)
[arXiv:1004.0476 [hep-th]].
%%CITATION = PRLTA,105,061602;%%

\bibitem{Bern:2010yg} 
  Z.~Bern, T.~Dennen, Y.~-t.~Huang and M.~Kiermaier,
 % ``Gravity as the Square of Gauge Theory,''
  Phys.\ Rev.\ D {\bf 82}, 065003 (2010)
  [arXiv:1004.0693 [hep-th]].
  %%CITATION = ARXIV:1004.0693;%%

  
\bibitem{BCDJR}
  Z.~Bern, J.~J.~M.~Carrasco, L.~J.~Dixon, H.~Johansson and R.~Roiban,
  %``The Complete Four-Loop Four-Point Amplitude in N=4 Super-Yang-Mills Theory,''
  Phys.\ Rev.\ D {\bf 82}, 125040 (2010)
  [arXiv:1008.3327 [hep-th]].
  %%CITATION = ARXIV:1008.3327;%%
  %46 citations counted in INSPIRE as of 20 Apr 2013  
  
\bibitem{N=4SG} 
Z.~Bern, S.~Davies, T.~Dennen and Y.~-t.~Huang,
  %``Absence of Three-Loop Four-Point Divergences in N=4 Supergravity,''
  Phys.\ Rev.\ Lett.\  {\bf 108}, 201301 (2012)
  [arXiv:1202.3423 [hep-th]];\\
  %%CITATION = ARXIV:1202.3423;%%
  %26 citations counted in INSPIRE as of 20 Apr 2013
  %
  Z.~Bern, S.~Davies, T.~Dennen and Y.~-t.~Huang,
  %``Ultraviolet Cancellations in Half-Maximal Supergravity as a Consequence of the Double-Copy Structure,''
  Phys.\ Rev.\ D {\bf 86}, 105014 (2012)
  [arXiv:1209.2472 [hep-th]].
  %%CITATION = ARXIV:1209.2472;%%
  %10 citations counted in INSPIRE as of 20 Apr 2013  
  
 
\bibitem{stringtheoryBCJ}
N.~E.~J.~Bjerrum-Bohr, P.~H.~Damgaard and P.~Vanhove,
%``Minimal Basis for Gauge Theory Amplitudes,''      
Phys.\ Rev.\ Lett.\  {\bf 103}, 161602 (2009)
[0907.1425 [hep-th]];\\
%%CITATION = PRLTA,103,161602;%%
%
S.~Stieberger,
  %``Open & Closed vs. Pure Open String Disk Amplitudes,''
  arXiv:0907.2211 [hep-th];\\
  %%CITATION = ARXIV:0907.2211;%%
  %99 citations counted in INSPIRE as of 25 May 2013
%
  C.~R.~Mafra and O.~Schlotterer,
  %``The Structure of n-Point One-Loop Open Superstring Amplitudes,''
  arXiv:1203.6215 [hep-th];\\
  %%CITATION = ARXIV:1203.6215;%%
  %5 citations counted in INSPIRE as of 25 May 2013
%
O.~Schlotterer and S.~Stieberger,
  %``Motivic Multiple Zeta Values and Superstring Amplitudes,''
  arXiv:1205.1516 [hep-th];\\
  %%CITATION = ARXIV:1205.1516;%%
  %11 citations counted in INSPIRE as of 25 May 2013
  %
%
J.~Broedel, O.~Schlotterer and S.~Stieberger,
  %``Polylogarithms, Multiple Zeta Values and Superstring Amplitudes,''
  arXiv:1304.7267 [hep-th].
  %%CITATION = ARXIV:1304.7267;%%
  %2 citations counted in INSPIRE as of 25 May 2013

  
\bibitem{E8} 
  N.~Marcus and J.~H.~Schwarz,
 % ``Three-Dimensional Supergravity Theories,''
  Nucl.\ Phys.\ B {\bf 228}, 145 (1983).
  %%CITATION = NUPHA,B228,145;%%    
  

\bibitem{HenrikYt} 
  Y.~-t.~Huang and H.~Johansson,
  %``Equivalent D=3 Supergravity Amplitudes from Double Copies of Three-Algebra and Two-Algebra Gauge Theories,''
  arXiv:1210.2255 [hep-th].
  %%CITATION = ARXIV:1210.2255;%%
  %7 citations counted in INSPIRE as of 06 Jun 2013
  
\bibitem{uniqueBLG}
Nagy, P.-A.\ 2007,  arXiv:0712.1398:\\
 G.~Papadopoulos,
  %``M2-branes, 3-Lie Algebras and Plucker relations,''
  JHEP {\bf 0805}, 054 (2008)
  [arXiv:0804.2662 [hep-th]];\\
  %%CITATION = ARXIV:0804.2662;%%
  %156 citations counted in INSPIRE as of 06 Jun 2013   
  J.~P.~Gauntlett and J.~B.~Gutowski,
  %``Constraining Maximally Supersymmetric Membrane Actions,''
  JHEP {\bf 0806}, 053 (2008)
  [arXiv:0804.3078 [hep-th]].
  %%CITATION = ARXIV:0804.3078;%%
  %150 citations counted in INSPIRE as of 06 Jun 2013
  
\bibitem{KLT} 
  H.~Kawai, D.~C.~Lewellen and S.~H.~H.~Tye,
  %``A Relation Between Tree Amplitudes of Closed and Open Strings,''
  Nucl.\ Phys.\ B {\bf 269}, 1 (1986).
  %%CITATION = NUPHA,B269,1;%%
  %331 citations counted in INSPIRE as of 20 Apr 2013    
  
\bibitem{Nicolai:1987kz} 
  H.~Nicolai,
  %``The Integrability Of N=16 Supergravity,''
  Phys.\ Lett.\ B {\bf 194}, 402 (1987);\\
  %%CITATION = PHLTA,B194,402;%%
  %79 citations counted in INSPIRE as of 03 Jun 2013
   H.~Nicolai and N.~P.~Warner,
  %``The Structure Of N=16 Supergravity In Two-dimensions,''
  Commun.\ Math.\ Phys.\  {\bf 125}, 369 (1989).
  %%CITATION = CMPHA,125,369;%%
  %35 citations counted in INSPIRE as of 03 Jun 2013
  
  \bibitem{Nicolai:1998gi} 
  H.~Nicolai and H.~Samtleben,
  %``Integrability and canonical structure of d = 2, N=16 supergravity,''
  Nucl.\ Phys.\ B {\bf 533}, 210 (1998)
  [hep-th/9804152].
  %%CITATION = HEP-TH/9804152;%%
  %21 citations counted in INSPIRE as of 03 Jun 2013  
  
\bibitem{DDM}
V.~Del Duca, L.~J.~Dixon and F.~Maltoni,
  %``New color decompositions for gauge amplitudes at tree and loop level,''
  Nucl.\ Phys.\ B {\bf 571}, 51 (2000)
  [hep-ph/9910563].
  %%CITATION = HEP-PH/9910563;%%  
  
\bibitem{KK} 
  R.~Kleiss and H.~Kuijf,
 % ``Multi - Gluon Cross-sections And Five Jet Production At Hadron Colliders,''
  Nucl.\ Phys.\ B {\bf 312}, 616 (1989).
  %%CITATION = NUPHA,B312,616;%%  
  
\bibitem{Kemal} 
  H.~Ita, and K.~Ozeren,
  %``Colour Decompositions of Multi-quark One-loop QCD Amplitudes,''
  JHEP {\bf 1202}, 118 (2012)
  [arXiv:1111.4193 [hep-ph]].
  %%CITATION = ARXIV:1111.4193;%%
  %8 citations counted in INSPIRE as of 29 Mar 2013
  
  
\bibitem{Bargheer:2010hn} 
  T.~Bargheer, F.~Loebbert and C.~Meneghelli,
  %``Symmetries of Tree-level Scattering Amplitudes in N=6 Superconformal Chern-Simons Theory,''
  Phys.\ Rev.\ D {\bf 82}, 045016 (2010)
  [arXiv:1003.6120 [hep-th]].
  %%CITATION = ARXIV:1003.6120;%%

 \bibitem{Gang} 
  D.~Gang, Y.-t.~Huang, E.~Koh, S.~Lee and A.~E.~Lipstein,
 % ``Tree-level Recursion Relation and Dual Superconformal Symmetry of the ABJM Theory,''
  JHEP {\bf 1103}, 116 (2011)
  [arXiv:1012.5032 [hep-th]].
  
\bibitem{SangminYt} 
  Y.~-t.~Huang and S.~Lee,
  %``A new integral formula for supersymmetric scattering amplitudes in three dimensions,''
  Phys.\ Rev.\ Lett.\  {\bf 109}, 191601 (2012)
  [arXiv:1207.4851 [hep-th]].
  %%CITATION = ARXIV:1207.4851;%%    

  
\bibitem{Feng:2010my} 
  B.~Feng, R.~Huang and Y.~Jia,
 % ``Gauge Amplitude Identities by On-shell Recursion Relation in S-matrix Program,''
  Phys.\ Lett.\ B {\bf 695}, 350 (2011)
  [arXiv:1004.3417 [hep-th]].
  %%CITATION = ARXIV:1004.3417;%%  
  
 \bibitem{RSVW}
  R.~Roiban, M.~Spradlin and A.~Volovich,
  %``On the tree level S matrix of Yang-Mills theory,''
  Phys.\ Rev.\ D {\bf 70} (2004) 026009
  [hep-th/0403190].
  %%CITATION = HEP-TH/0403190;%%
  %121 citations counted in INSPIRE as of 10 Mar 2013


\bibitem{Lee:2010du} 
  S.~Lee,
  %``Yangian Invariant Scattering Amplitudes in Supersymmetric Chern-Simons Theory,''
  Phys.\ Rev.\ Lett.\  {\bf 105}, 151603 (2010)
  [arXiv:1007.4772 [hep-th]].
  %%CITATION = ARXIV:1007.4772;%%  
  
 \bibitem{Cachazo:2013iaa} 
  F.~Cachazo, S.~He and E.~Y.~Yuan,
  %``Scattering in Three Dimensions from Rational Maps,''
  arXiv:1306.2962 [hep-th].
  %%CITATION = ARXIV:1306.2962;%% 
 \bibitem{Vaman:2010ez} 
  D.~Vaman and Y.~-P.~Yao,
  %``Constraints and Generalized Gauge Transformations on Tree-Level Gluon and Graviton Amplitudes,''
  JHEP {\bf 1011}, 028 (2010)
  [arXiv:1007.3475 [hep-th]].
  %%CITATION = ARXIV:1007.3475;%%
  %17 citations counted in INSPIRE as of 05 Apr 2013  
  
 
 \bibitem{LoopBCJnumerators} 
  J.~J.~Carrasco and H.~Johansson,
  %``Five-Point Amplitudes in N=4 Super-Yang-Mills Theory and N=8 Supergravity,''
  Phys.\ Rev.\ D {\bf 85}, 025006 (2012)
  [arXiv:1106.4711 [hep-th]];\\
  %%CITATION = ARXIV:1106.4711;%%
  %36 citations counted in INSPIRE as of 22 Apr 2013
  %  
   R.~H.~Boels, B.~A.~Kniehl, O.~V.~Tarasov and G.~Yang,
  %``Color-kinematic Duality for Form Factors,''
  JHEP {\bf 1302}, 063 (2013)
  [arXiv:1211.7028 [hep-th]];\\
  %%CITATION = ARXIV:1211.7028;%%
  %
  J.~J.~M.~Carrasco, M.~Chiodaroli, M.~Gunaydin and R.~Roiban,
  %``One-loop four-point amplitudes in pure and matter-coupled N <= 4 supergravity,''
  JHEP {\bf 1303}, 056 (2013)
  [arXiv:1212.1146 [hep-th]];\\
  %%CITATION = ARXIV:1212.1146;%%
  %
  R.~H.~Boels, R.~S.~Isermann, R.~Monteiro and D.~O'Connell,
  %``Colour-Kinematics Duality for One-Loop Rational Amplitudes,''
  JHEP {\bf 1304}, 107 (2013)
  [arXiv:1301.4165 [hep-th]];\\
  %%CITATION = ARXIV:1301.4165;%%
  %
    N.~E.~J.~Bjerrum-Bohr, T.~Dennen, R.~Monteiro and D.~O'Connell,
  %``Integrand Oxidation and One-Loop Colour-Dual Numerators in N=4 Gauge Theory,''
  arXiv:1303.2913 [hep-th];\\
  %%CITATION = ARXIV:1303.2913;%%
  %3 citations counted in INSPIRE as of 22 Apr 2013
  %
  Z.~Bern, S.~Davies, T.~Dennen, Y.~-t.~Huang and J.~Nohle,
  %``Color-Kinematics Duality for Pure Yang-Mills and Gravity at One and Two Loops,''
  arXiv:1303.6605 [hep-th]\,.
  %%CITATION = ARXIV:1303.6605;%%
  %1 citations counted in INSPIRE as of 22 Apr 2013
 
   \bibitem{N>=4SG} 
  Z.~Bern, C.~Boucher-Veronneau and H.~Johansson,
  %``N >= 4 Supergravity Amplitudes from Gauge Theory at One Loop,''
  Phys.\ Rev.\ D {\bf 84}, 105035 (2011)
  [arXiv:1107.1935 [hep-th]];\\
  %%CITATION = ARXIV:1107.1935;%%
  %31 citations counted in INSPIRE as of 23 Jun 2013
  %
  C.~Boucher-Veronneau and L.~J.~Dixon,
  %``N >- 4 Supergravity Amplitudes from Gauge Theory at Two Loops,''
  JHEP {\bf 1112}, 046 (2011)
  [arXiv:1110.1132 [hep-th]].
  %%CITATION = ARXIV:1110.1132;%%
  %23 citations counted in INSPIRE as of 23 Jun 2013
 
\bibitem{allic}
  Allic Sivaramakrishnan, to appear
  
\bibitem{deWit:1992up} 
  B.~de Wit, A.~K.~Tollsten and H.~Nicolai,
 % ``Locally supersymmetric D = 3 nonlinear sigma models,''
  Nucl.\ Phys.\ B {\bf 392}, 3 (1993)
  [hep-th/9208074].
  %%CITATION = HEP-TH/9208074;%%\end{thebibliography}

 \bibitem{ZZ} 
  A.~B.~Zamolodchikov and A.~B.~Zamolodchikov,
  %``Factorized s Matrices in Two-Dimensions as the Exact Solutions of Certain Relativistic Quantum Field Models,''
  Annals Phys.\  {\bf 120}, 253 (1979).
  %%CITATION = APNYA,120,253;%%
  %1077 citations counted in INSPIRE as of 04 Jun 2013  
 
\bibitem{ArkaniHamed:2012nw} 
  N.~Arkani-Hamed, J.~L.~Bourjaily, F.~Cachazo, A.~B.~Goncharov, A.~Postnikov and J.~Trnka,
  %``Scattering Amplitudes and the Positive Grassmannian,''
  arXiv:1212.5605 [hep-th].
  %%CITATION = ARXIV:1212.5605;%%
  %21 citations counted in INSPIRE as of 04 Jun 2013 
  
\bibitem{ABJMBCFW} 
  A.~Brandhuber, G.~Travaglini and C.~Wen,
  %``A note on amplitudes in N=6 superconformal Chern-Simons theory,''
  JHEP {\bf 1207}, 160 (2012)
  [arXiv:1205.6705 [hep-th]];\\
  %%CITATION = ARXIV:1205.6705;%%
  %10 citations counted in INSPIRE as of 04 Jun 2013
  % 
  S.~Caron-Huot and Y.~-t.~Huang,
  %``The two-loop six-point amplitude in ABJM theory,''
  JHEP {\bf 1303}, 075 (2013)
  [arXiv:1210.4226 [hep-th]].
  %%CITATION = ARXIV:1210.4226;%%
  %6 citations counted in INSPIRE as of 04 Jun 2013
 
  \bibitem{CachazoZ} 
  F.~Cachazo, L.~Mason and D.~Skinner,
  %``Gravity in Twistor Space and its Grassmannian Formulation,''
  arXiv:1207.4712 [hep-th].
  %%CITATION = ARXIV:1207.4712;%%
  %7 citations counted in INSPIRE as of 10 Mar 2013
 
\end{thebibliography}
\end{document}